\documentclass[bimj,fleqn]{w-art}
\usepackage{times}
\usepackage{w-thm}
\usepackage[authoryear]{natbib}
\usepackage{color}
\theoremstyle{plain}

\theoremstyle{definition}

\usepackage[]{graphicx}
\usepackage{float}
\usepackage[flushleft]{threeparttable}
\usepackage{array,multirow}
\usepackage{subfigure}
\usepackage{booktabs}
\chardef\bslash=`\\ 

\begin{document}
\DOIsuffix{bimj.200100000}
\Volume{52}
\Issue{61}
\Year{2010}
\pagespan{1}{}
\keywords{Correlated prior process; Cubic B-splines; Piecewise functions; Survival analysis; Weibull distribution.\\
\noindent \hspace*{-4pc} {\small\it}\\
\hspace*{-4pc} {\small\it}\\[1pc]
\noindent\hspace*{-4.2pc} Supporting Information for this article is available from the author or on the WWW under\break \hspace*{-4pc} \underline{http://dx.doi.org/10.1022/bimj.XXXXXXX} (please delete if not
applicable)
}  

\title[Bayesian regularization for flexible baseline hazard functions]{Bayesian regularization for flexible baseline hazard functions in Cox survival models}
\author[Elena L\'azaro {\it{et al.}}]{Elena L\'azaro\footnote{Corresponding author: {\sf{e-mail: elena.lazaro@uv.es}}, Phone: +34-665-813-113}\inst{,1}}
\address[\inst{1}]{Department of Statistics and Operation Research, University of Valencia, 46100 Burjassot, Spain}
\author[]{Carmen Armero\inst{1}}
\author[]{Danilo Alvares\inst{2}}
\address[\inst{2}]{Department of Statistics, Pontificia Universidad Cat{\'o}lica de Chile, 7820436 Macul, Chile}
\Receiveddate{zzz} \Reviseddate{zzz} \Accepteddate{zzz}

\begin{abstract}

\noindent Fully Bayesian methods for Cox models specify a model  for the baseline hazard function. Parametric approaches generally provide monotone estimations. Semi-parametric choices allow for more flexible patterns but they can suffer from overfitting and instability. Regularization methods through  prior distributions with correlated structures usually give reasonable answers to these types of  situations.

We discuss Bayesian regularization for   Cox survival models defined via flexible baseline hazards specified by a mixture of piecewise constant functions  and by a cubic B-spline function. For those ``semi-parametric'' proposals, different prior scenarios    ranging   from prior independence to particular correlated structures are discussed in a real study with micro-virulence data and in an extensive simulation scenario that includes  different data sample and time axis partition sizes in order to capture risk variations. The posterior distribution of the parameters was approximated using Markov chain Monte Carlo methods. Model selection was performed in accordance with the Deviance Information Criteria and the Log Pseudo-Marginal Likelihood.

The results obtained reveal that, in general, Cox models present great robustness in covariate effects and survival estimates independent of the baseline hazard specification. In relation to the ``semi-parametric'' baseline hazard specification, the B-splines hazard function is less dependent on the regularization process than the piecewise specification because it demands a smaller time axis partition to estimate a similar behaviour of the risk.  \\

\end{abstract}

\maketitle                   






\section{Introduction} \label{sec:1}

The Cox proportional hazards model~\citep{cox1972} is the most popular regression model in survival analysis. It expresses the  hazard function $h(t)$ of the survival time of
 each individual in the target population as the product of a common baseline hazard function $h_0(t)$, which determines the shape of $h(t)$, and an exponential
 regression term which includes the relevant covariates.Baseline hazard misspecification can imply a loss of valuable information that is necessary
   to fully report the estimation of the outcomes of interest, such as probabilities  or survival curves~\citep{royston2011}. This issue is especially important in survival
   studies where $h_0(t)$ represents the natural course of a disease or an infection,  or even the control group when comparing  several treatments.

The frequentist estimation of the Cox model focuses on  the   regression coefficients $\boldsymbol{\beta}$, which can be obtained
 without specifying a model for  $h_0(t)$ by using the partial likelihood methodology~\citep{cox1972, houwelingen2014a}. Frequentist Cox
 can also provide a point estimation of $h_0(t)$ by means of the Breslow estimator by plugging the estimate $\hat{\boldsymbol \beta}$ into
 $\boldsymbol \beta$ and point estimations of the survival function via analogues of the Nelson-Aalen  and the Kaplan-Meier estimators~\citep{houwelingen2014a}. Uncertainty
 about these estimates is assessed through confidence intervals which rely on  asymptotics~\citep{andersen1982, tsiatis1981}.

Bayesian analysis of the Cox model  needs to specify a model  for $h_0(t)$~\citep{christensen2011}.
It provides a natural framework to jointly analyse all  the uncertainties  in  the statistical modelling, $h_0(t)$  and   $\boldsymbol{\beta}$, by means of its joint posterior
 distribution. This posterior contains all the relevant information from the study and it is usually the starting   point for the subsequent  estimation and prediction of the outcomes
  of interest.    In this regard, Bayesian inference, unlike frequentist statistics, does not generally use asymptotic arguments to assess the variability of the estimates~\citep{ibrahim2005}. 
Baseline hazard functions can be defined through   parametric or   semi-parametric approaches.    Parametric models give
  restricted shapes which do not allow for the presence of irregular behaviours~\citep{dellaportas1993, kim2000}. Semi-parametric choices
   result in   flexible baseline  shapes~\citep{sahu1997,ibrahim2005} but they may  suffer from overfitting and instability~\citep{breiman1996}. Regularization methods
   modify the  estimation procedures to solve these types of  problems. Frequentist regularization  introduces some changes in  the likelihood function. Bayesian reasoning
   accounts for this issue through  prior distributions.

   In  fully Bayesian studies, the joint posterior distribution   is obtained via Bayes' theorem from the likelihood function and the prior distribution. This is why the prior
    can be considered as the element that regularizes the likelihood   and the reason why the  elicitation of prior distributions  is relevant, particularly in   survival analysis
     when   $h_0(t)$  is defined in terms of flexible modelling. The selection of different baseline hazard functions implies different likelihood specifications  and  different
     prior distributions,  which for a given $h_0(t)$ can range from prior independence to some particular correlated prior distributions in order to avoid overfitting.
     
The prior distribution is a fundamental element of Bayesian methodology that serves as a starting point for any Bayesian study. In general terms, prior distributions can be non-informative (or almost) or informative.
     Non-informative distributions try to play a neutral role in the inferential process and give full prominence to the data. Informative prior distributions are relevant in the statistical procedure, especially in studies with little data.  In these cases, it is especially important to add sensitivity analyses to the study in order to check the robustness of the results with regard to the elicited prior distributions. A non-robust prior distribution can be the source of important biases in the results~\citep{berger1994, ibrahim2011}.

Regularization methods   originated in mathematical settings and
were fruitfully and widely disseminated to the world of statistics, providing many different approaches and concepts~\citep{girosi1993, benner2010}. All of them share the general and easy idea of
combining the aim of simultaneously looking for a function that is close to the data and also smooth. The statistical
background on the subject, Bayesian and mainly frequentist, is so extensive that reviewing and understanding the concepts, issues and relationships
 within each statistical approach is beyond the scope of this paper (see~\cite{bickel2006} for an up-to-date review).

We have a twofold objective in this paper: to assess the role of the   specification of  $h_0(t)$ and to discuss  the effect of the Bayesian regularization    in the case of  semi-parametric  modelling of  $h_0(t)$. We consider two   flexible specifications for $h_0(t)$ that allow for multimodal patterns: a mixture of piecewise constant functions~\citep{sahu1997} and a cubic  B-spline   function~\citep{hastie2009}. A Weibull baseline hazard
distribution, the usual parametric proposal  for $h_0(t)$, is also included for  comparison purposes. The baseline risk functions with which we work in this
paper, as well as the different prior distributions considered, are methodological proposals known in Bayesian literature that, as far as we know, have not been compared to date.
The novelty of our work lies in this comparison, which we carry out through different criteria of goodness for the estimated models.

Piecewise constant functions for  $h_0(t)$  have a long tradition in Bayesian survival~\citep{kalbfleisch1973,sahu1997}. Relevant proposals that induce correlated structures in the
 subsequent   prior distribution for the coefficients of the piecewise functions are based on discrete time martingale processes, Gamma process priors, and
  random-walk priors~\citep{ibrahim2005}. Cubic B-spline functions for  $h_0(t)$ are far more recent. They come from the world of generalized
  additive models~\citep{hastie2009} and are widely used in  spatial and spatio-temporal analysis. Their use in survival settings was
  proposed by~\cite{cai2002}, ~\cite{fahrmeir2003} and ~\cite{Sharef2010}
   by means of first or second random walk smoothness priors with Gaussian errors. Other flexible  models for baseline hazard
    functions are based on low-rank thin plate linear splines~\citep{murray2016}, truncated basis splines~\citep{crainiceanu2005}, M-splines~\citep{ramsay1988}   or the popular
     P-splines~\citep{durban2015}, particular B-splines with penalties in the frequentist setting.

The remainder of this article is organized as follows. Section~\ref{sec:2} introduces  Weibull, piecewise constant and B-spline baseline hazard functions for the Cox model as well as
the most common   prior distributions for these scenarios. Section~\ref{sec:3} explores non-penalized frequentist and Bayesian estimation with
piecewise constant and cubic B-spline functions and discusses Bayesian   regularization for   $h_0(t)$  for  a real microbial virulence study. Section~\ref{sec:4} explores various simulation scenarios to  compare the behaviour of the different $h_0(t)$ and
 prior distributions. These last two sections deal with  regularization in the semi-parametric  settings with regard to  different partitions of the time axis in which a mixture of piecewise
 constant and cubic B-spline functions are defined. The article ends with some general remarks and conclusions.


\section{Cox proportional hazards model} \label{sec:2}

Let $T_i$ be the random variable that accounts for the observed event time for  individual $i$, $i=1,\ldots,n$. It is defined as $T_i =\mbox{min}(T_i^*, C_i)$, the minimum between the
 true failure time for individual $i$, $T_i^*$, and the right-censoring time, $C_i$, determined by the end of the study (administrative censoring). The   event indicator
  $\delta_i =I(T_i^* \leq C_i)$ is 1 if the survival time is observed, and 0 otherwise. We assume that $T_i^*$  is a continuous random variable with survival function, $S_i(t)=P(T_i^* >t)$, and
   hazard function $h_i(t)$,  $\forall t \geq 0$, which represents the instantaneous rate of occurrence of the event.

The Cox proportional hazards model for $T_i^*$ expresses the hazard function for individual $i$  in the form
\begin{align}
\label{eqn:bayesiancox}
h_{i}(t \mid  h_{0}, \boldsymbol{x}_{i},   \boldsymbol{\beta}) &=h_{0}(t)\,\mbox{exp}\{ \boldsymbol{x}_{i}^{\prime}\,\boldsymbol{\beta}\}, 					
\end{align}

\noindent where $\boldsymbol{x}_i$ is a vector of $J$ covariates,    $ \boldsymbol{\beta}$ is the vector of  regression coefficients,  and
$h_{0}(t)$ is  the baseline hazard function.

\subsection{Baseline hazard function}

We discuss three different proposals for $h_0(t)$, a Weibull hazard function  and two semi-parametric
ones, namely a mixture of piecewise constant functions  and a cubic  B-spline function. \vspace*{0.15cm}\\

\noindent \textbf{\textit{Weibull function}} \vspace*{0.1cm}\\
The most popular parametric model for $h_0(t)$ is  the  Weibull distribution, $\mbox{We}(\alpha,\lambda)$, with shape parameter $\alpha>0$ and
  scale   $\lambda>0$,  and baseline  hazard function
\begin{equation}
\label{eqn:weibull}
h_{0}(t \mid  \alpha, \lambda)=\lambda\, \alpha\, t^{\alpha-1},\,\, t>0.
\end{equation}

This is a traditional model for survival data in biometrical applications. It is highly suitable thanks  to its computational simplicity, especially in
small-sample settings, but it has no flexibility to represent risks away from monotonicity~\citep{lee2016}.    \vspace*{0.1cm}\\

 \noindent \textbf{\textit{Mixture of piecewise constant functions}} \vspace*{0.1cm}\\
Piecewise functions are defined by polynomial functions.  They generate a flexible framework for modelling survival data  with a long tradition~\citep{henschel2009,ibrahim2005} in the Bayesian literature as   alternative models to Weibull $h_0(t)$.  The overall shape of the baseline
  hazard function does not have to be imposed in advance as is the case with the parametric models.

We  assume   a finite partition of the time axis with knots $c_0 \leq c_1\leq \ldots \leq c_K$, where $c_{0}\,=\,0$, and $c_{K}$ are usually taken as the last observed survival
 or censoring time. The  hazard function   is a mixture of piecewise constant functions defined as
\begin{equation}
\label{eqn:piecewise}
h_{0}(t \mid \boldsymbol{\varphi})= \sum_{k=1}^K\, \varphi_{k} \,I_{(c_{k-1},c_{k}]}(t), \,\, t>0,
\end{equation}
where $\boldsymbol \varphi =  (\varphi_1, \ldots, \varphi_K)$,   $I_{(c_{k-1},c_{k}]}(t)$ is the indicator function
 defined as 1 when $t\in (c_{k-1},c_{k}]$ and 0 otherwise. This baseline hazard function is usually known as the
\textit{piecewise constant} (\textit{PC} from now on). \vspace*{0.1cm}\\

 \noindent \textbf{\textit{Cubic B-spline functions}} \vspace*{0.1cm}\\
We  assume the same  finite partition of the time axis as specified for the \textit{PC} baseline hazard function.
The  spline function for the  baseline hazard function is usually defined in logarithmic scale \citep{murray2016} to accommodate normality and positivity for the
subsequent selection of prior distributions. It is   defined as
\begin{equation}
\label{eqn:Bsplines}
 \mbox{log}(h_{0}(t \mid \boldsymbol \gamma))= \displaystyle     \sum_{k=1}^{K+3}\gamma_{k}\,B_{(k,4)}(t), \,\, t>0,
\end{equation}
\noindent where $\boldsymbol \gamma=(\gamma_1, \ldots, \gamma_{K+3})$, $\{B_{(k,4)}(t), k=1,\ldots, K+3\}$ is a cubic basis of B-splines
with boundary knots $c_0$ and $c_K$  and internal knots $c_{k},\,k=1,..,K-1$ defined recursively by means of   the
  de Boor   formula~\citep{deboor1978} as
 \begin{equation}
 \label{eqn:defBsplines}
B_{(k,4)}(t)=\frac{t-\tau_k}{\tau_{k+3}-\tau_k}\,B_{(k,3)}(t)+ \frac{\tau_{k+4}-t}{\tau_{k+4}-\tau_{k+1}}\,B_{(k+1,3)}(t), \,\,k=1,\ldots, K+3,
 \end{equation}

 \noindent where $B_{(k,1)}(t)=1$ if $\tau_k \leq t \leq \tau_{k+1}$, $k=1,2,\ldots, K$ and zero otherwise. It is worth noting that the definition of this B-spline function needs
 augmentation of the original  knot sequence $\boldsymbol c=(c_0, c_1, \ldots, c_K)$ to $\boldsymbol \tau$, defined as~\citep{hastie2009}
\begin{align}
\label{eqn:augmented}
  \tau_1 \leq \ldots \leq \tau_4 \leq c_0; \,\,\,\,
 \tau_{j+4} = c_{j}, \,j=1,2,\ldots, K-1;\,\,\,\,
  c_{K} \leq \tau_{K+4} \leq \ldots \leq \tau_{K+7}.
\end{align}

This modelling strategy is known as a \textit{piecewise cubic B-spline} function ($PS$ from now on). Note that functions in hazard (\ref{eqn:piecewise}) are B-spline functions of order 1.

\subsection{Bayesian inferential process}
\subsubsection*{Regularization}

\textit{PC} and \textit{PS} baseline hazard functions  can accommodate  different shapes depending on the particular characteristics of the partition of the time axis. This is a relevant issue
with a great amount of research activity:~\cite{breslow1974} considered various failure times as end points of intervals;~\cite{kalbfleisch1973} supported the theory that the grid
should be selected independently of the data; ~\cite{murray2016} proposed equally-spaced partitions;~\cite{henschel2009} fixed the intervals assuming the condition that all the intervals contain comparable information, i.e. a similar number of events; and~\cite{lee2016}  avoided reliance on fixed partitions of the time scale by introducing the number of splits as a parameter to be estimated. When $K$ is large, the model has so many parameters that it could suffer from overfitting problems. On the contrary, choices of $K$ that are too small will lead to poor model fitting.   When using a shrinkage or regularization procedure, the effect of increasing
$K$ often diminishes. Regularization processes in the Bayesian setting are usually carried out  by means of informative prior distributions that restrict the freedom of the parameters.

The elicitation of   prior distributions for \textit{PC} and \textit{PS} baseline hazard functions    includes different prior distribution proposals for the coefficients
$\boldsymbol \varphi$ and $\boldsymbol \gamma$ in (\ref{eqn:piecewise}) and (\ref{eqn:Bsplines}), respectively. They range from a default situation of prior independence among all the coefficients
to a correlated  prior distribution  that accounts for shape restrictions in order to    avoid overfitting and strong irregularities in the estimation process.

We consider  four  prior scenarios for $h_0(t)$ defined in terms of a mixture of piecewise constant functions based on different correlation patterns among the coefficients associated
with the piecewise functions.\vspace*{-0.2cm}\\

\noindent  \textbf{\textit{Scenario PC1.}} Independent gamma prior distributions
  \begin{equation}
    \label{pc1}
\pi(\varphi_{k})= \mbox{Ga}(\eta_{k},\psi_{k}),  \,\,k = 1,2,\ldots,K.
\end{equation}

This is the most flexible and general prior scenario. A common selection is $\eta_{k}\,=\,\psi_{k}\,=\,0\mbox{.}01$.\vspace*{-0.2cm}\\

\noindent  \textbf{\textit{Scenario PC2.}}  Independent gamma  prior distributions
\begin{equation}
\label{pc2}
\pi(\varphi_{k})= \mbox{Ga}(w_{0}\,\eta_{0}\,(c_{k}- c_{k-1}), w_{0}\,(c_{k}- c_{k-1})), \,\, k=1,\ldots,K.
\end{equation}

All these marginal prior distributions  share the same prior expectation, $\eta_{0}$, but the prior variance of each $\varphi_k$ is inversely proportional to the corresponding
interval length, $c_{k}-c_{k-1}$. The selection $w_0=\,0.01$  is  a usual value which provides the prior distribution with a high level of  uncertainty. We will assume
 the \textit{ad hoc} proposal by~\cite{christensen2011} for the elicitation of $\eta_0$ that considers $\eta_0= 0.69315/\tilde{t}$, where $\tilde{t}$ is the  median survival time of the reference group. \vspace*{-0.2cm}\\

\noindent  \textbf{\textit{Scenario PC3.}} Correlated conditional gamma prior distributions
\begin{equation}
\label{pc3}
\pi(\varphi_{k}\mid \varphi_1,\ldots, \varphi_{k-1})=\mbox{Ga} (\eta_{k},\,\eta_{k}/\varphi_{k-1}), \,\, k=2,\ldots,K.
\end{equation}

This prior is based on a discrete-time martingale process~\citep{sahu1997} which correlates the $\varphi$'s
 of adjacent intervals with E$(\varphi_k \mid \varphi_1,\ldots, \varphi_{k-1})=\varphi_{k-1}$ and
  Var$(\varphi_k \mid \varphi_1,\ldots, \varphi_{k-1})=\varphi_{k-1}^2/\eta_k$. The parameter  $\eta_k$ is
   very important because it   controls the level of smoothness, which decreases  as $\eta_k$ reaches zero.  A common elicitation is
    $\eta_{k }=0.01, \,\, k= 2,\ldots,K$ and $\pi(\varphi_{1})= \mbox{Ga} (0.01,0.01)$.\vspace*{-0.2cm}\\

\noindent \textbf{\textit{Scenario PC4.} } Correlated conditional normal prior distributions for the $\varphi$ coefficients in a logarithmic scale
\begin{equation}
\label{pc4}
\pi(\mbox{log}\,(\varphi_{k})\mid \varphi_{1},\ldots,\varphi_{k-1})= \mbox{N}(\mbox{log}\,(\varphi_{k-1}),\sigma^{2}_{\varphi}), \,\,k=2,\ldots,K,
\end{equation}
\noindent with $\pi(\mbox{log}\,(\varphi_{1}))=\mbox{N}(0,\,\sigma^{2}_{\varphi})$. This is also a proposal based on a discrete-time martingale process. It  comes from
the areas of  spatial statistics~\citep{banerjee2014} and Bayesian B-splines~\citep{lang2004}, where it is better known as  a first-order random walk.   Correlation between
 the $\mbox{log}(\varphi_{k})$   corresponding to neighbouring intervals is   expressed assuming conditional normal prior distributions. \vspace*{0.1cm}\\

Non-informative prior distributions for  $\sigma^2_{\varphi}$ have generally been taken as  inverse gamma distributions, $ \mbox{IG}(\nu_0, \nu_0)$, with small
 values for $\nu_0$. However,   some research   questions the role of these distributions for describing lack of prior information.~\cite{gelman2006} proposed the
  use of proper uniforms and  half-t distributions for the standard deviations  as sensible choices, which   were understood  as reference models to
  be used as a standard of comparison or a starting point of the inferential process~\citep{bernardo1979}.
We also considered different prior specifications for the coefficients of the   \textit{PS} modelling of baseline hazard functions  that follow  the    idea of smoothing
 its level of flexibility and prevent overfitting. These scenarios are not a mere repetition of those considered for \textit{PC} baseline hazard functions. They have been chosen
  because they are  usual proposals in the statistical literature regarding cubic B-splines specifications.\vspace*{-0.2cm}\\

\noindent\textbf{\textit{Scenario PS1.}} Independent normal prior distributions
\begin{equation}
\label{ps1}
\pi(\gamma_{k})= \mbox{N}(0,\sigma^{2}_{k}), \,\, k=1,\ldots,K+3.
\end{equation}

This is the simplest scenario, similar to \textit{PC1},  in which $\gamma_{k}$ are considered as independent and normally distributed with a known  variance.\vspace*{-0.2cm}\\

\noindent \textbf{\textit{Scenario PS2.}}  Hierarchical normal prior distributions
\begin{equation}
\label{ps2}
\pi(\gamma_{k} \mid \sigma^2_{\gamma}) = \mbox{N}(0,\sigma^2_{\gamma}), \,\, k=1,\ldots,K+3,
\end{equation}

\noindent where $\sigma^2_{\gamma}$ is the common  variance population. As mentioned previously, a usual choice for the hyperprior distribution for $\sigma^2_{\gamma}$ is an inverse
gamma distribution or also a proper uniform distribution~\citep{gelman2006}.\vspace*{-0.2cm}\\

\noindent \textbf{\textit{Scenario PS3.}}  Correlated conditional normal prior distributions defined as
\begin{equation}
\label{ps3}
\pi(\gamma_{k}\mid \gamma_{1},\ldots,\gamma_{k-1})=\mbox{N} (\gamma_{k-1},\sigma^{2}_{\gamma}),\,\, k=2,\ldots,K+3,
\end{equation}

\noindent and  based on a first-order   Gaussian  random walk which involves an intrinsic Gaussian Markov random field as the conditional
   joint prior distribution for the spline coefficients given $\sigma^{2}_{\gamma}$. This proposal
   comes from the so-called Bayesian P-splines~\citep{lang2004,fahrmeir2011}. It has  been widely used in  Bayesian  spatial statistics~\citep{banerjee2014}, where it is usually expressed
   in terms of  conditional distributions in the form
\begin{equation}
    \label{ps3.2}
\pi(\gamma_{k}\mid  \boldsymbol{\gamma}_{-k})=\mbox{N}\left(\frac{1}{2}(\gamma_{k-1}+\gamma_{k+1}),2\sigma^{2}_{\gamma}\right),\,\, k=2,\ldots,K+3,
\end{equation}

\noindent where $\boldsymbol{\gamma}_{-k}$ denotes all spline coefficients except $\gamma_k$. Popular marginal prior
distribution choices for $\sigma_{\gamma}$ that try to be as neutral as possible are  Ga$(1, 0.0005)$ ~\citep{lang2004} and Ga(0.001, 0.001) as a default option in the software \verb"BayesX"~\citep{belitz2015}.
This scenario is  analogous to \textit{Scenario PC4}. Consequently, all the discussion regarding the
 elicitation of the prior distribution for the variance $\sigma^{2}_{\gamma}$ (precision or
 standard deviation $\tau_{\gamma}$  and $\sigma_\gamma$, respectively) also applies here.



\subsubsection*{Posterior distribution} \label{sec:2.2}

We considered a prior independent  scenario between the parameters in $h_0(t)$
and the regression coefficients associated to  covariates.  We also reckoned prior independence between the regression coefficients within a \textit{non-informative} scenario, with normal
distributions centred at zero and a wide known variance:
\begin{align}
\pi(h_{0}, \boldsymbol{ \beta}) &=\pi(h_{0})\, \pi(\boldsymbol \beta)  =  \pi(h_{0})\, \mbox{$\prod$}_{j=1}^{J}\,\mbox{N}(\beta_{j} \mid 0, \sigma_{j}^2),
 \end{align}

\noindent where $\pi(h_0)$ is the prior distribution of all  parameters and hyperparameters in    $h_0(t)$. The model needs to be  fed with data $\mathcal{D}= \{(t_i, \delta_i, \boldsymbol x_i), i=1, \ldots, n\}$, where $t_i$ is the observed survival  time   for the $i$th individual, $\delta_i$ is the
 indicator taking 1 if the event has occurred and 0 otherwise, and $\boldsymbol x_i$
are the  subsequent  covariates.

Bayes' theorem combines prior knowledge and experimental information in  the posterior distribution
 $$\pi(h_{0}, \boldsymbol{ \beta} \mid \mathcal{D}) \propto \mathcal{L}(h_{0}, \boldsymbol{\beta})\,   \pi(h_{0}, \boldsymbol{ \beta}),$$

\noindent where $ \mathcal{L}(h_{0}, \boldsymbol{ \beta})$ is the likelihood function of  $(h_{0}, \boldsymbol{ \beta})$ given by~\cite{ibrahim2005} as
\begin{align}
\label{eqn:2}
\mathcal{L}(h_{0},\boldsymbol{\beta})\,&=
 \,\prod_{i=1}^n h_0(t_{i})^{\delta_{i}}\,\mbox{exp}\{-H_{0}(t_i)\} [\mbox{exp}\{\boldsymbol{x}_{i}^{\prime}\,\boldsymbol{\beta}\}]^{\delta_{i}}\mbox{exp}\{\mbox{exp}\{\boldsymbol{x}_{i}^{\prime}\,\boldsymbol{\beta}\}\},
\end{align}

\noindent with  $H_{0}(t) = \int_{0}^{t}\,h_{0}(u)\,\mbox{d}u $   as the   cumulative baseline hazard function.

In the case of a Weibull hazard baseline function, the  cumulative baseline hazard function is
 $H_0(t)=\lambda t^{\alpha}, t>0.$ When the baseline function is defined via a mixture of piecewise constant functions,  as in  (\ref{eqn:piecewise})
$$H_0(t)= \mbox{$\sum$}_{m=1}^{k-1}\,\varphi_{m}(c_m-c_{m-1})+\varphi_k(t-c_{k-1}),\,\, c_{k-1} \leq t <c_k,\,\, k=1,\ldots,K.$$

The expression of the cumulative baseline hazard for    $h_0(t)$ defined in (\ref{eqn:Bsplines}) in logarithmic scale in terms of cubic B-spline functions  needs to take into account some additional  properties of B-splines~\citep{deboor1978, sherar2004}. In particular,
\begin{align}
\label{eqn:3}
\,\int_0^{t}\, \sum_{k=1}^{K+3}\gamma_{k}\,B_{(k,4)}(u) \mbox{d}u =\, \sum_{k=1}^{K+4}\,\phi_k\,B_{(k,5)}(t),
\end{align}
\noindent with $\phi_1=0$, and $\phi_{m+1}=\frac{\tau_{m+1}-\tau_5}{4}\, \sum_{j=1}^{m}\,\gamma_j,\,m=1,2,\ldots, K+3.$ Note that B-splines of order 5 need to add two additional nodes to the augmented   knot sequence $\boldsymbol \tau$ in (\ref{eqn:augmented}).


\section{An experiment on microbial virulence} \label{sec:3}

\subsection{Virulence data and modelling}

A dataset involving a virulence assay is taken into account to explore the baseline hazard specifications discussed above.  The data came from an experiment designed to assess the effect of the use of a cauliflower by-product infusion treatment in \textit{Salmonella enterica serovar} Typhimurium (\textit{S.} Typhimurium)   virulence behaviour. \textit{S.} Typhimurium  is   one of the most usual serotypes related to salmonellosis outbreaks and cauliflower by-product infusion treatment is an alternative preservation treatment against it.

One and three exposures to the  treatment were evaluated. A pathogen \textit{S.} Typhimurium ($ST$) population non-exposed to the treatment was considered as
the control group. The nematode \textit{Caenorhabditis elegans (C. elegans)} was used as a host model to quantify the virulence of the
 pathogen.  \textit{ST}  non-treated (\textit{ST0}), \textit{ST}   treated once (\textit{ST1}), and \textit{ST}
 treated three times (\textit{ST3}) was the source of nutrition of 250 synchronized young adult nematodes  kept  in identical environmental conditions
  throughout  their lifespan (approximately three weeks at the most). Virulence for each worm  was defined in terms of   their  survival time
  (see~\cite{sanzpuig2017} for more details about the validation and special conditions of the study). Most of the data
were fully observed. Only five survival times were
right-censored due to the accidental death of the individuals when they were being transferred  from one plate to another.

\begin{figure}[H]
 \centering
 {\includegraphics[width=5cm, height=5cm, angle=270]{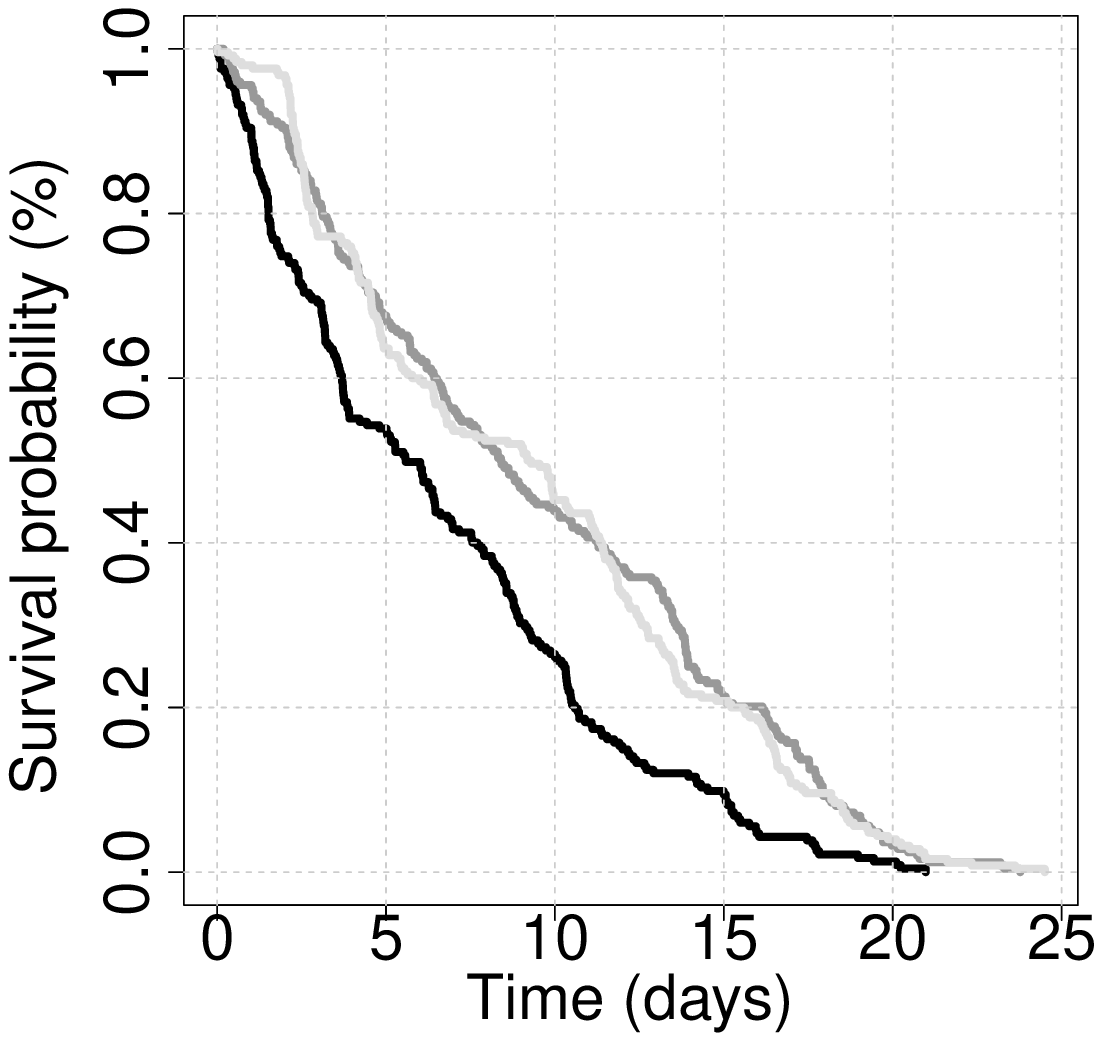}}
 \caption{Kaplan-Meier survival curve, in days, for individuals  fed on a) \textit{ST0} (black), b) \textit{ST1} (dark gray),  and c) \textit{ST3} (gray).}
 \label{figure:Kaplan-Meier}
 \end{figure}

Figure~\ref{figure:Kaplan-Meier} shows a Kaplan-Meier curve, in days, for each of the  \textit{ST} populations considered. Individuals fed
on \textit{ST0} (the control group) showed a survival curve that was lower over time in relation to the ones fed on \textit{ST1} and \textit{ST3}, with a median survival
time of  $5.58$ days versus $8.40$ and $9.24$, respectively. The \textit{ST1} and \textit{ST3} groups exhibit similar trajectories which cross at certain time points, thus confirming a similar behaviour.

 \begin{center}
 \textbf{FIGURE 1 AROUND HERE}
 \end{center}

Virulence for the $i$-th worm was modelled by means of the   Cox proportional hazards model
\begin{align}
\label{eqn:bayesiancox2}
h_{i}(t \mid  h_{0}, \boldsymbol{x}_{i},   \boldsymbol{\beta})
&=	h_{0}(t)\,\mbox{exp}\{\beta_\texttt{1} \,I_\texttt{1}(i)+\beta_\texttt{3} \,I_\texttt{3}(i)\},							
\end{align}

\noindent where $I_\texttt{1}(i)$  and $I_\texttt{3}(i)$ are indicator  variables for groups \textit{ST1} and \textit{ST3}, respectively.  It is important to
 highlight that $h_i(t \mid \cdot)=h_0(t)$ in the case of \textit{ST0}, which acts as the control group, $h_i(t\mid \cdot)=h_0(t)\,\mbox{exp}\{\beta_\texttt{1} \,I_\texttt{1}(i)\}$ when it is
  \textit{ST1}, and $h_i(t \mid \cdot)=h_0(t)\,\mbox{exp}\{\beta_\texttt{3} \,I_\texttt{3}(i)\}$ when the group is \textit{ST3}.

We considered a Weibull model for $h_0(t)$ as well as $PC$ and $PS$ baseline hazard functions based on four different partitions of the time axis with number of knots  $K$ = 5, 10, 25 and  40. All these partitions were chosen following   the proposal  by \cite{murray2016} based on selecting  intervals  with the same length. The last knot in all  $PC$ and $PS$ models is $24\mbox{.}50$ days, which was the longest survival time observed.

\subsection{Posterior inferences}
We carried out all Bayesian survival inferential processes derived from the combination of the  generic specifications of the baseline  hazard function
 above with the different prior scenarios and number of knots ($K$ = 5, 10, 25 and  40) for $PC$ and $PS$ models. The joint posterior distribution for
 each model was approximated using  the \texttt{JAGS} software~\citep{plummer2003}. For each estimated model, we ran three parallel chains with 50,000 iterations and a burn-in of 5,000. Chains were also thinned by storing every 5th iteration   to reduce autocorrelation in the sample. Convergence to the joint posterior distribution was guaranteed with a potential scale reduction factor close to 1 and an effective number of independent simulation draws greater than 100.

\subsection{Model selection, hazard ratios and baseline hazard-survival function}

Deviance information criterion (DIC)~\citep{spiegelhalter2002}  and  log  pseudo-marginal likelihood (LPML)~\citep{geisser1979} were considered for model selection. DIC measures the information on a model by means
of its deviance  penalized with regard to its complexity. Additionally, from the DIC computation we derived the effective number of parameters (pD) to evaluate the model complexity~\citep{spiegelhalter2002}. LPML is based on predictive criteria. It combines, on a logarithmic scale, the conditional predictive ordinate value (CPO) associated with
observations of each individual~\citep{gelfand1996}. Smaller values for DIC are preferred, while  larger LPML values   indicate better predictive performance. pD is interpreted together with DIC, as a complementary criterion.

As a rule of thumb, if two models differ in the DIC by more than 3, the one with the smaller DIC is preferred
as the best fitting~\citep{spiegelhalter2002}. In the case of LPML, there is no rule of thumb about how much this difference should be~\citep{bogaerts2017}. However, the LPML statistics from two competing models, $\mbox{LPML}_{1}$ and $\mbox{LPML}_{2}$, can be used to compute what has been termed a ``pseudo Bayes factor'' (PBF), which roughly indicates which model is superior at predicting the observed data: $\mbox{PBF}_{12}\,=\, \exp(\mbox{LPML}_{1} - \mbox{LPML}_{2})$~\citep{hanson2007, branscum2008, zhao2014}. We interpret the PBF following the guidelines proposed by \citet{jeffreys1961} and \citet{kass1995}; thus, a $\mbox{PBF}_{12}$ above 3 denotes there is substantial evidence in favour of model 1.

Table~\ref{tab:1} shows the DIC, the pD and LPML values  of the estimated  models. Based on the DIC and LPML values, \textit{PS}  models exhibit   better behaviour than  Weibull or \textit{PC} specifications. The Weibull model shows the worst preformance even if showing the lowest complexity (as measured by pD value). An increase in the number of knots for \textit{PC} models  generally results in a clear improvement in the modelling (from $K\,=\,5$ to $K\,=\,25$), since increasing $K$ up to 40 does not substantially improve goodness of fit while meaningfully increasing model complexity. Differences in DIC and PSB that are higher than 3 favour models with $K\,\leq\,25$. This fact is more relevant with  correlated prior distributions, especially for scenario $PC4$. $PS$ (regardless of the number of knots and prior setting) are always the best models, showing no relevant differences between their DIC and LPML (PBF) values. Thus, PS models with $K=5$ show similar performance to their $K=10$, $K=25$ and $K=40$ counterparts. In relation to the pD values, the complexity of the models is clearly influenced by prior specification. $PC4$ and $PS3$ models (above all for $K$ = 25 and $K$ = 40) show a prior-induced parameter reduction (the true parameters (not considering hyperparameters) for $PC$ and $PS$ models can be estimated as $K$+2 and $K$+5, respectively); hence they show an improvement in model complexity with respect to their counterparts.


 \begin{center}
 \textbf{TABLE 1 AROUND HERE}
 \end{center}

Below we focus on the posterior stability of the posterior distribution for the hazard ratios as well as the   behaviour of  the subsequent marginal posterior distribution for
    the baseline hazard function, which reflects the natural course of the infection, and  the survival function.\vspace*{0.1cm}

\subsubsection*{\textbf{Hazard ratios}}

\begin{figure}[H]
\begin{flushleft}

{\includegraphics[width=3.6cm, height=3.6cm, angle=270]{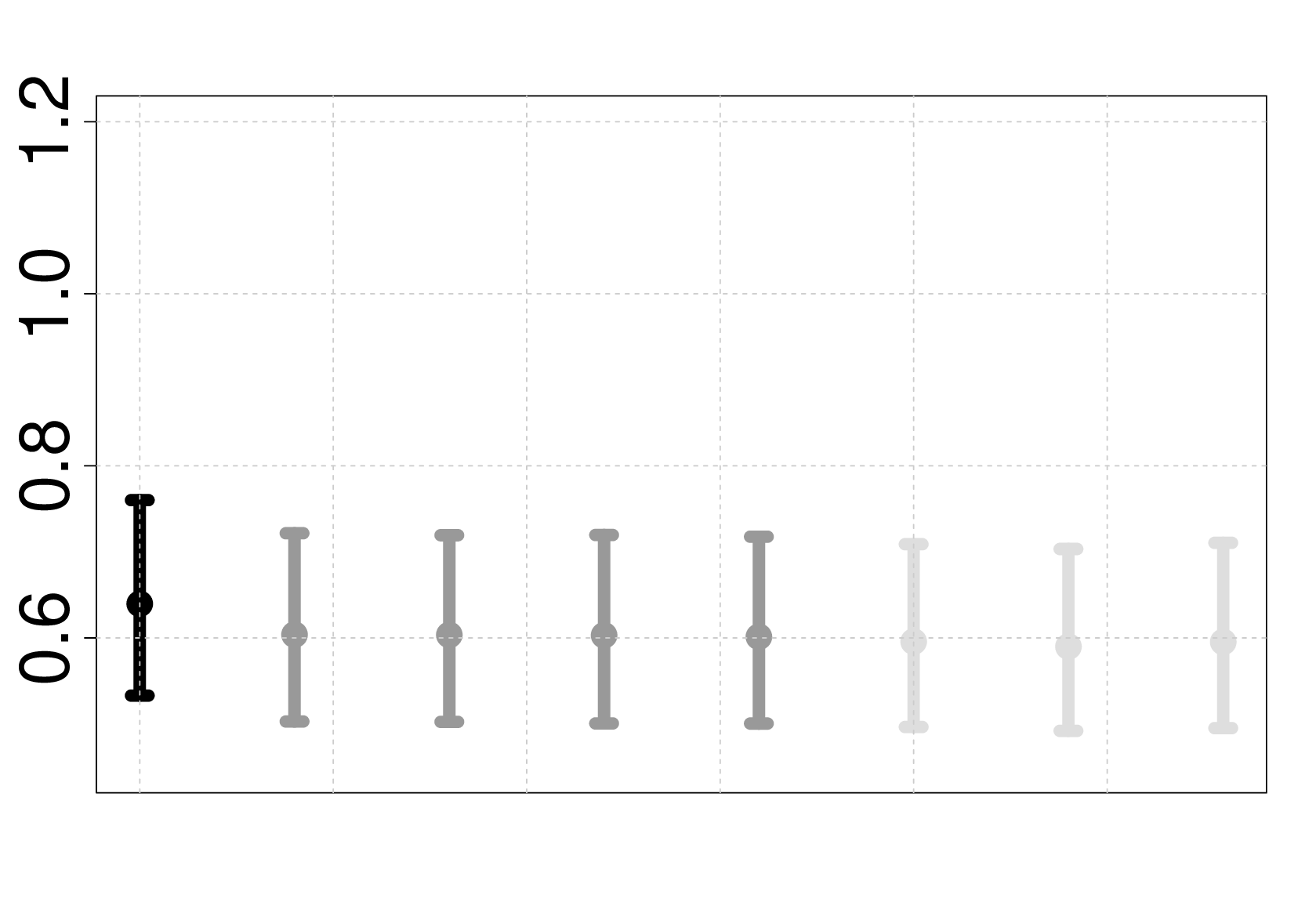}}\hspace*{-.15cm}
{\includegraphics[width=3.6cm, height=3.6cm, angle=270]{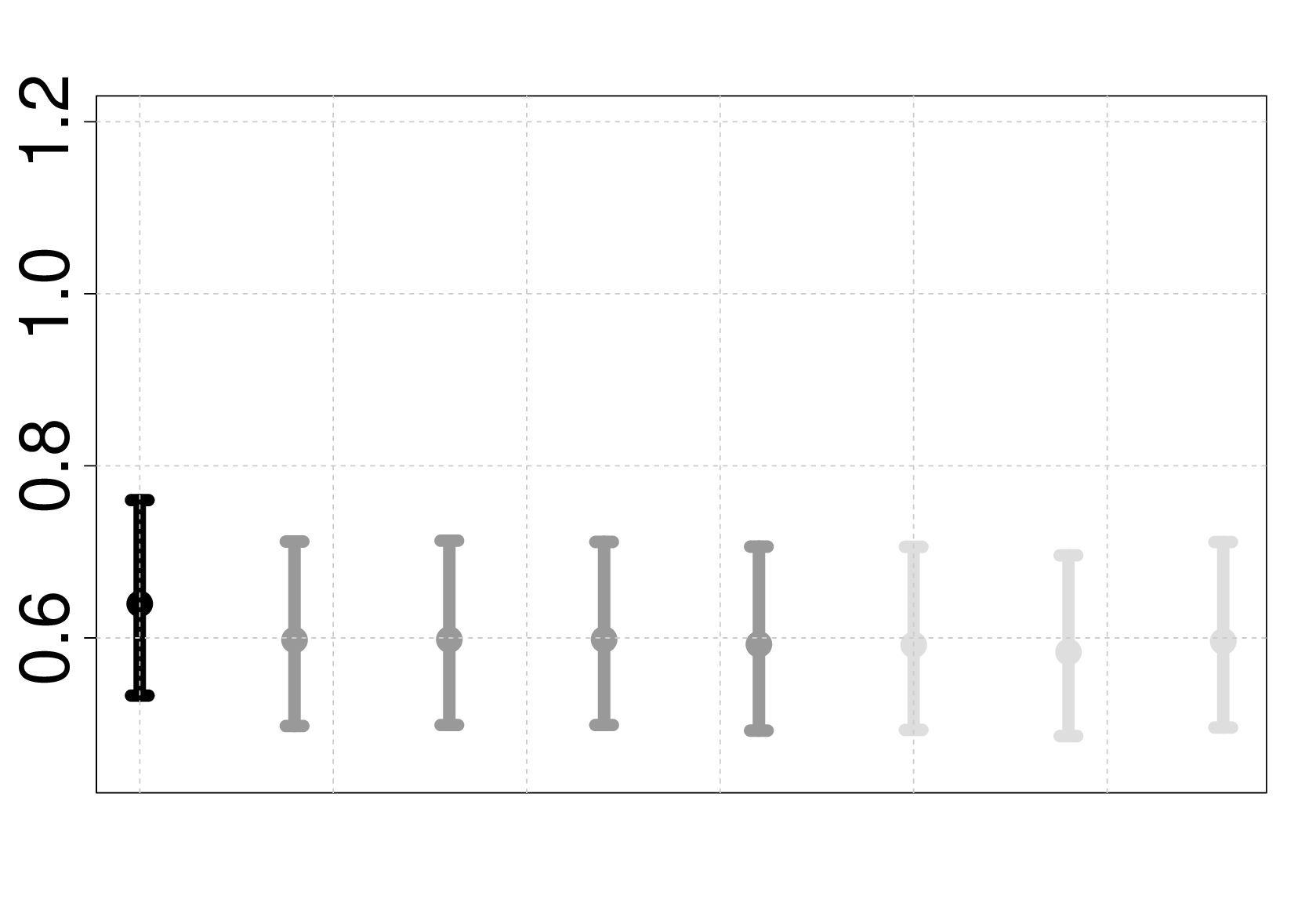}}\hspace*{-0.15cm}
{\includegraphics[width=3.6cm, height=3.6cm, angle=270]{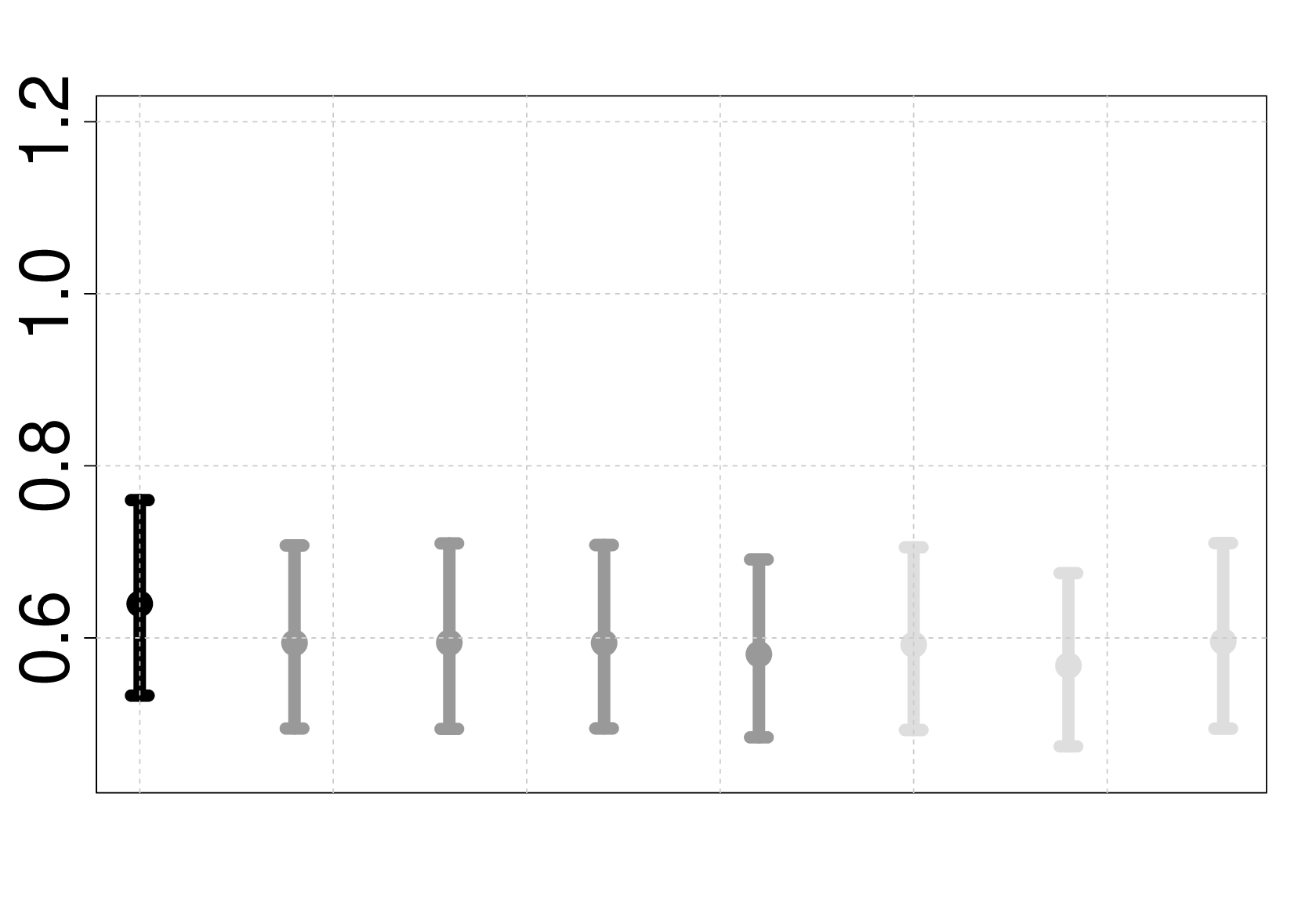}}\hspace*{-0.15cm}
{\includegraphics[width=3.6cm, height=3.6cm, angle=270]{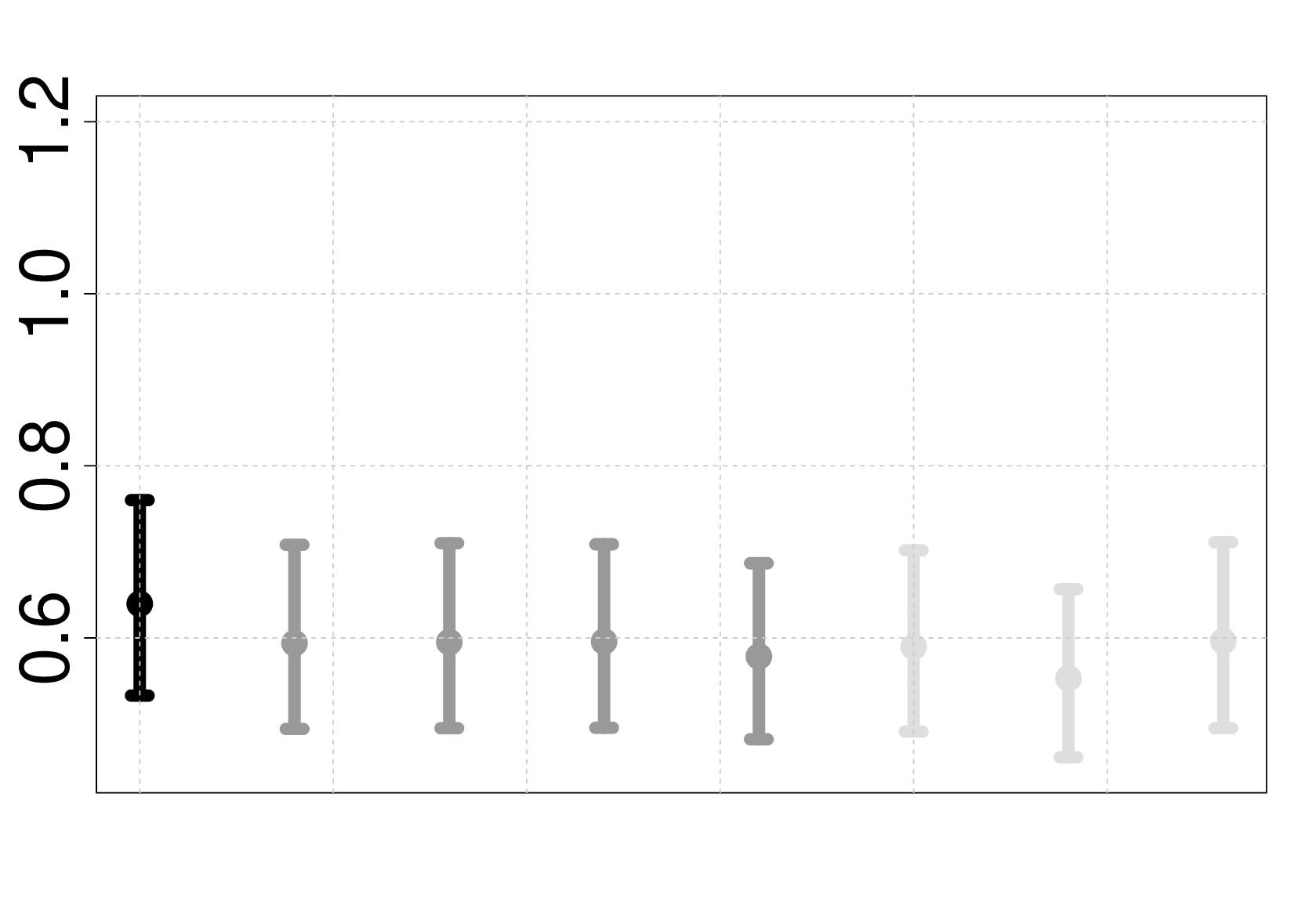}}\\
\vspace{-0.55cm}
{\includegraphics[width=3.6cm, height=3.6cm, angle=270]{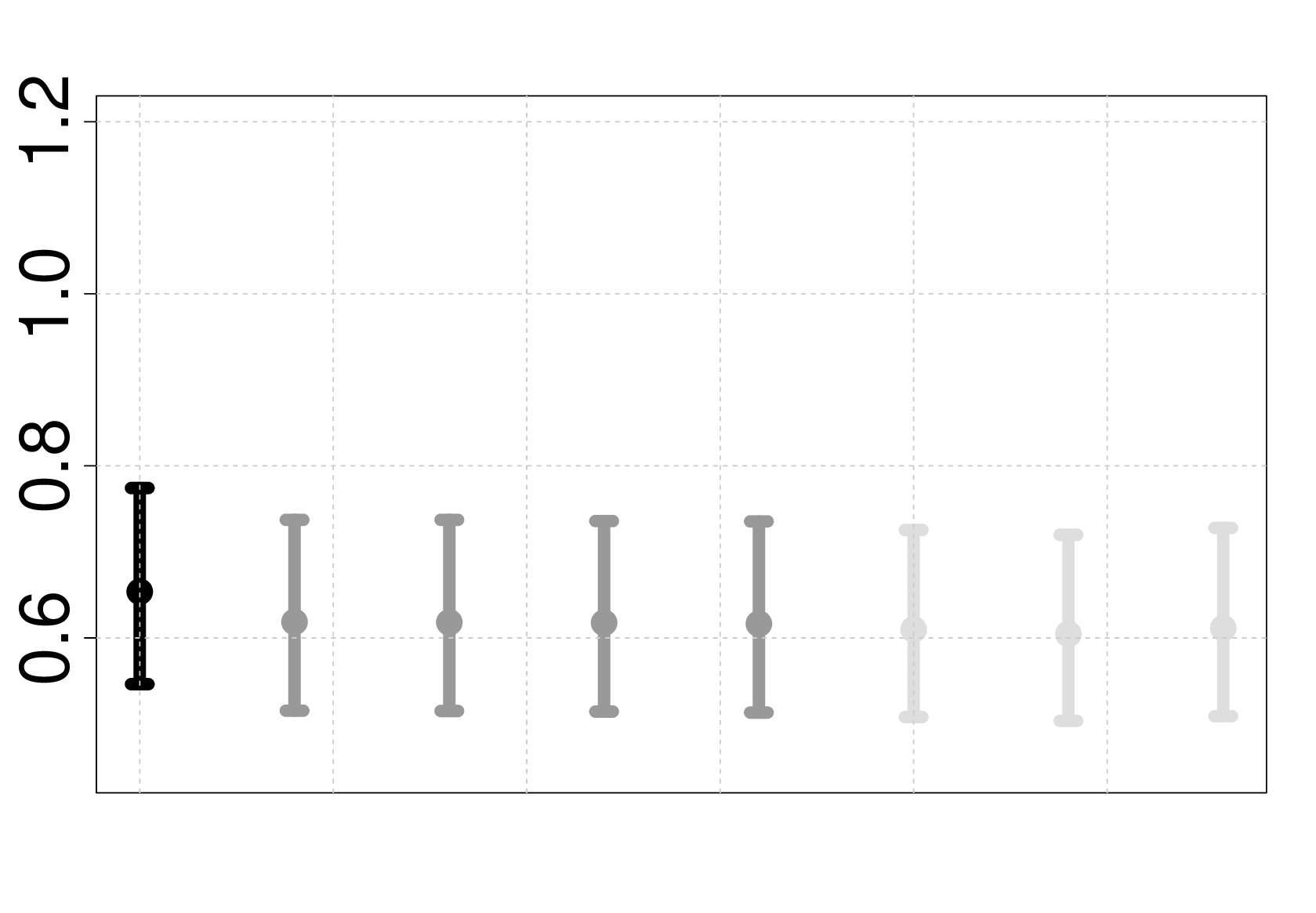}}\hspace*{-.15cm}
{\includegraphics[width=3.6cm, height=3.6cm, angle=270]{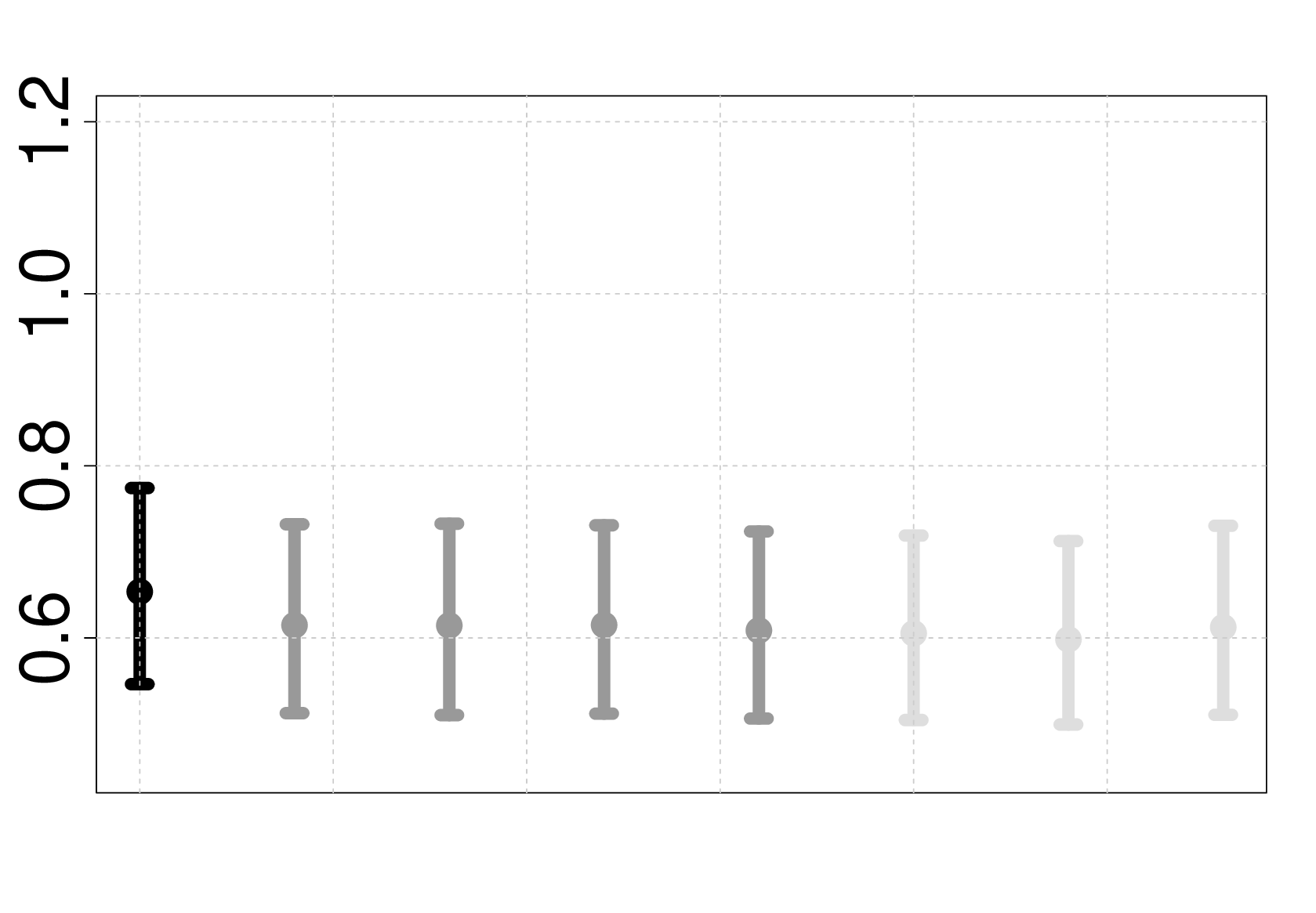}}\hspace*{-0.15cm}
{\includegraphics[width=3.6cm, height=3.6cm, angle=270]{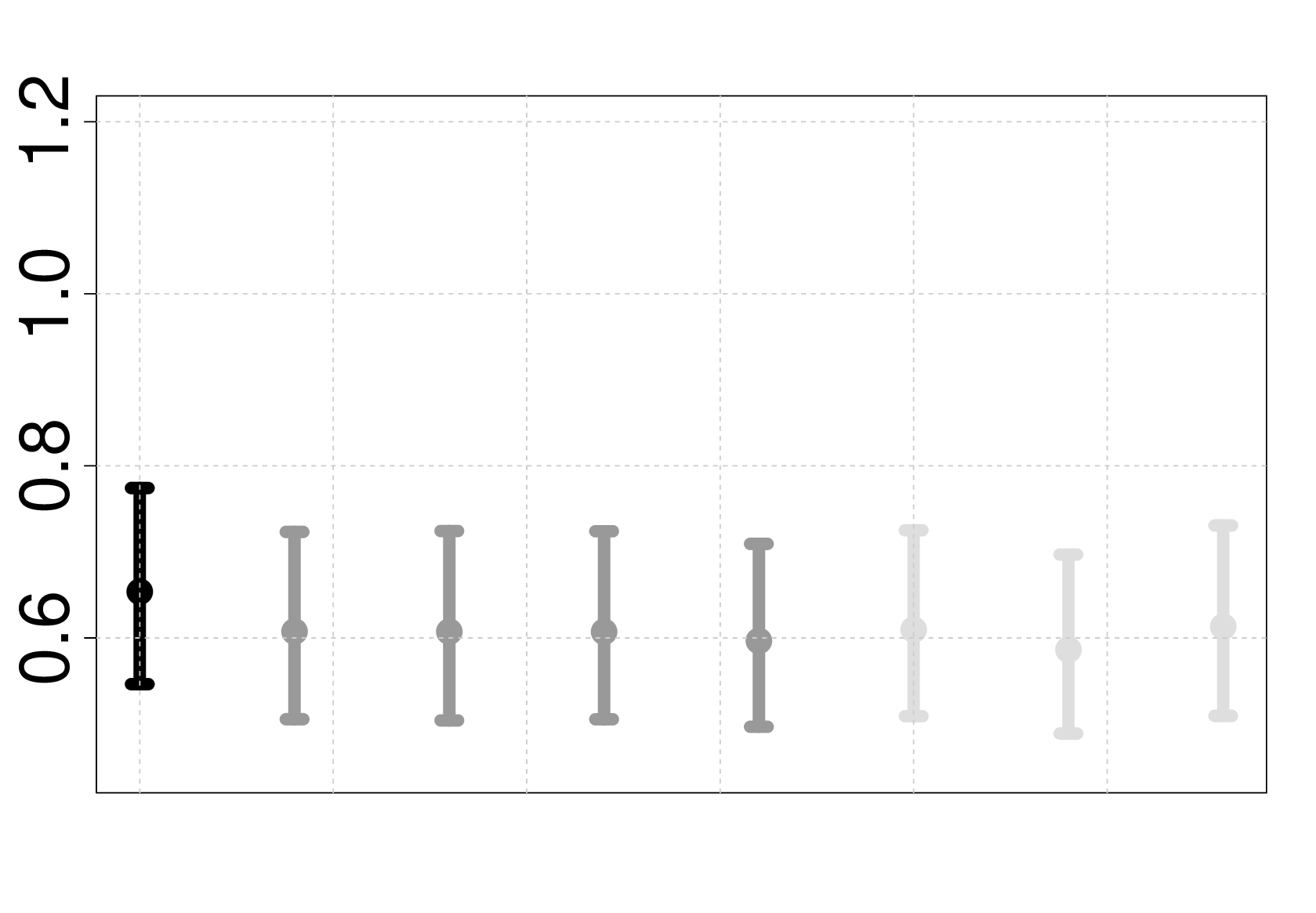}}\hspace*{-0.15cm}
{\includegraphics[width=3.6cm, height=3.6cm, angle=270]{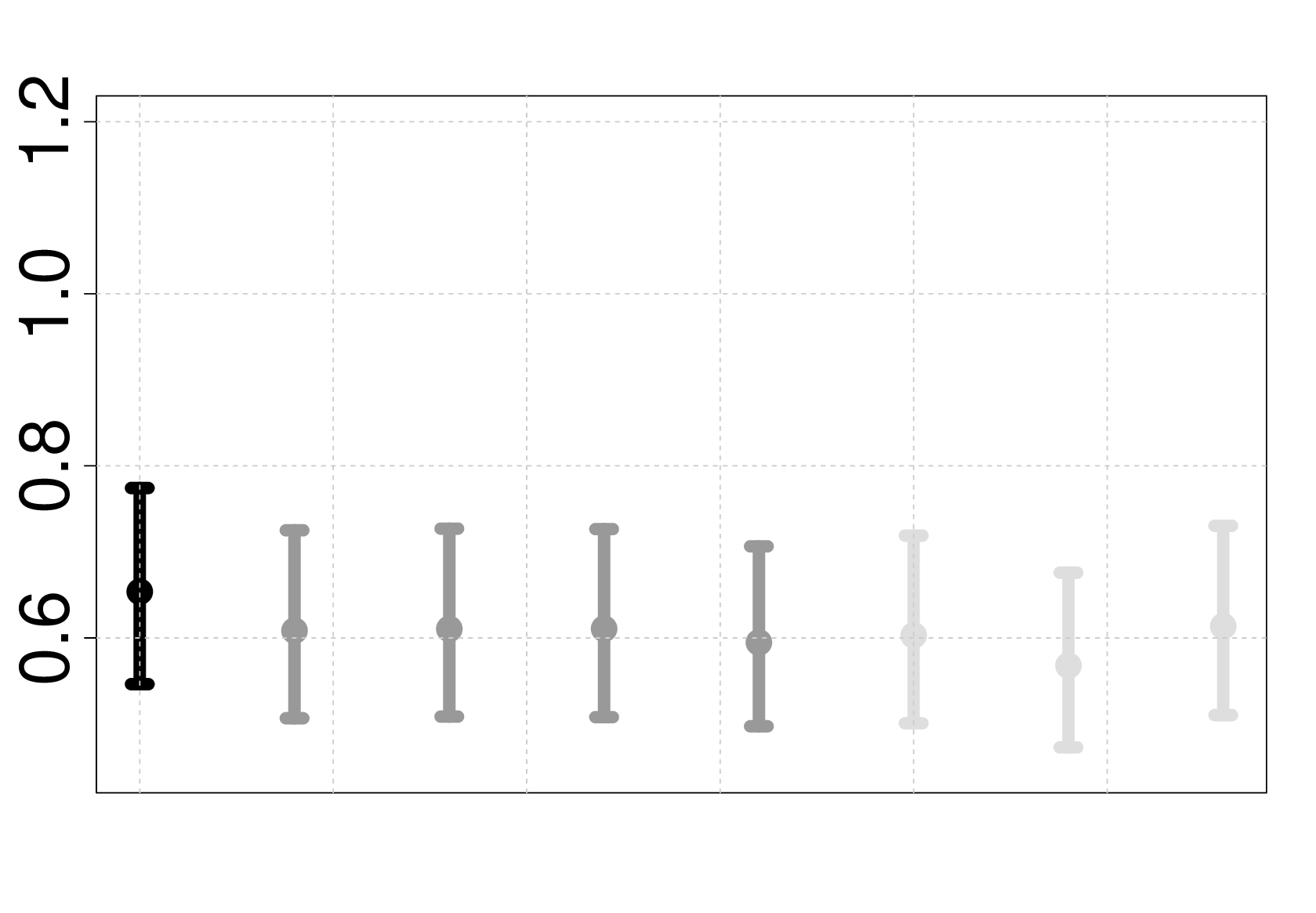}}\\
\vspace{-0.65cm}
 \subfigure[\textit{K}=5] {\includegraphics[width=3.6cm, height=3.6cm, angle=270]{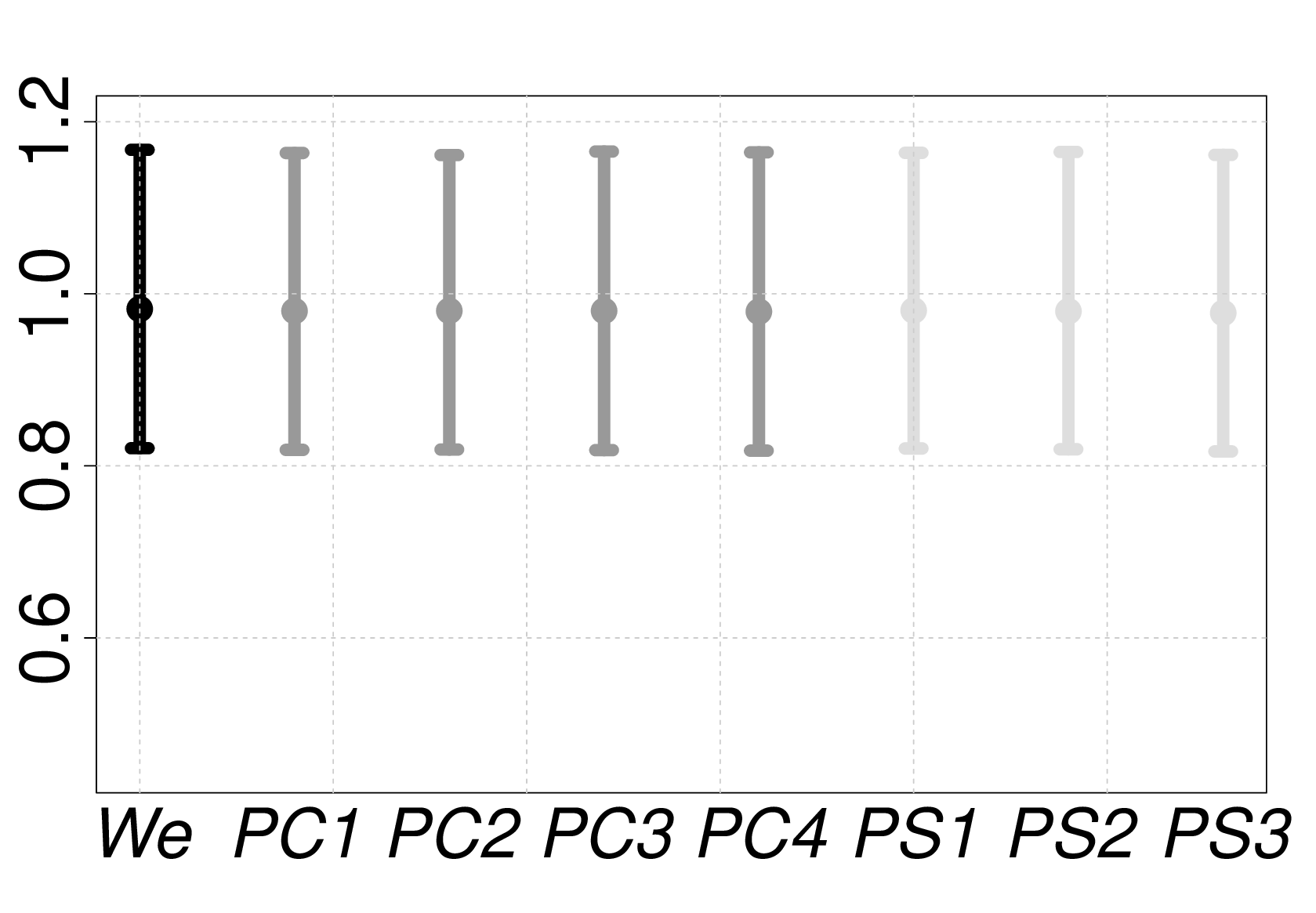}}\hspace*{-.15cm}
 \subfigure[\textit{K}=10]{\includegraphics[width=3.6cm, height=3.6cm, angle=270]{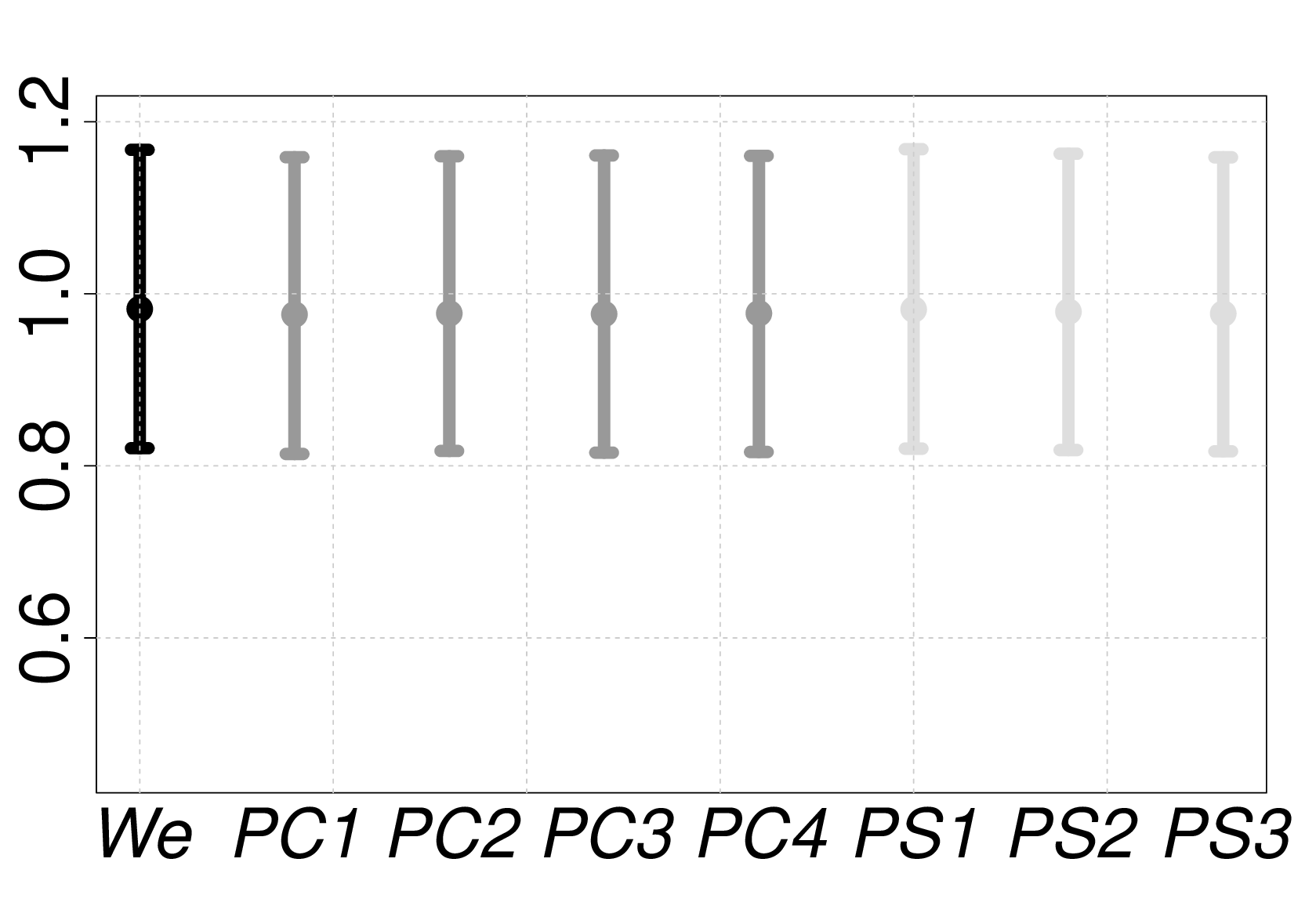}}\hspace*{-0.15cm}
 \subfigure[\textit{K}=25]{\includegraphics[width=3.6cm, height=3.6cm, angle=270]{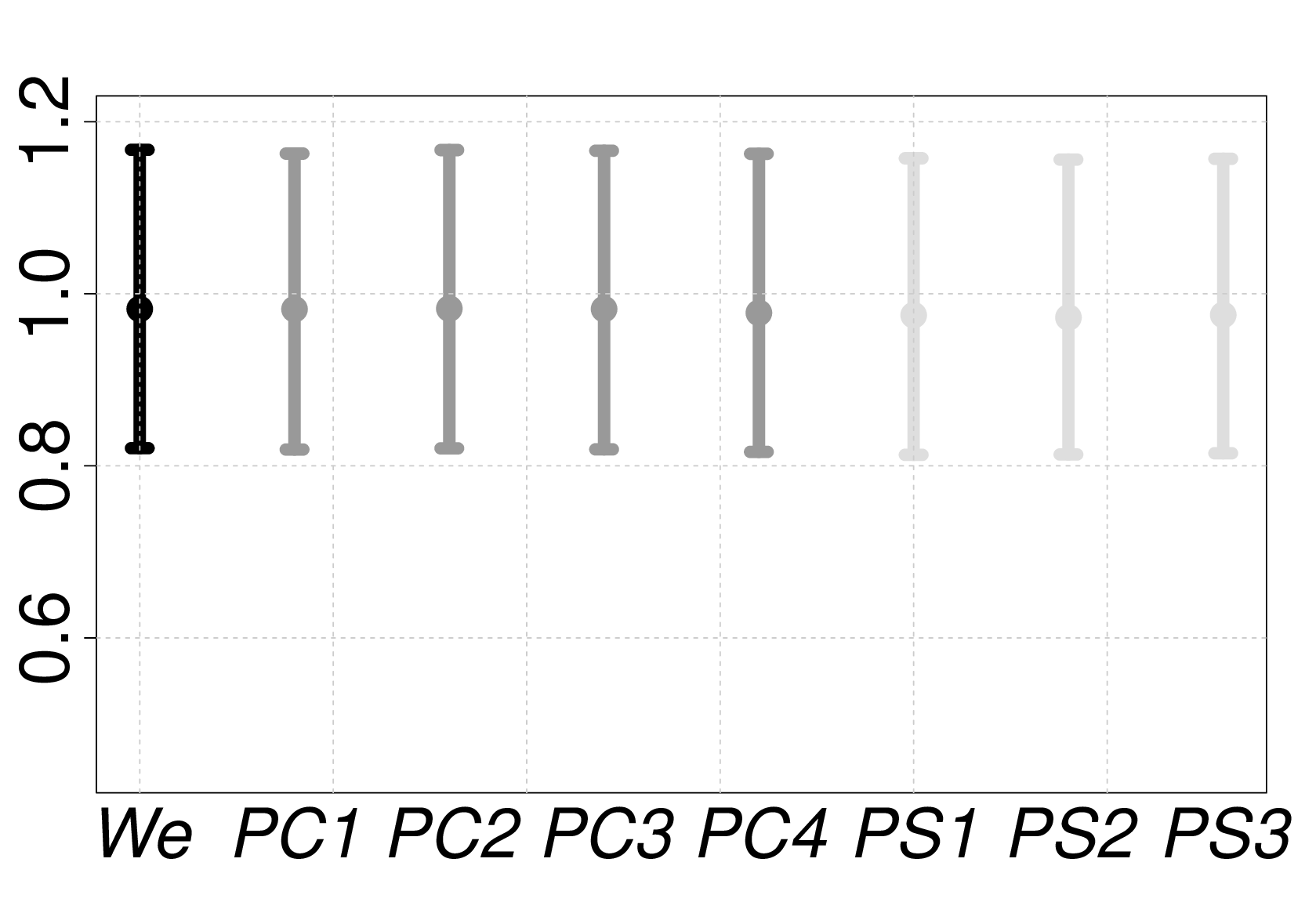}}\hspace*{-0.15cm}
 \subfigure[\textit{K}=40]{\includegraphics[width=3.6cm, height=3.6cm, angle=270]{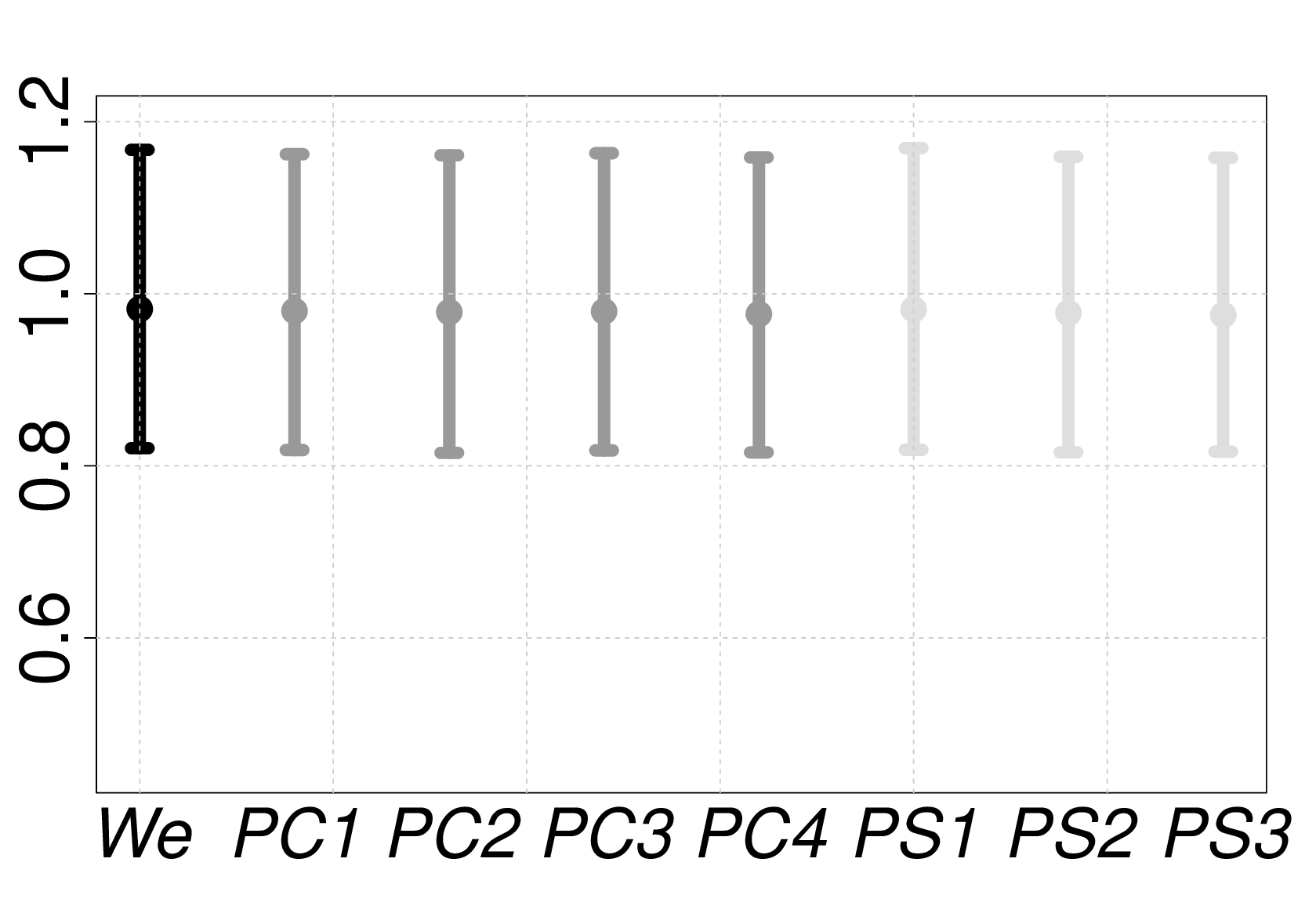}}\\
\vspace{-0.30cm}
 \caption{Posterior mean and 95\% credible interval for the hazard ratios, HR$_{ST1}$ (row one), HR$_{ST3}$ (row two)
 and HR$_{ST1/ST3}$ (row three),  for all survival models under evaluation.}
 \label{fig:2}
 \end{flushleft}
 \end{figure}

Discrepancies  between the posterior marginal distributions for the regression coefficients and for any of their corresponding derived quantities, such as hazard ratios, are a result of the different modelling of $h_0(t)$. Figure~\ref{fig:2} shows the posterior mean and a 95\% credible interval for the hazard ratios of interest HR$_{ST1}$, HR$_{ST3}$, HR$_{ST1/ST3}$ (computed as $\pi(\exp(\beta_1)\mid \mathcal{D})$, $\pi(\exp(\beta_3)\mid \mathcal{D})$ and $\pi(\exp(\beta_1-\beta_3)\mid \mathcal{D})$) with regard to the different specifications of the baseline hazard function,  prior scenarios and   number of knots for \textit{PC} and \textit{PS} models. HR$_{ST1}$ and HR$_{ST3}$ posterior distributions behave in a similar way, with values below 0 indicating efficacy in bacterium virulence reduction.  HR$_{ST1/ST3}$ posterior distributions are centred at approximately 1, pointing to similar efficacy for both treatments. We observe great internal robustness in the results of the $PS$ models and the $PC$ models. Weibull estimated coefficients  are also quite similar to those obtained from $PC$ and $PS$ models.\\

 \begin{center}
 \textbf{FIGURE 2 AROUND HERE}
 \end{center}

\subsubsection*{\textbf{Baseline hazard and baseline survival functions}}

\begin{figure}[H]
\begin{flushleft}
 \subfigure[\textit{We}] {\includegraphics[width=3.1cm, height=3.8cm, angle=270]{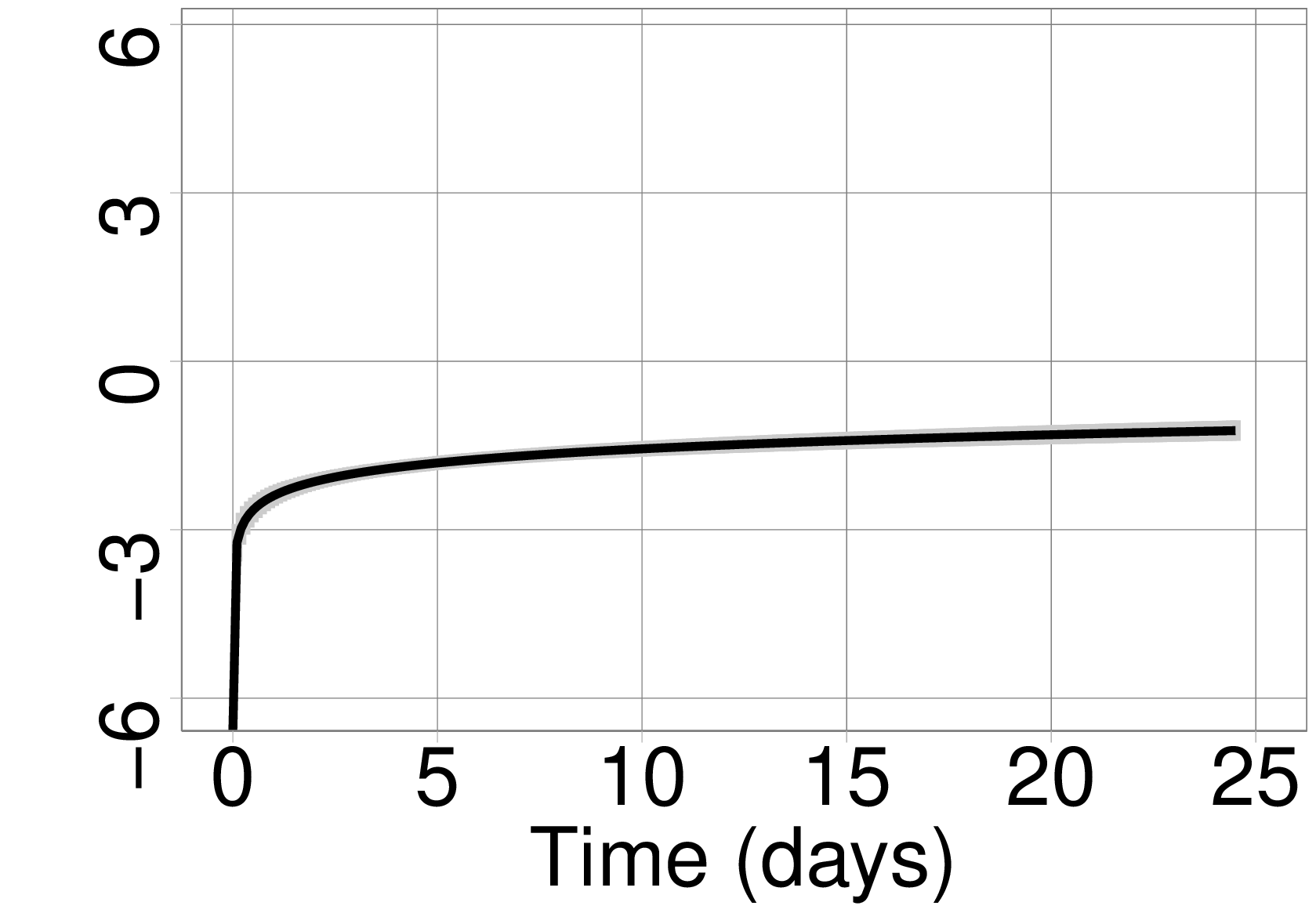}}\\\vspace{-0.30cm}
 \subfigure[\textit{PC1}] {\includegraphics[width=3.1cm, height=3.8cm, angle=270]{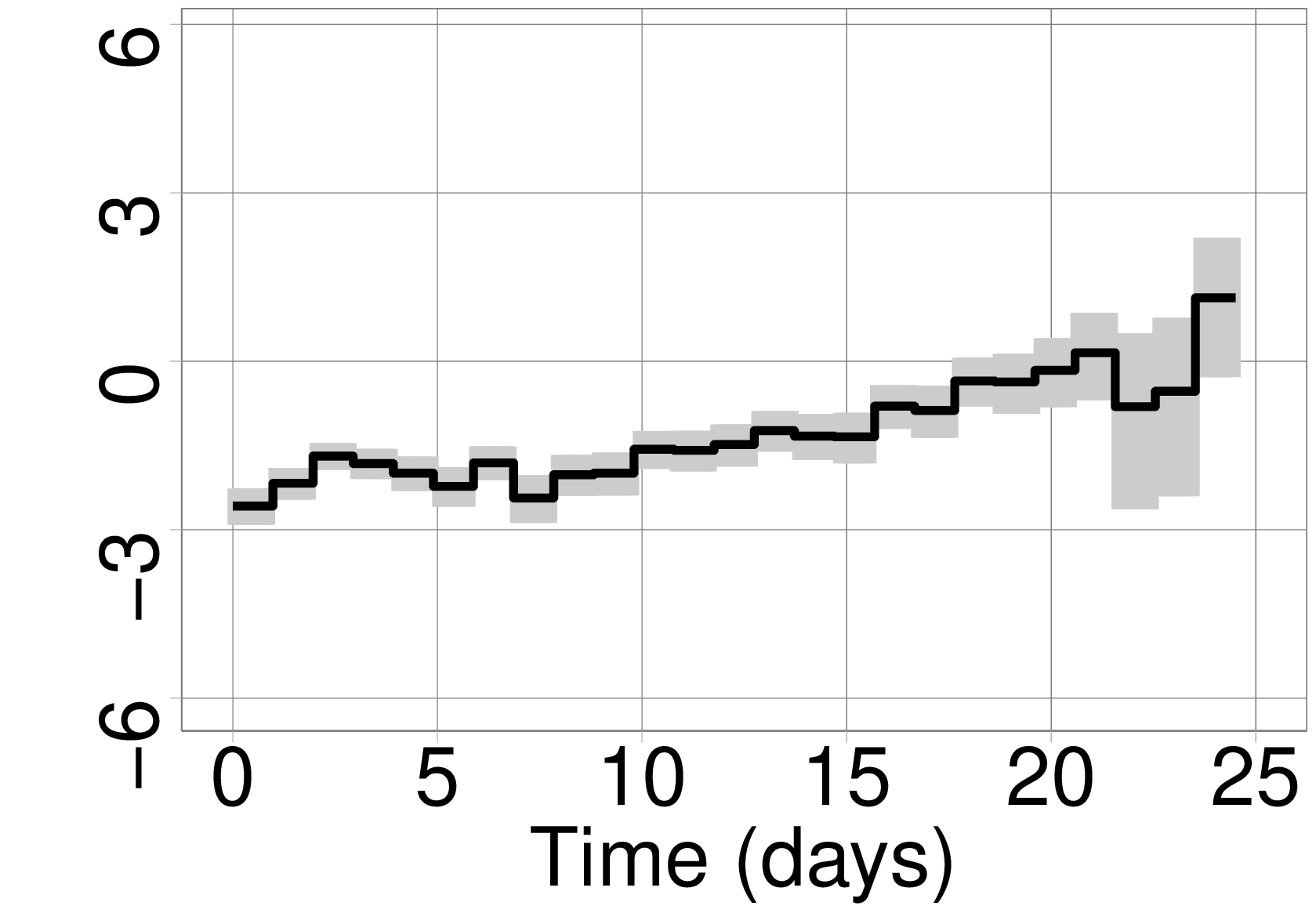}}\hspace*{-.05cm}
 \subfigure[\textit{PC2}] {\includegraphics[width=3.1cm, height=3.8cm, angle=270]{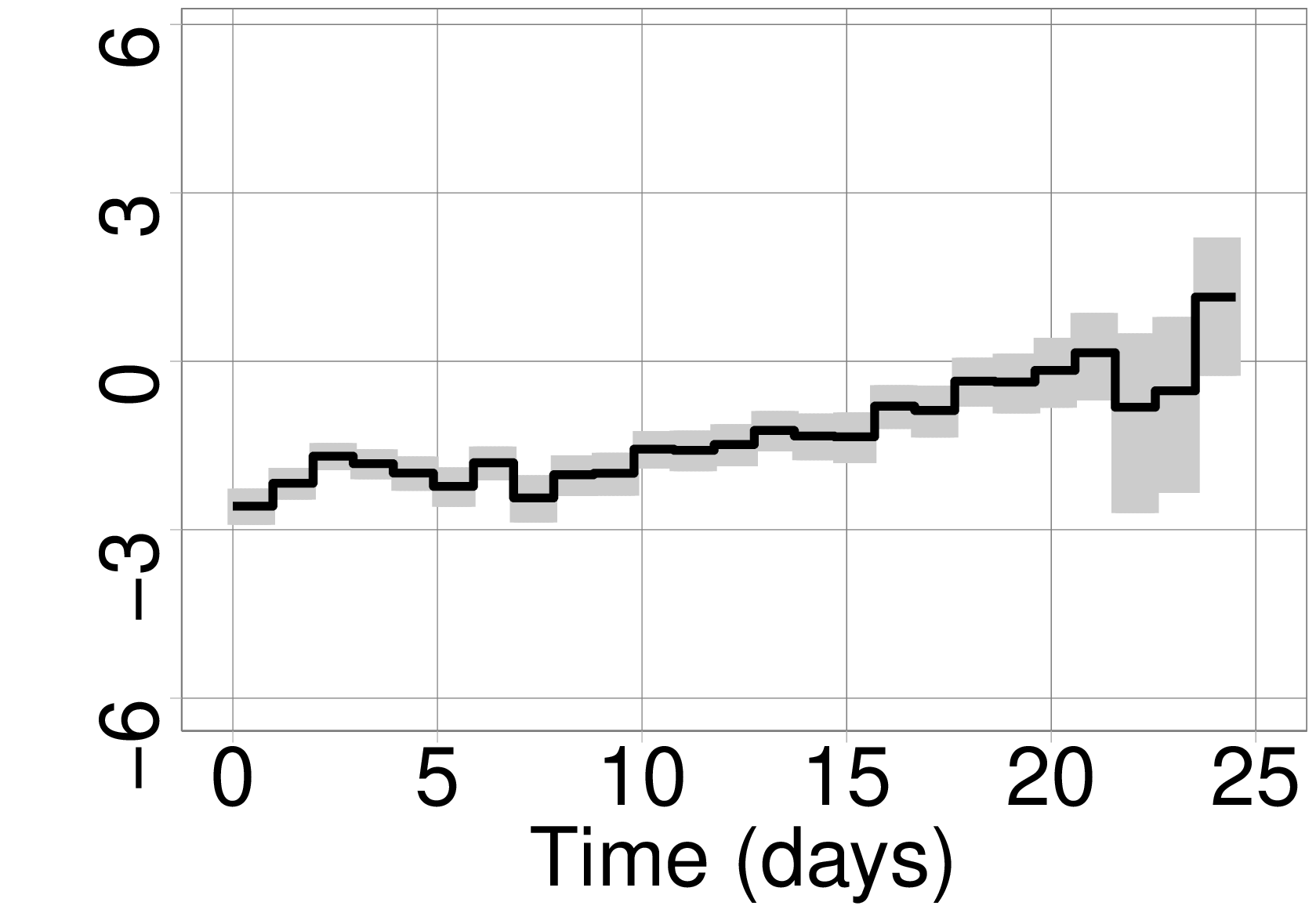}}\hspace*{-0.05cm}
 \subfigure[\textit{PC3}] {\includegraphics[width=3.1cm, height=3.8cm, angle=270]{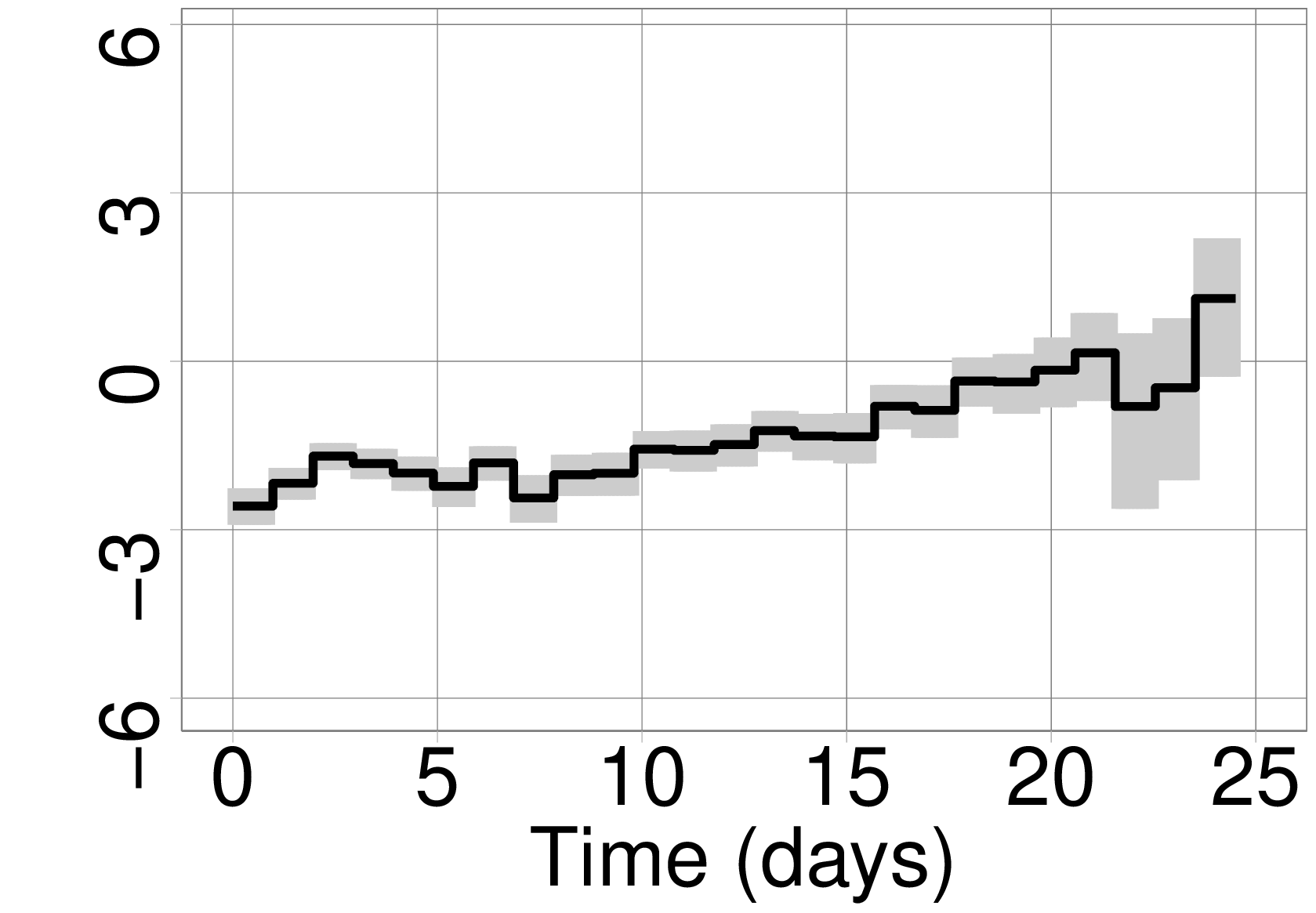}}\hspace*{-0.05cm}
 \subfigure[\textit{PC4}] {\includegraphics[width=3.1cm, height=3.8cm, angle=270]{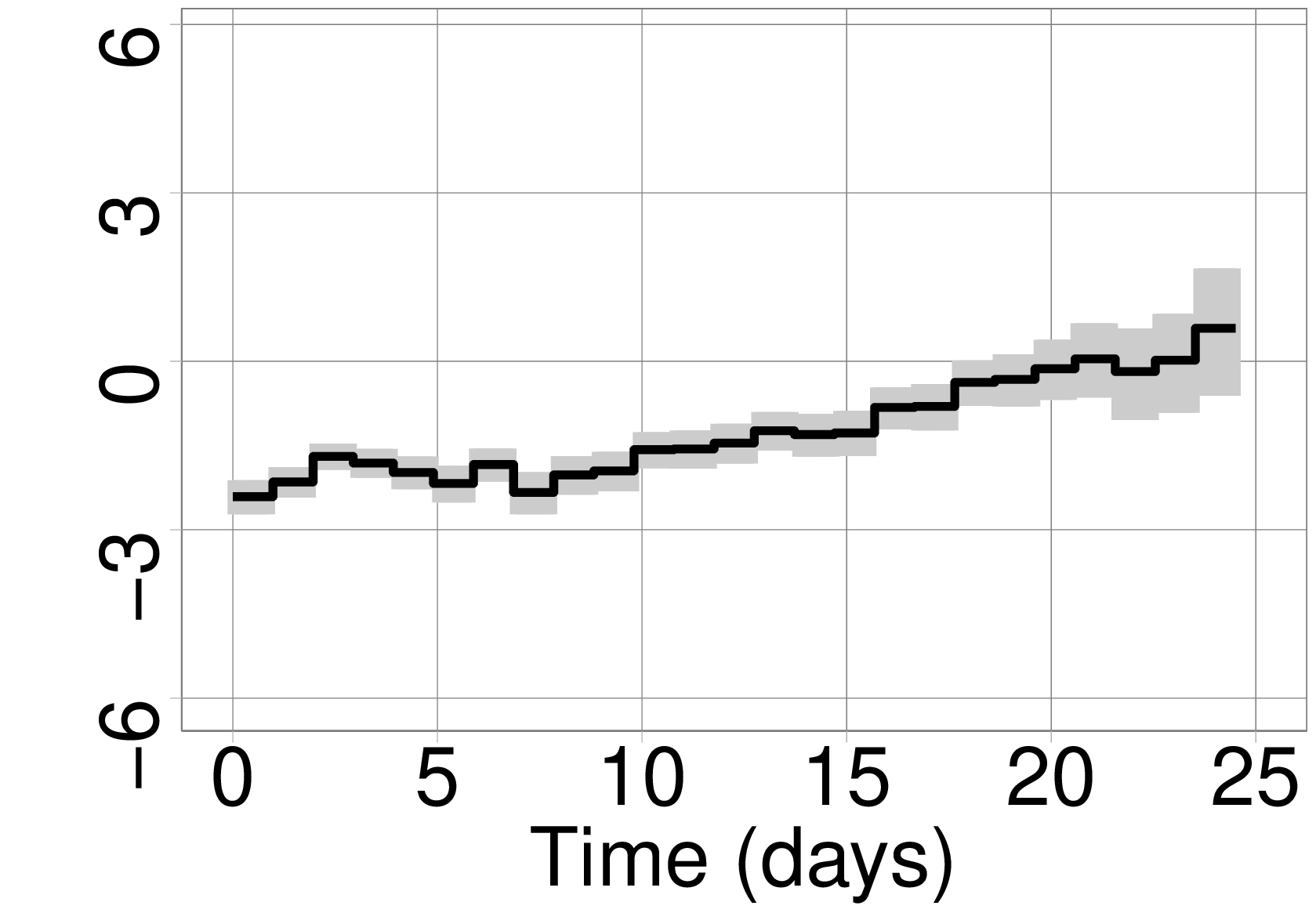}}\\\vspace{-0.30cm}
 \subfigure[\textit{PS1}] {\includegraphics[width=3.1cm, height=3.8cm, angle=270]{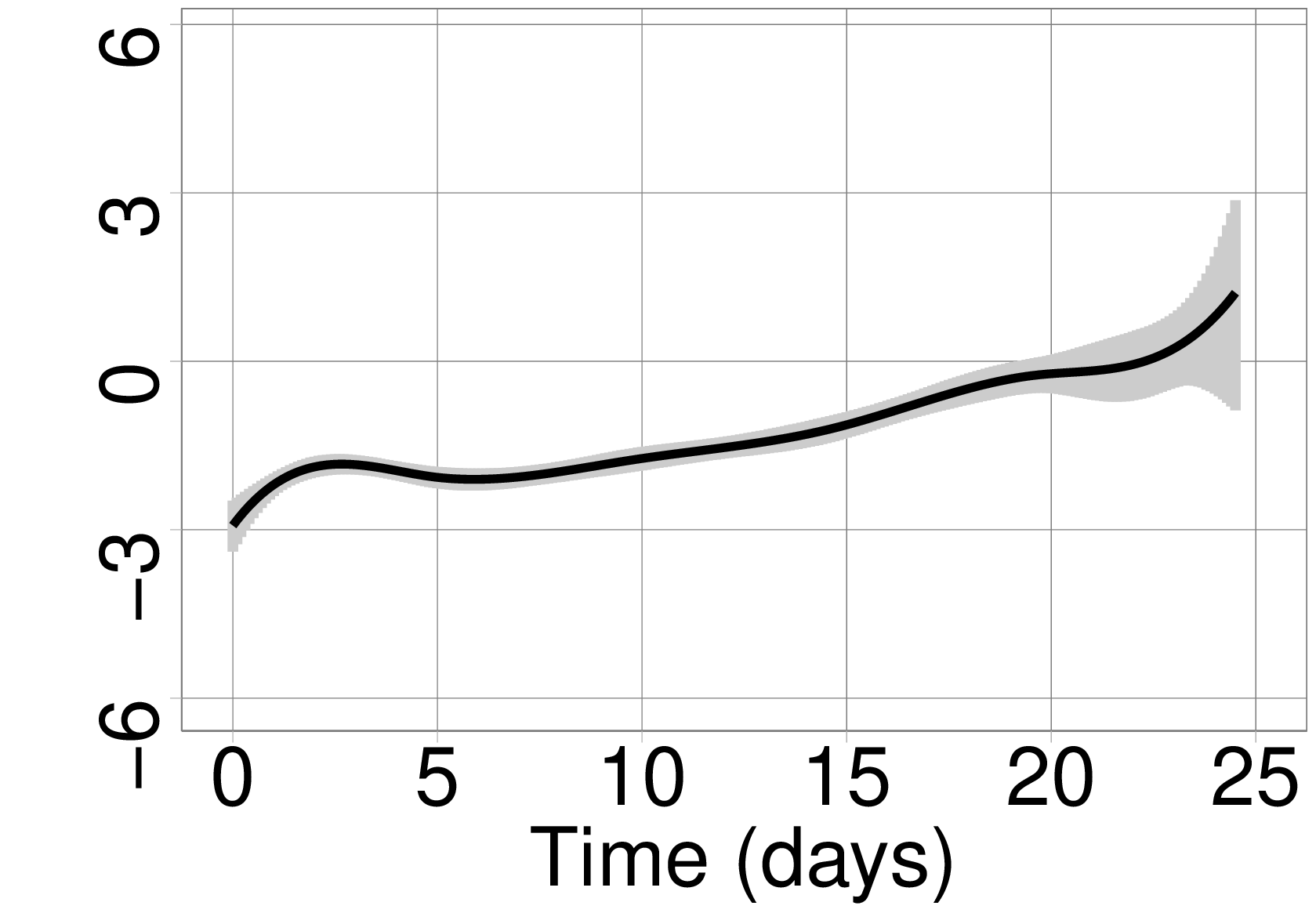}}\hspace*{-0.05cm}
 \subfigure[\textit{PS2}] {\includegraphics[width=3.1cm, height=3.8cm, angle=270]{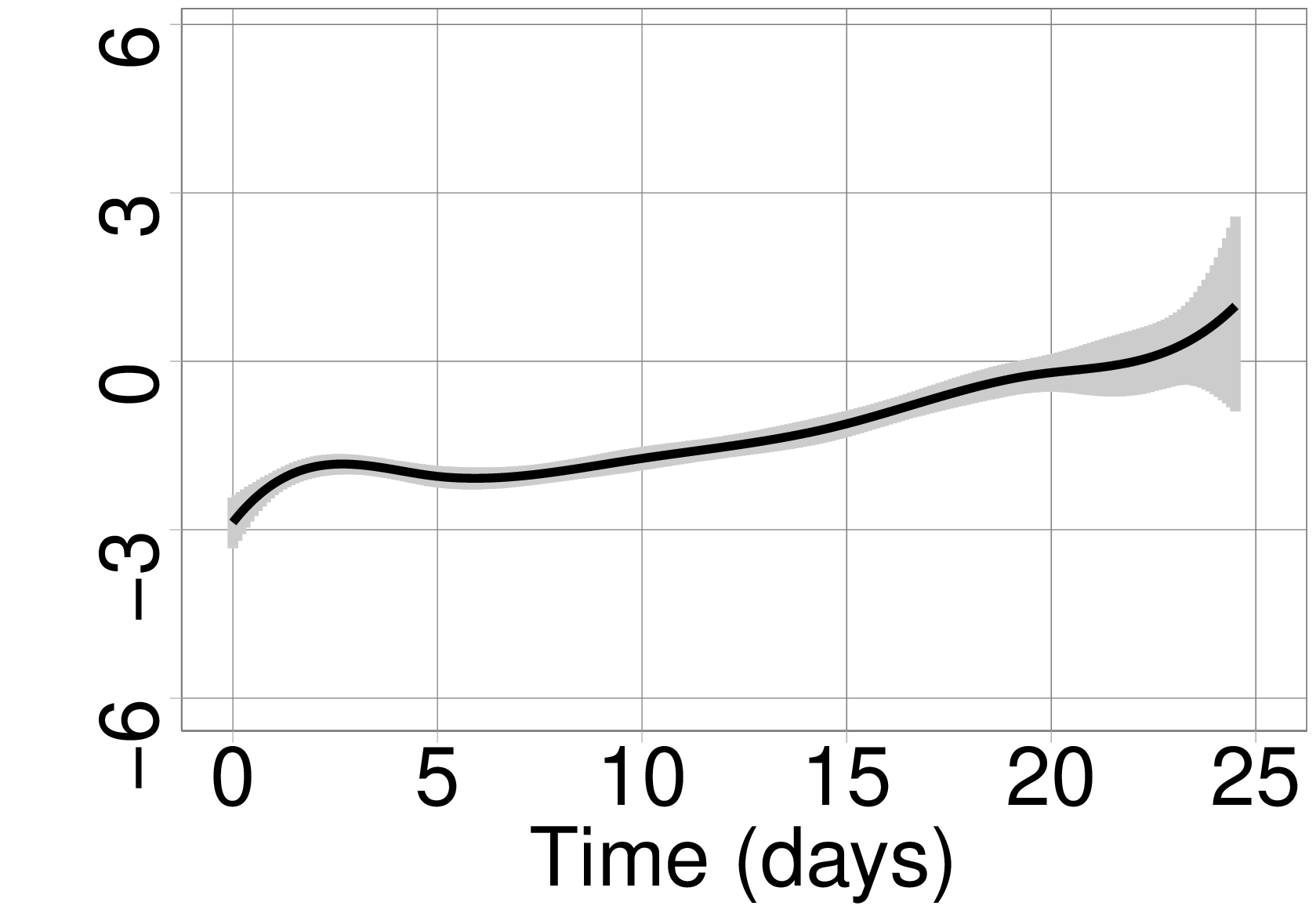}}\hspace*{-0.05cm}
 \subfigure[\textit{PS3}] {\includegraphics[width=3.1cm, height=3.8cm, angle=270]{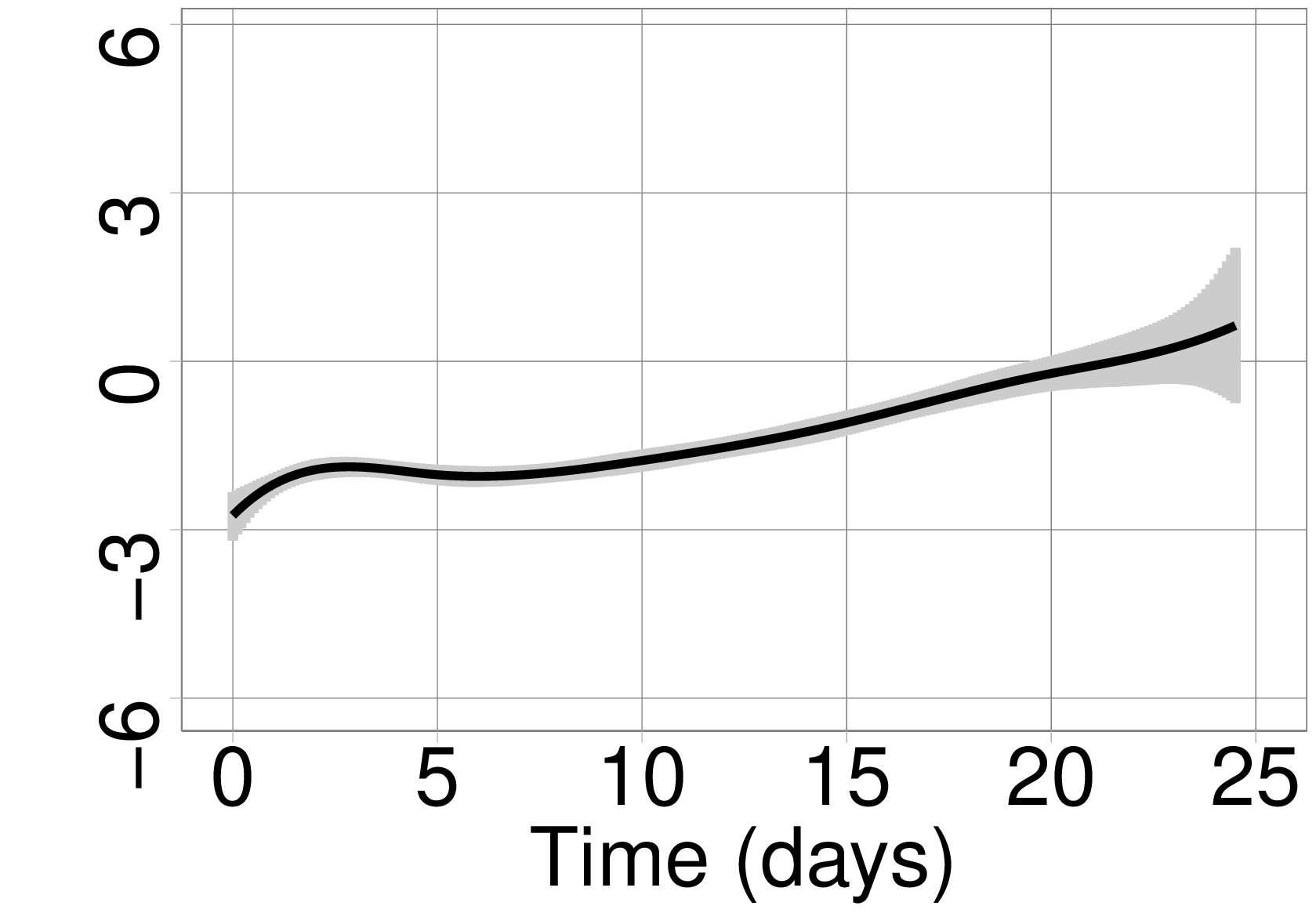}}
 \vspace{-0.30cm}
 \caption{Posterior  mean and 95\% credible interval for the log baseline hazard function  under Weibull (row one), $PC$ (row two) and $PS$ (row three) scenarios. $PC$ and $PS$ models are estimated with $K=25$ and $K=5$  knots, respectively.}
 \label{fig:3}
 \end{flushleft}
 \end{figure}

\begin{figure}[H]
\begin{flushleft}
 \subfigure[\textit{We}] {\includegraphics[width=3.1cm, height=3.8cm, angle=270]{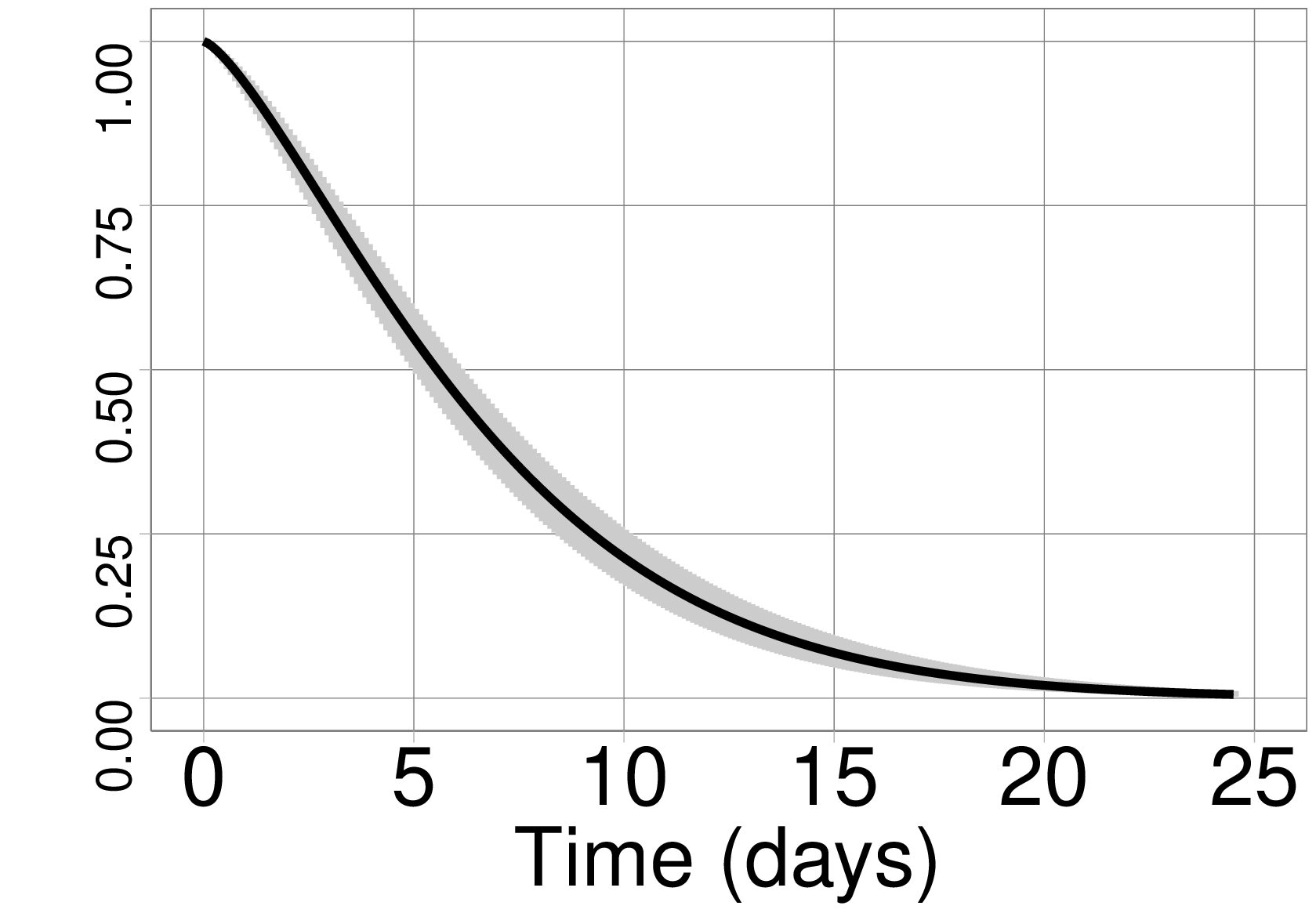}}\\\vspace{-0.30cm}
 \subfigure[\textit{PC1}] {\includegraphics[width=3.1cm, height=3.8cm, angle=270]{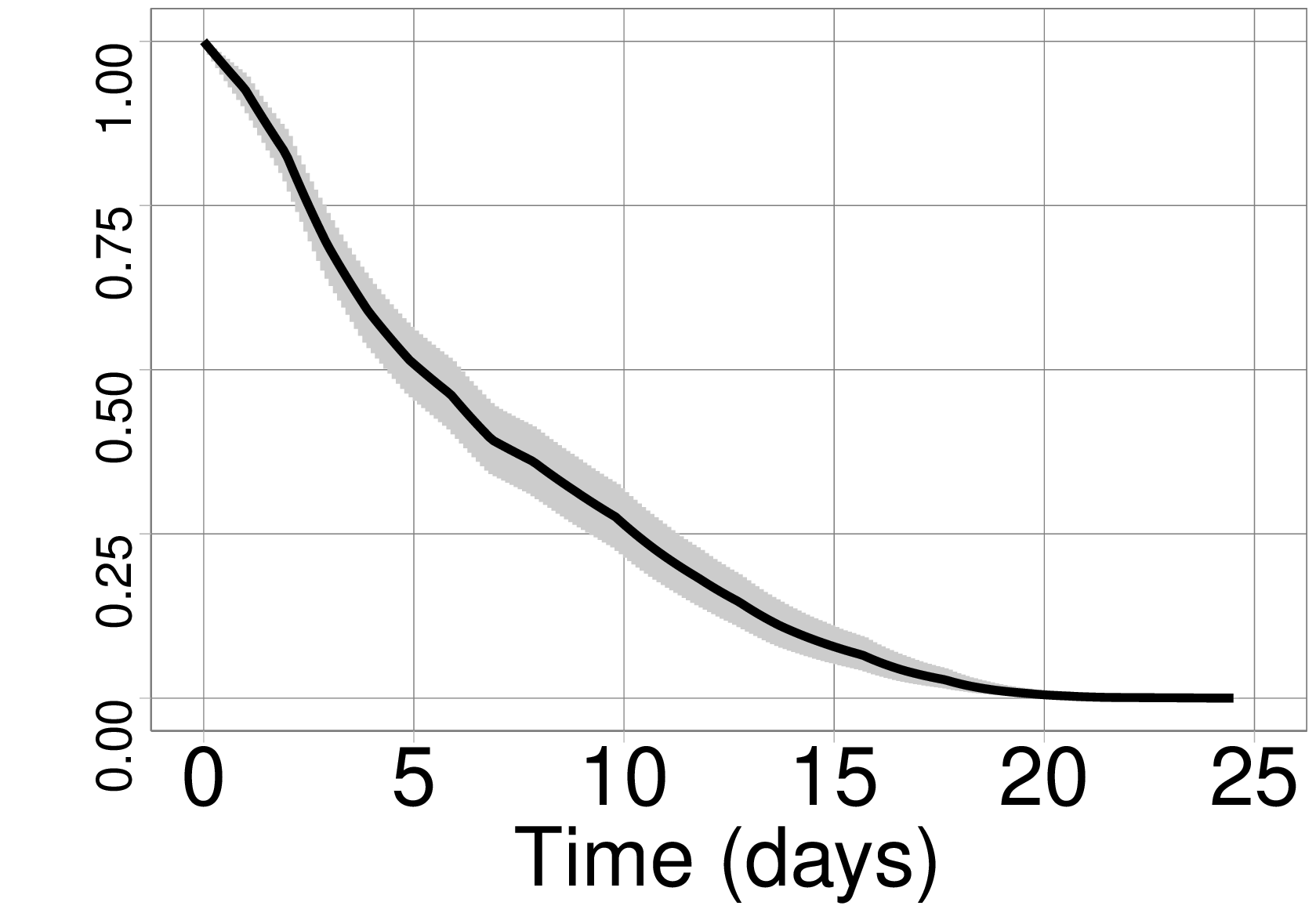}}\hspace*{-.05cm}
 \subfigure[\textit{PC2}] {\includegraphics[width=3.1cm, height=3.8cm, angle=270]{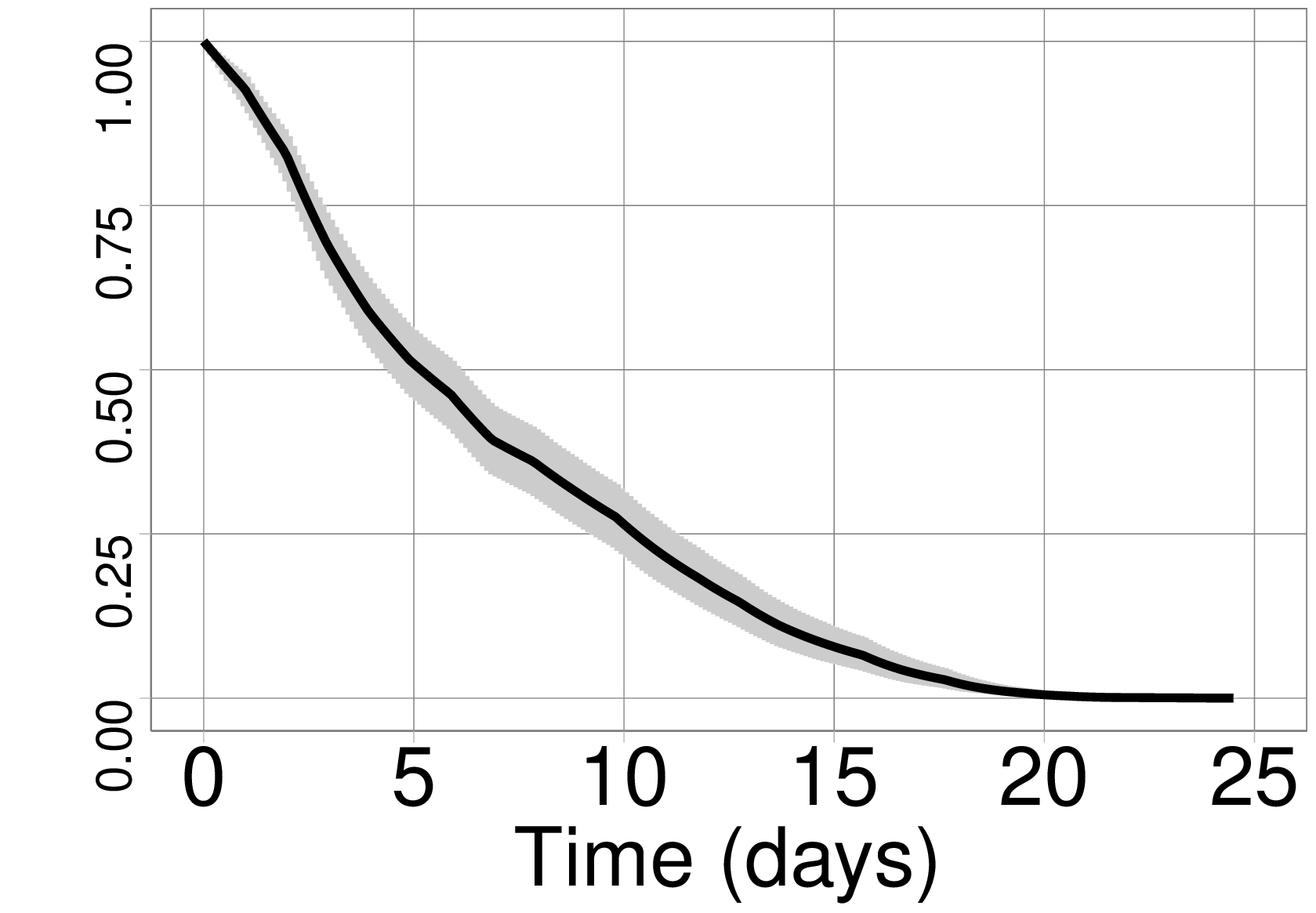}}\hspace*{-0.05cm}
 \subfigure[\textit{PC3}] {\includegraphics[width=3.1cm, height=3.8cm, angle=270]{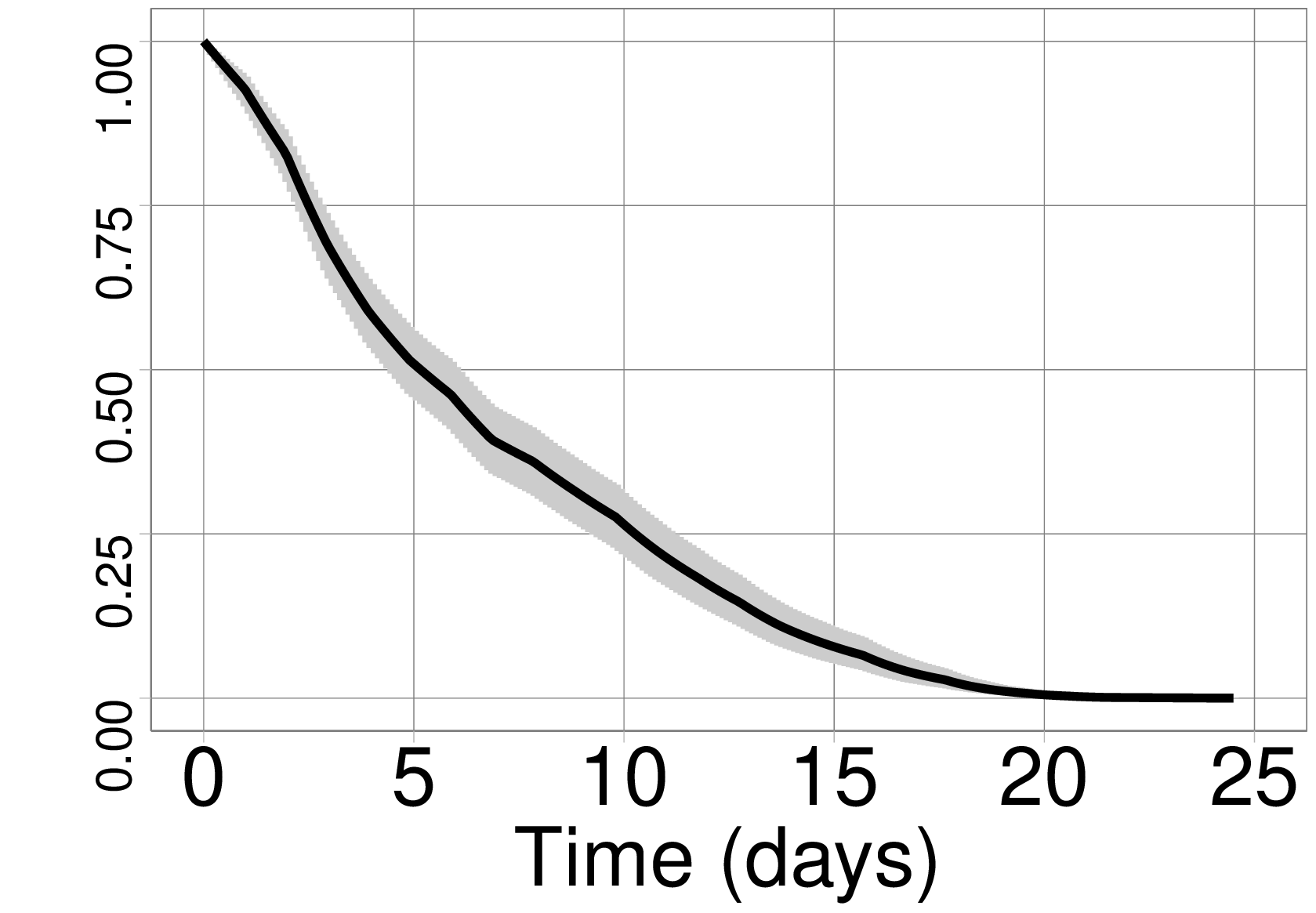}}\hspace*{-0.05cm}
 \subfigure[\textit{PC4}] {\includegraphics[width=3.1cm, height=3.8cm, angle=270]{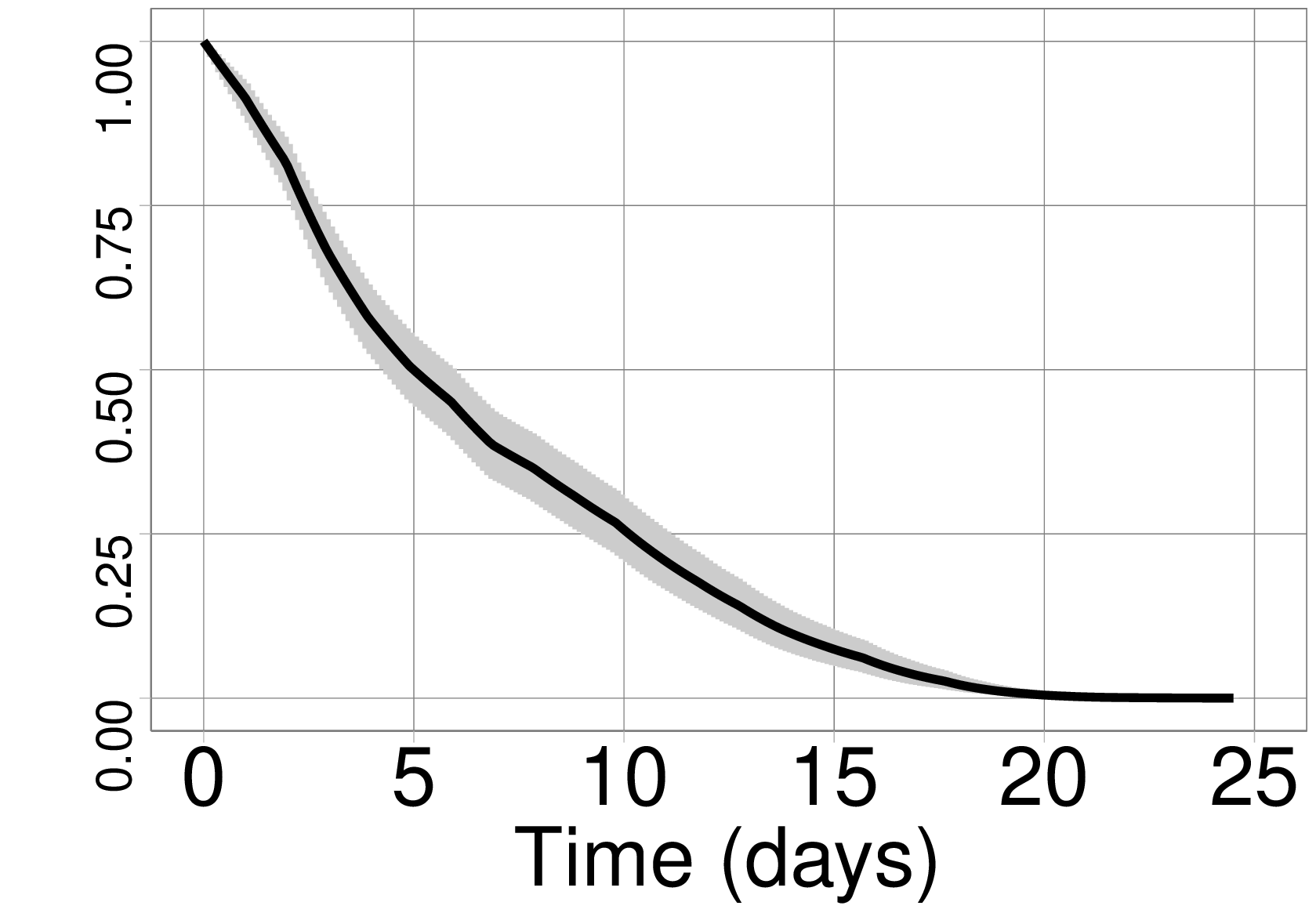}}\\\vspace{-0.30cm}
 \subfigure[\textit{PS1}] {\includegraphics[width=3.1cm, height=3.8cm, angle=270]{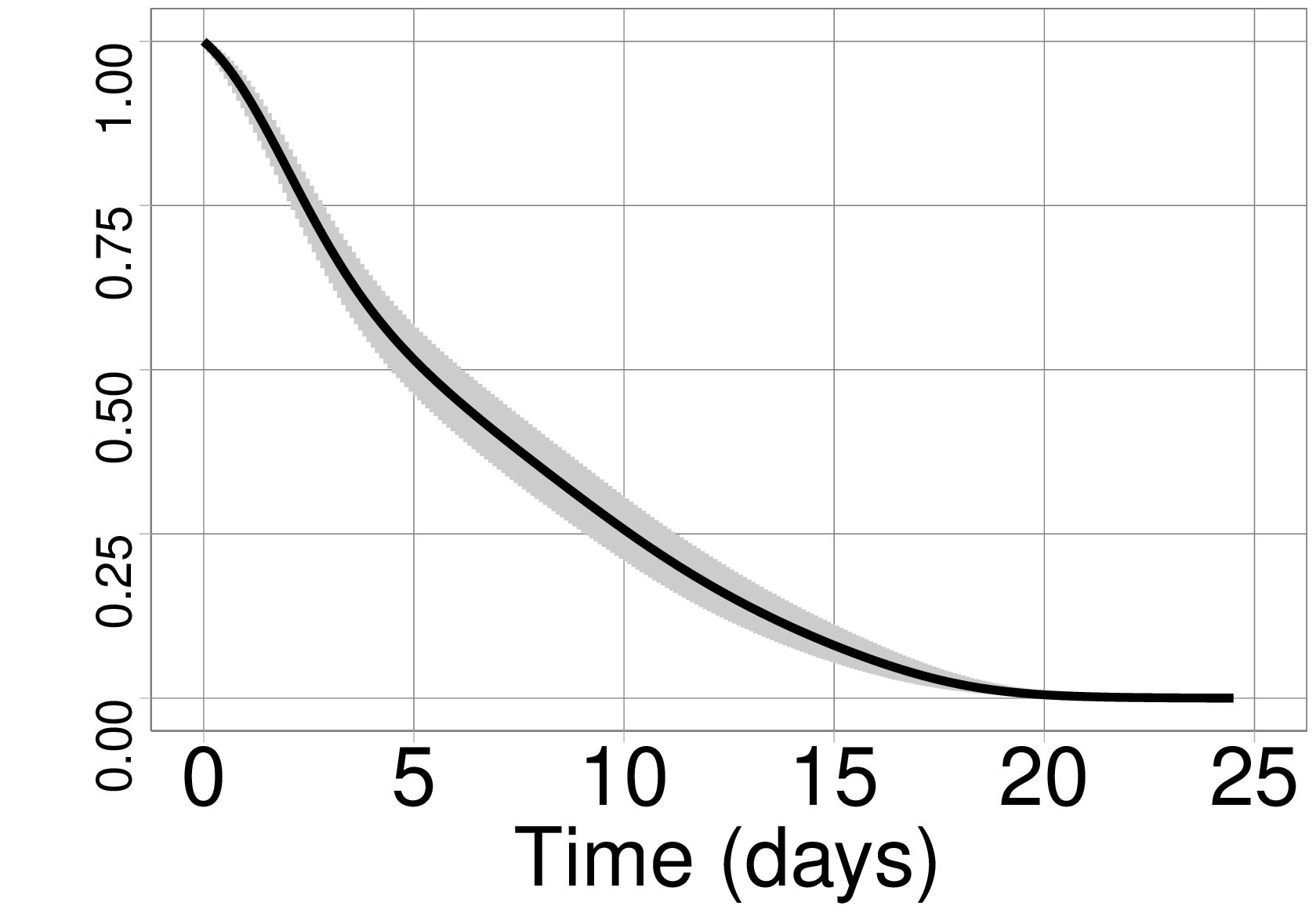}}\hspace*{-0.05cm}
 \subfigure[\textit{PS2}] {\includegraphics[width=3.1cm, height=3.8cm, angle=270]{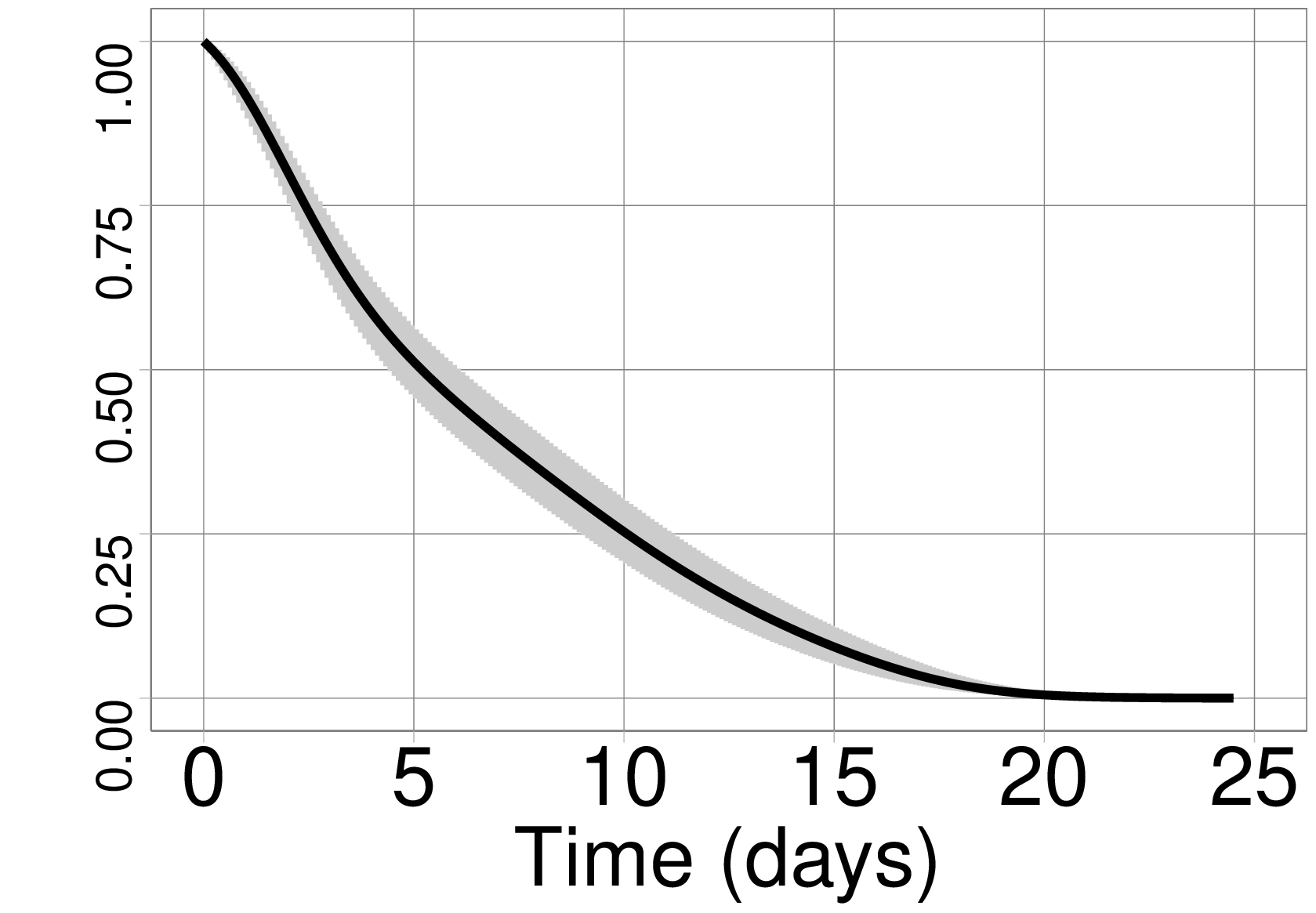}}\hspace*{-0.05cm}
 \subfigure[\textit{PS3}] {\includegraphics[width=3.1cm, height=3.8cm, angle=270]{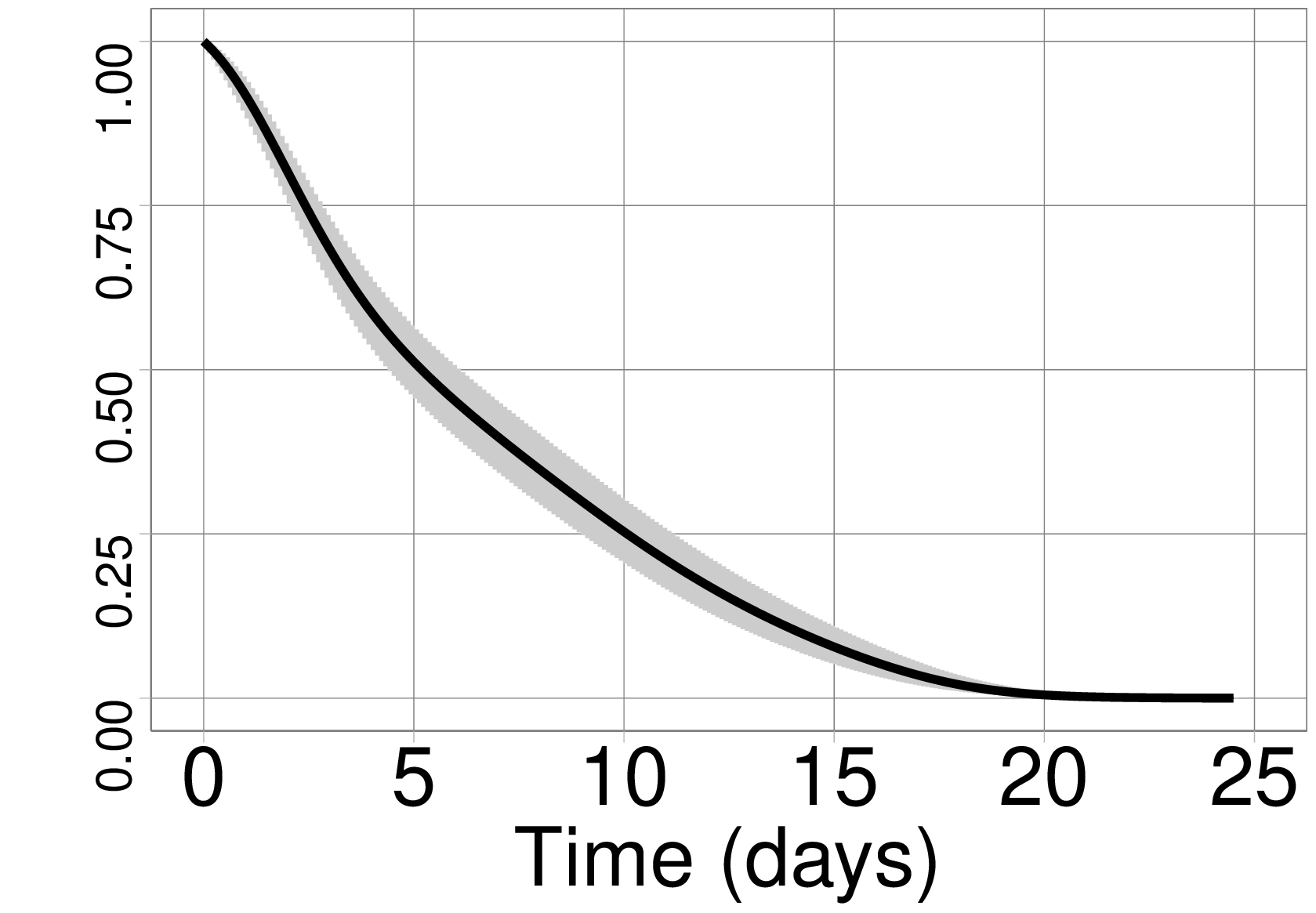}}\vspace{-0.30cm}
 \caption{Posterior  mean and 95\% credible interval for the baseline survival function  under Weibull (row one), $PC$ (row two) and $PS$ (row three) scenarios. $PC$ and $PS$ models are estimated with $K=25$ and $K=5$  knots, respectively.}
 \label{fig:4}
 \end{flushleft}
 \end{figure}

We now discuss the posterior distribution for $h_0(t)$ and  the survival function of the different models in the study. Models with $K=25$ knots were selected for $PC$ specification given that the $PC4$ $(K=25)$ showed the best performance based on DIC and LPML. For $PS$ specification, $PS1$ under $K=40$ showed the best performance based on DIC, but it was dismissed since it presented clear signs of overfitting and instability in the baseline hazard value associated to the last interval. Thus, models under $K=5$ were selected because the $PS1$ was the best model according to the two selection scores and it also shows similar values of pD to those of its counterparts ($PS2$ and $PS3$ under $K=5$). Figures~\ref{fig:3} and ~\ref{fig:4}  are a matrix of graphs for illustrating baseline hazard (logarithmic scale) and survival functions posterior distributions under $We$ (row one), PC $(K=25)$ (row two) and $PS (K=5)$ (row three) models.



 Baseline hazard  estimates are sensitive to their specification and their implicit regularization. The $We$ model  displays an increasing monotone  behaviour. \textit{PC} models report a general increasing trend  with different ups and downs. They show wider credible intervals in regions with very little data. The $PC4$ model evidences that Bayesian regularization not only smooths the  posterior mean but also reduces the uncertainty of
 the estimate. \textit{PS} models present a more flexible baseline hazard than $PC$'s and a regularization effect is mainly observed only in uncertainty estimates. On the contrary, estimates of posterior distribution $\pi(S_{0}(t) \mid \mathcal{D})$, which is encapsulated in the unit interval, are robust to baseline hazard function specification and differences between the different modelling proposals are imperceptible.\\

\begin{center}
 \textbf{FIGURES 3 AND 4 AROUND HERE}
 \end{center}

\subsection{Frequentist and Bayesian Cox model}
Although it is not a main objective of the article, we have performed a comparison of Bayesian Cox models against their frequentist counterparts. The comparison considered the three generic baseline hazard specifications $We$, $PC$ and $PS$ to be baseline hazard functions based on the four different partitions of the time axis exploited earlier ($K$ = 5, 10, 25, and  40). For the $PC$ and $PS$ models, we only considered models $PC1$ and $PS1$ due to their ``non-informative'' nature in prior specification. Frequentist Cox with Weibull baseline hazard was estimated   through the \texttt{survreg} function of the  \texttt{survival}  library. Results for the   Cox $PC$ and $PS$ models were obtained
  by the  \texttt{mexhaz}  function of the  \texttt{mexhaz} library, which uses the equivalence between  $PC$ models   and Poisson regression models \citep{holford1980, laird1981}.

 \begin{center}
 \textbf{TABLE 2 AROUND HERE}
 \end{center}
 
Table~\ref{tab:2} refers to the estimation of the hazard ratios HR$_{ST1}$ and HR$_{ST3}$. Bayesian reasoning provides the corresponding  posterior mean and
 95$\%$ credible interval. Frequentist statistics includes maximum likelihood estimates  and   95\% confidence intervals. Both estimation procedures are very stable, with   similar results for $PC$ and $PS$ models.


\section{Simulation study} \label{sec:4}

We continue with the  exploration of  the impact of the baseline hazard specification in the whole inferential process, specifically  the posterior estimates of the regression coefficients as well as  the posterior  for the hazard and survival function. We conduct three simulation studies (based on three different $h_0(t)$ definitions) to assess the performance of the Weibull, \textit{PC} and \textit{PS} definitions. $PC$ and $PS$ are also discussed with regard to   different  partitions of the time axis.

\subsection{Simulation scenarios}

Three  simulation scenarios  were generated  from a  CPH model  with different specifications for $h_0(t)$ as described below.\vspace{0.1cm}\\

\noindent \textbf{\textit{Scenario 1}}. A Weibull distribution with an increasing hazard function ($\alpha=1.5$ and $\lambda=0.5$).\vspace{0.15cm}\\
\noindent \textbf{\textit{Scenario 2}}. A mixture of five piecewise functions
\begin{equation*}
h_{0}(t \mid \boldsymbol{\varphi})= \sum_{k=1}^5\, \varphi_{k} \,I_{(c_{k-1},c_{k}]}(t), \,\, t>0,
\end{equation*}

\noindent where $\varphi_1=0.5$ in $0<t\leq0.2$,  $\varphi_2=2.5$ in $0.2<t\leq0.4$, $\varphi_3=0.5$ in $0.4<t\leq0.6$, $\varphi_4=1$ in $0.6<t\leq0.8$,  and $\varphi_5=1.5$ in $t>0.8$.\vspace{0.15cm}\\
\noindent \textbf{\textit{Scenario 3}}. A mixture of two Weibull distributions

\begin{align*}
 \hspace*{-0.5cm}h_{0}(t \mid \alpha_1,\alpha_2,\lambda_1,\lambda_2)=
  \dfrac{\lambda_1\,\alpha_1\,t^{\alpha_{1}-1}\,p\,\mbox{exp}\{-\lambda_1\,t^{\alpha_1}\}+\lambda_2\,\alpha_2\,t^{\alpha_{2}-1}\,(1-p)\,\mbox{exp}\{-\lambda_2\,t^{\alpha_2}\}}{p\,\mbox{exp}\{-\lambda_1\,t^{\alpha_1}\}+(1-p)\,\mbox{exp}\{-\lambda_2\,t^{\alpha_2}\}},\,\,t>0
\end{align*}
\noindent  with shape $\alpha_{1}=3$, $\alpha_{2}=1$,  scale $\lambda_{1}=\lambda_{2}=0.5$, and mixing probability parameter $p=0.2$.\\

These scenarios included an indicator covariate with regression coefficient $\beta=1$. Data were assigned to each  group according to a Bernoulli distribution with probability 0.5. We considered   right censoring at time $C_R$. It was previously fixed for each scenario from the condition  $S_{0}(C_{R})=0.3$ for the baseline survival function.  Each scenario was replicated   $R=100$ times for sample sizes of $N=100$ and $N=300$.

All the simulated dataset were analysed via each of the stated modellings discussed in Section 2. The estimation of the \textit{PC} and \textit{PS} models   was based on two different partitions of the time axis with   $K$ = 5 and 15 knots with intervals of the same length (\citep{murray2016}). The last knot in all models corresponds to the previously referred censored time ($C_{R}$), which is the longest survival time observed.

\subsection{Generating survival times}

We follow the inversion method~\citep{bender2005,austin2012,crowther2013} to simulate survival data for \textit{Scenarios 1} and \textit{2}. This method is based on  the relationship between the cumulative distribution function (CDF)   of a survival random variable   and    a standard uniform random variable. It can be directly applied when the subsequent CDF  has a closed form expression and can be directly inverted and   easily implemented with \texttt{R}~\citep{r2013} packages \texttt{simsurv}~\citep{brilleman2017} and \texttt{SimSCRPiecewise}~\citep{chapple2016}. The inversion method for  \textit{Scenario 3} is not directly suitable. The subsequent cumulative hazard function cannot  be directly inverted and we have used iterative root-finding techniques~\citep{crowther2013} to solve it. This procedure is implemented for the \texttt{R} software~\citep{r2013} in the \texttt{simsurv}~\citep{brilleman2017} package. Further details of the inversion method and its corresponding extension to simulate complex baseline hazard functions are described in the supporting information.

\subsection{Posterior inferences}

Each simulation dataset was used to estimate all the    survival models with all the    specifications of $h_0(t)$ and the different prior scenarios in Section 2. Posterior distributions were approximated by  \texttt{JAGS} software~\citep{plummer2003} based  on    three parallel chains with 20,000 iterations each plus another 2,000   for the burn-in period. Moreover, the chains were additionally thinned by storing every 10th draw  to reduce   autocorrelation in the sequences. Convergence  of the chains to the posterior distribution  was guaranteed by monitoring in all inferences to ensure that the potential scale reduction factor was close to 1 and the effective number of independent simulation draws was greater than 100.

\subsection{Regression coefficients and baseline hazard function}
We considered $R=100$ replicas of each inferential process and, consequently, we constructed 100 approximate random samples of the   posterior distribution for $\beta$.    Let $\{\beta^{(1)}_{(r)}, \ldots, \beta^{(N)}_{(r)} \}$ be the approximate MCMC sample of size $N$ of the posterior marginal distribution for   $\beta$ corresponding to the replica $r$.

The stability of the posterior distribution for the regression coefficients were assessed by means of the following measures:
\begin{itemize}
\item \textbf{Bias}: Difference between the average of the posterior sample means of the  replicas and the true regression coefficient, $(\sum_{r=1}^{R} \,\bar{\beta}_{(r)} /R)-\beta$, where $\bar{\beta}_{(r)}$  is the sample mean of the posterior sample corresponding to the replica $r$.
\item \textbf{Standard error (SE)}: Square root $\sqrt{\sum_{r=1}^{R}\, s_{(r)}^2 / R}$ of the average of the   posterior variances $s_{(r)}^2$ of the replicas.
\item \textbf{Standard deviation (SD)}: Standard deviation of the set $\{\bar{\beta}_{(1)}, \ldots, \bar{\beta}_{(R)} \}$  that includes the posterior sample  mean of the regression coefficient of all  replicas.
\item \textbf{Coverage probability (CP)}: Proportion of the $R=100$  95$\%$ credible intervals  which contain the true value of the regression coefficient.
\end{itemize}

The performance of the set of models considered was also evaluated   in terms of the posterior   baseline hazard estimates (logarithmic transformation). For the posterior sample of each replica we  construct an approximate posterior sample of the log baseline hazard function at each time, whose average can be used as a point estimate  of the true baseline hazard at that time.  We then    merge the information of all the replicas to obtain a global estimation, log$(\widehat{h}_0(t))$, by calculating their average.   This procedure is also useful for   extracting information about   the posterior variability and constructing, for example,      95$\%$ credible intervals for the posterior of the baseline hazard at each time.

 The accuracy of the estimation was measured through the difference  between the posterior estimation of the baseline hazard and the true hazard function. A general measure that accounts for this difference over the time period of the study is   the root-mean squared deviation (RMSD), computed as
\begin{equation}
\mbox{RMSD}\,=\,\sqrt{\dfrac{\sum_{m=1}^{M} [\mbox{log}(\widehat{h}_{0}(t_{m}))-\mbox{log}(h_{0}(t_{m}))]^{2} }{M}},
\end{equation}
\noindent    a discrete approximation  based on the idea of the Riemann sums to approach  an integral. At this point, we would like to note that we have used a wide  partition of the time axis,  with knots spaced at $0.01$ time points  from 0 to the maximum time value of each scenario. This maximum time value is determined by the corresponding censoring time ($C_{R}$).\\

\begin{center}
\textbf{TABLES 3, 4, 5 AROUND HERE}\\
\end{center}

Tables~\ref{tab:3}, \ref{tab:4} and \ref{tab:5}  display the values of the average, bias, SE, SD and CP (related to $\beta$ and RMSD (related to log($h_{0}(t)$)) referring to the three simulation scenarios. In relation to the $\beta$ estimate, the $We$ model is very stable for the three scenarios and the effect of $N$ is not appreciated. $PC$ and $PS$ models approximate the regression coefficient quite well, which is slightly affected by the number of knots ($K$) and the sample size ($N$).


Under \textit{Scenario 1}, the $We$ models provide the closest fit to the true function with the lowest RMSD values. \textit{PS} models are generally  better than \textit{PC}'s,  which show the worst performance,  possibly because of their non-continuous behaviour. Under \textit{Scenario 2},  $PC4$ models (for $N = 100$ and $N = 300$) provide the closest fit to the true function with the lowest RMSD values, thereby underlining  the relevance of  sensitivity to prior scenarios. \textit{PS} models also seem to capture the behaviour of the true function, on the whole, showing RMSD values lower than the \textit{PC1}, \textit{PC2}, \textit{PC3} models. The $We$ models present the highest RMSD. Under \textit{Scenario 3}, $PS$ models provide the lowest RMSD values as a general rule. \textit{PS3} specification shows the lowest values for all $K$ configurations. The $We$ models present higher RMSD estimates in relation to $PS$'s. Between \textit{PC}'s, $PC4$ specification improves the RMSD values of its $PC$ counterparts. For all scenarios, the prior distribution has a strong effect on the baseline hazard estimation of \textit{PC} models.

\begin{figure}[H]
\centering
 \subfigure[{\scriptsize\textit{We}. RMSD=0.039 }] {\includegraphics[width=3.1cm, height=3.8cm, angle=270]{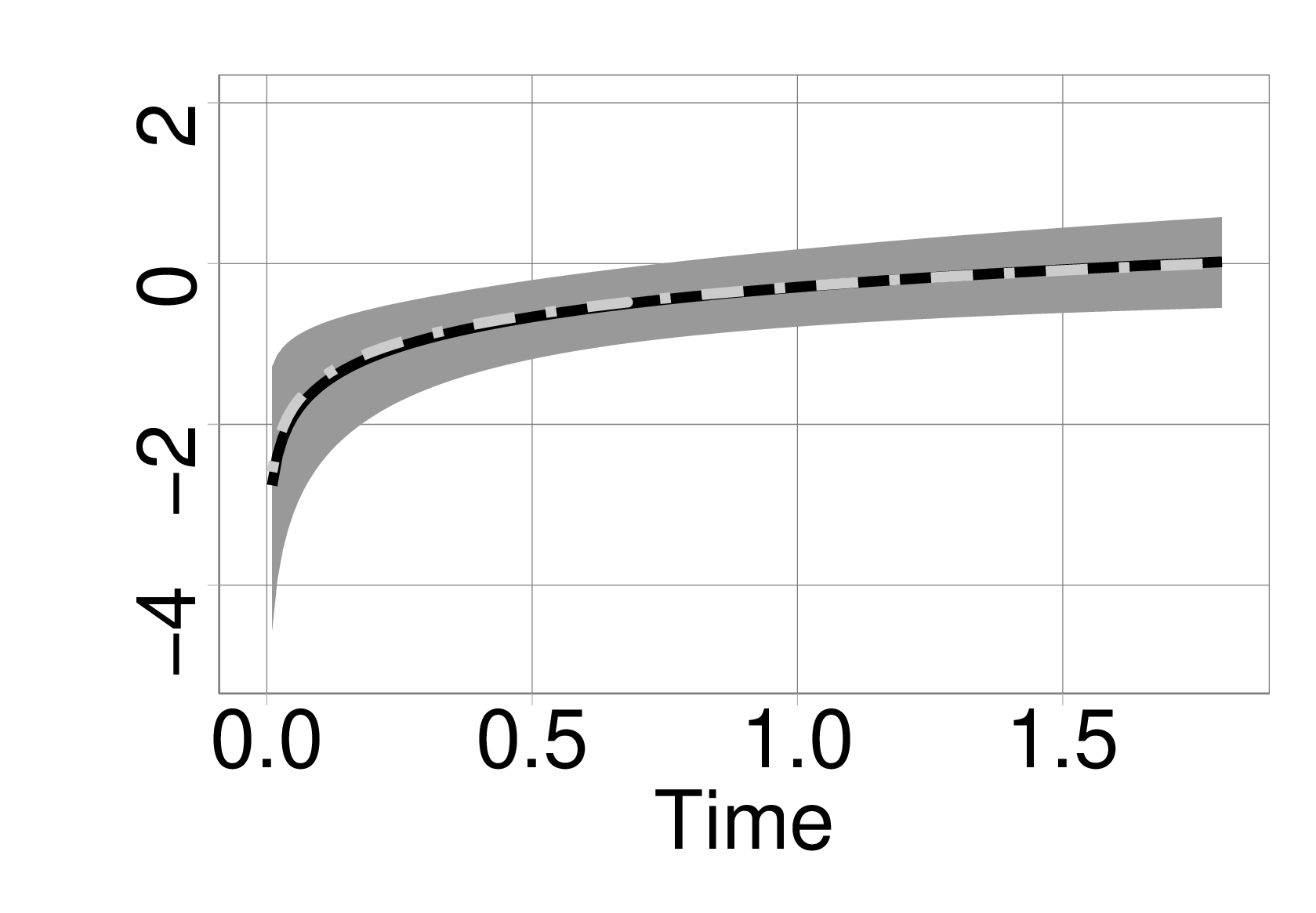}}\hspace*{-0.05cm}
 \subfigure[{\scriptsize $PC1$ ($K=5$). RMSD=0.205}] {\includegraphics[width=3.1cm, height=3.8cm, angle=270]{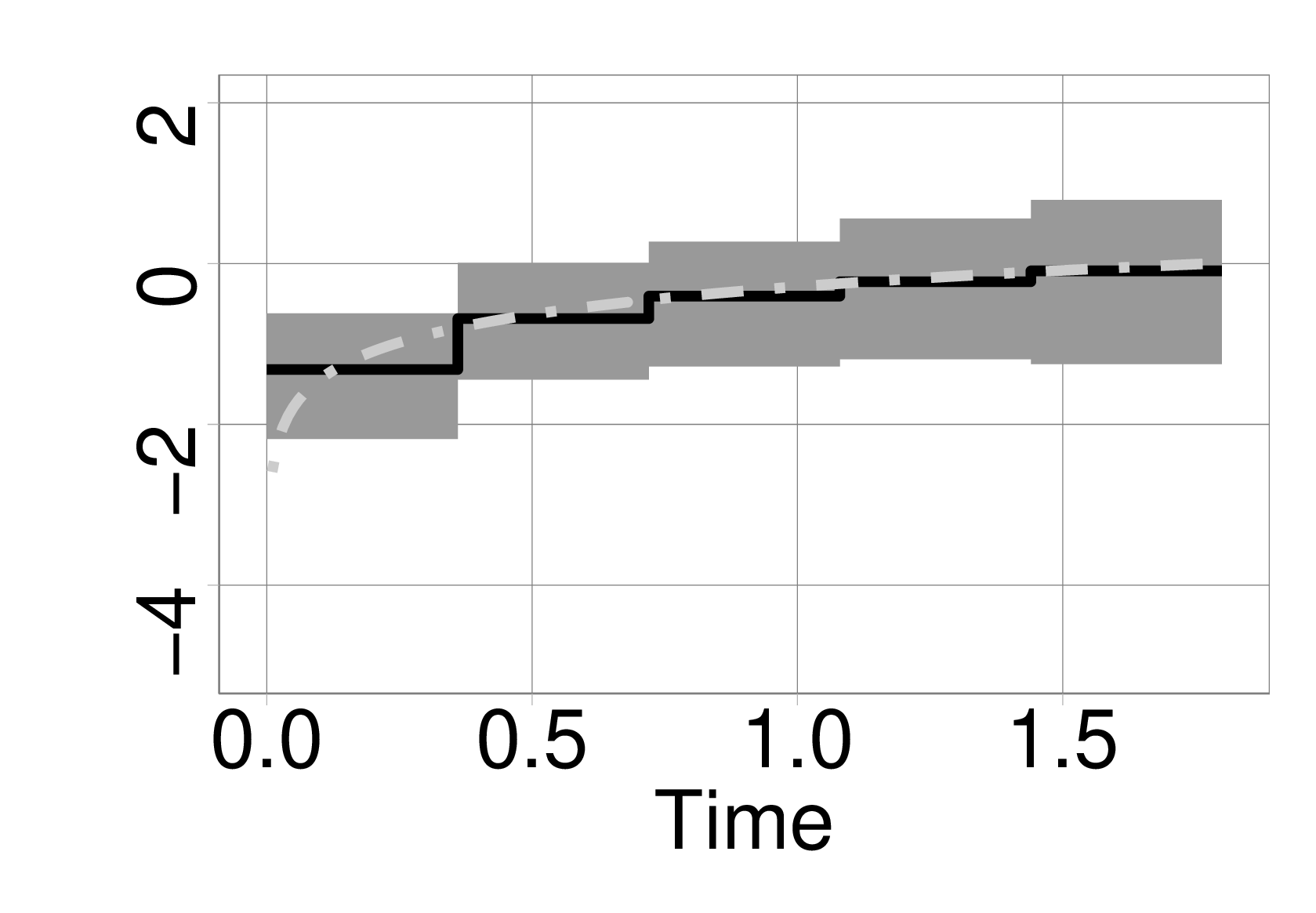}}\hspace*{-0.05cm}
 \subfigure[{\scriptsize \textit{PS2}($K=5$). RMSD=0.095}] {\includegraphics[width=3.1cm, height=3.8cm, angle=270]{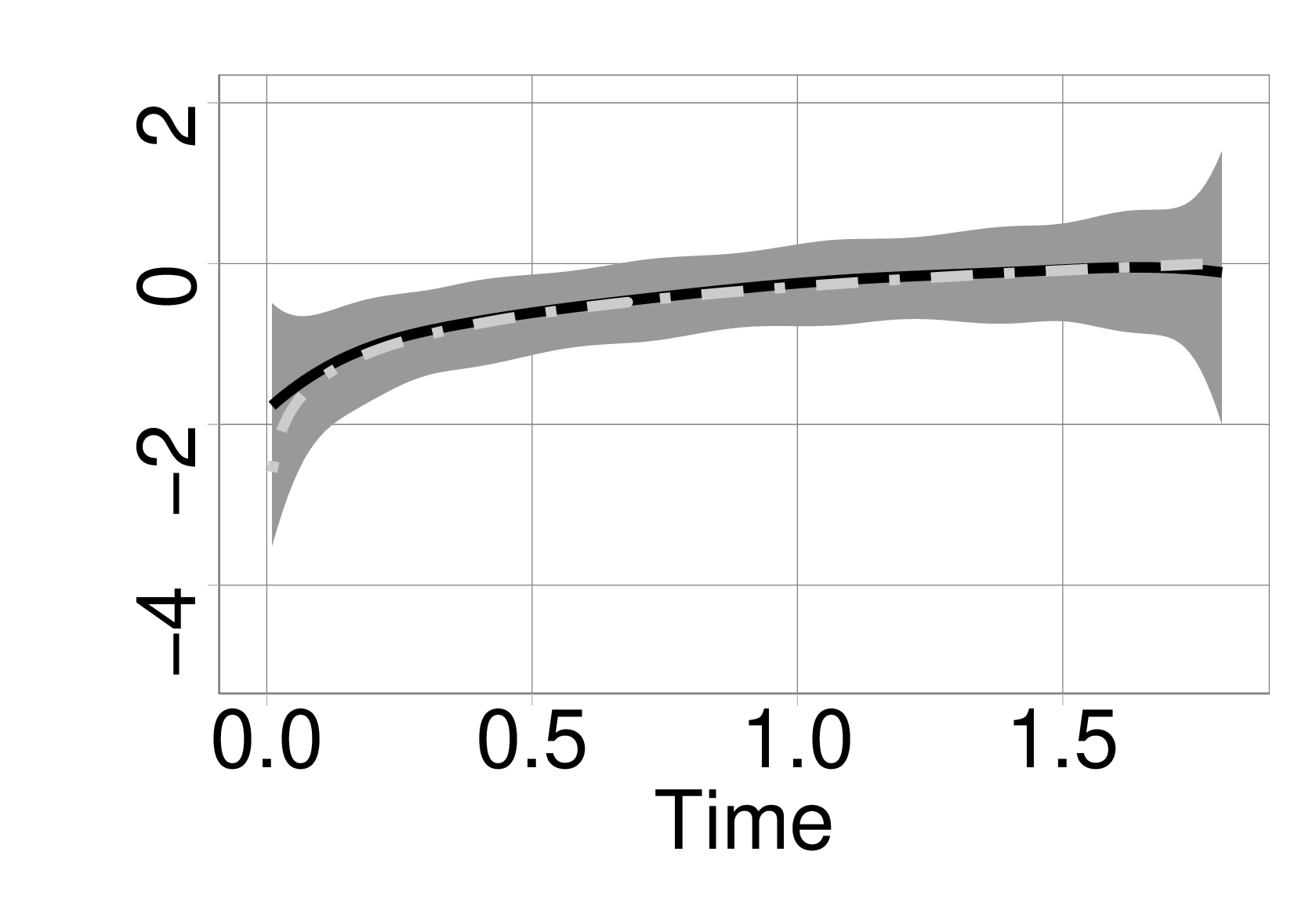}}\\\vspace{-0.30cm}
 \subfigure[{\scriptsize \textit{We}. RMSD=0.007}] {\includegraphics[width=3.1cm, height=3.8cm, angle=270]{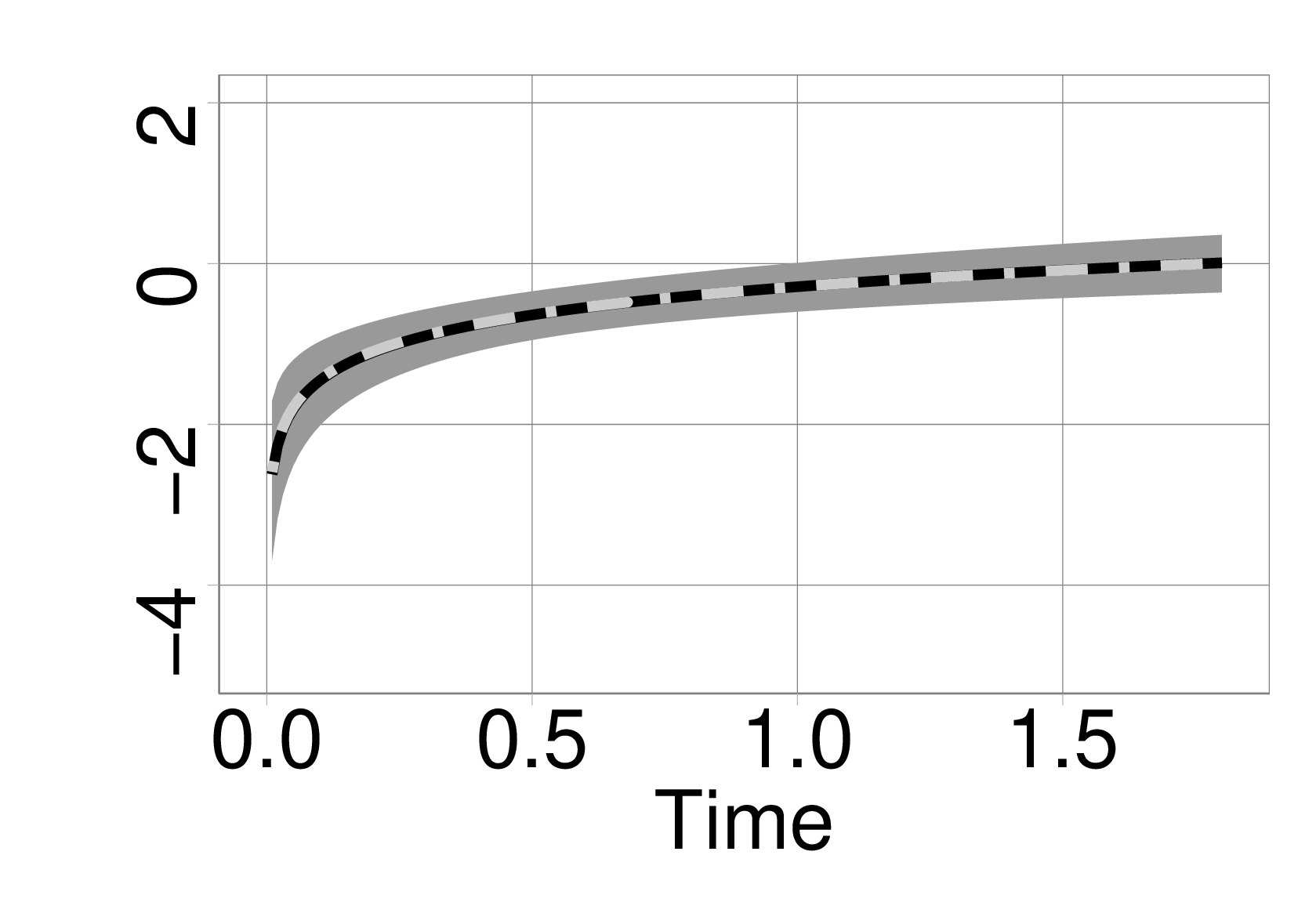}}\hspace*{-0.05cm}
 \subfigure[{\scriptsize \textit{PC3} ($K=15$). RMSD=0.130}] {\includegraphics[width=3.1cm, height=3.8cm, angle=270]{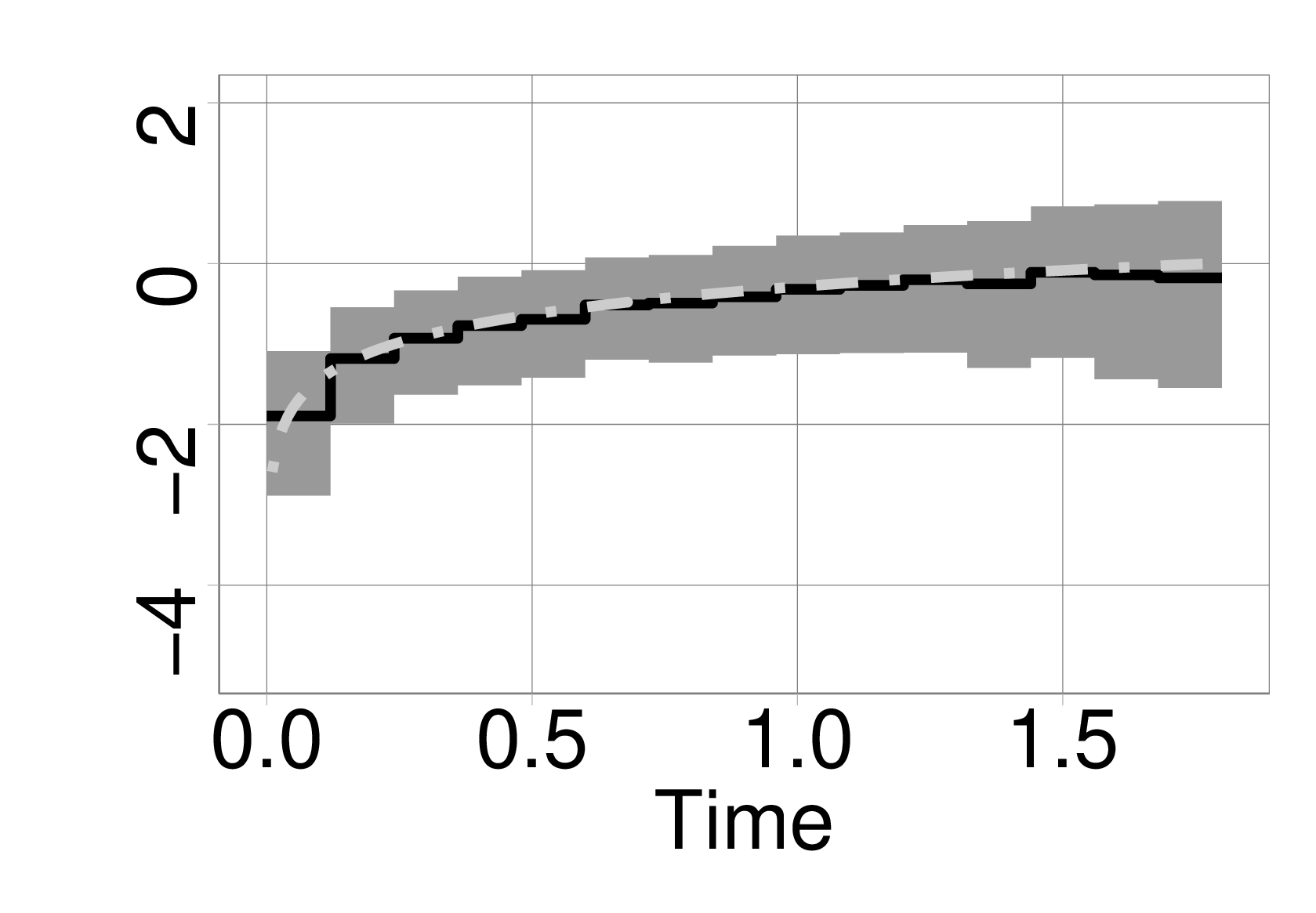}}\hspace*{-0.05cm}
 \subfigure[{\scriptsize \textit{PS1}($K=5$). RMSD=0.063}] {\includegraphics[width=3.1cm, height=3.8cm, angle=270]{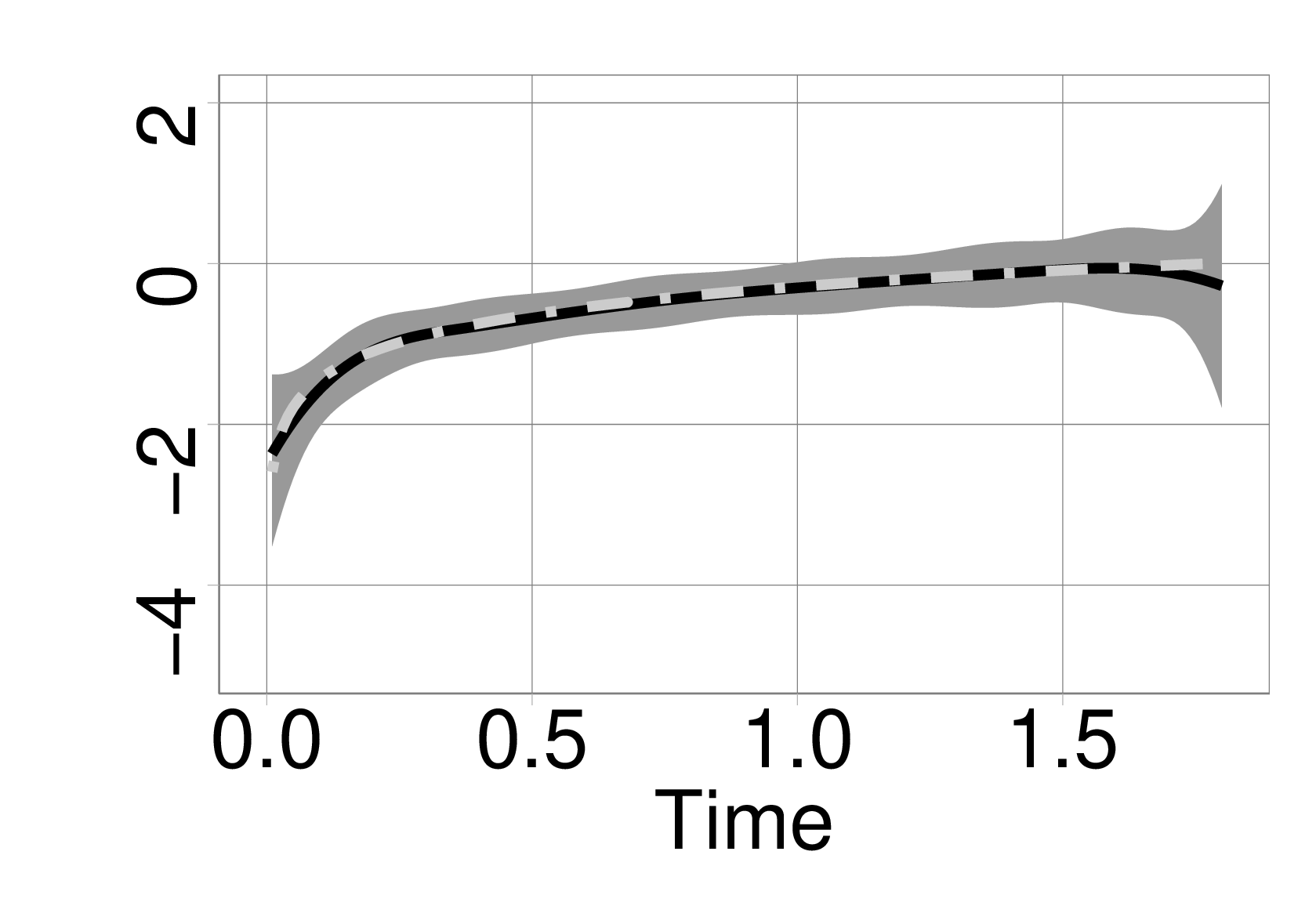}}\\\vspace{-0.30cm}

 \caption{Average replica pointwise of the posterior approximate means of the log-baseline hazard estimate (black solid line), average replica of the posterior 95\% credible intervals (dark grey area), and true log-baseline hazard function (grey dash-dotted line) in the simulated \textit{Scenario 1} under the $We$,  $PC1$ ($K=5$), $PS2$ ($K=5$) for $N$=100 (row 1) and under the $We$,  $PC3$ ($K=15$), $PS1$ ($K=5$) for $N$ = 300 (row 2).}
 \label{fig:5}
 \end{figure}
\begin{figure}[H]
\centering
 \subfigure[{\scriptsize\textit{We}. RMSD=0.626 }] {\includegraphics[width=3.1cm, height=3.8cm, angle=270]{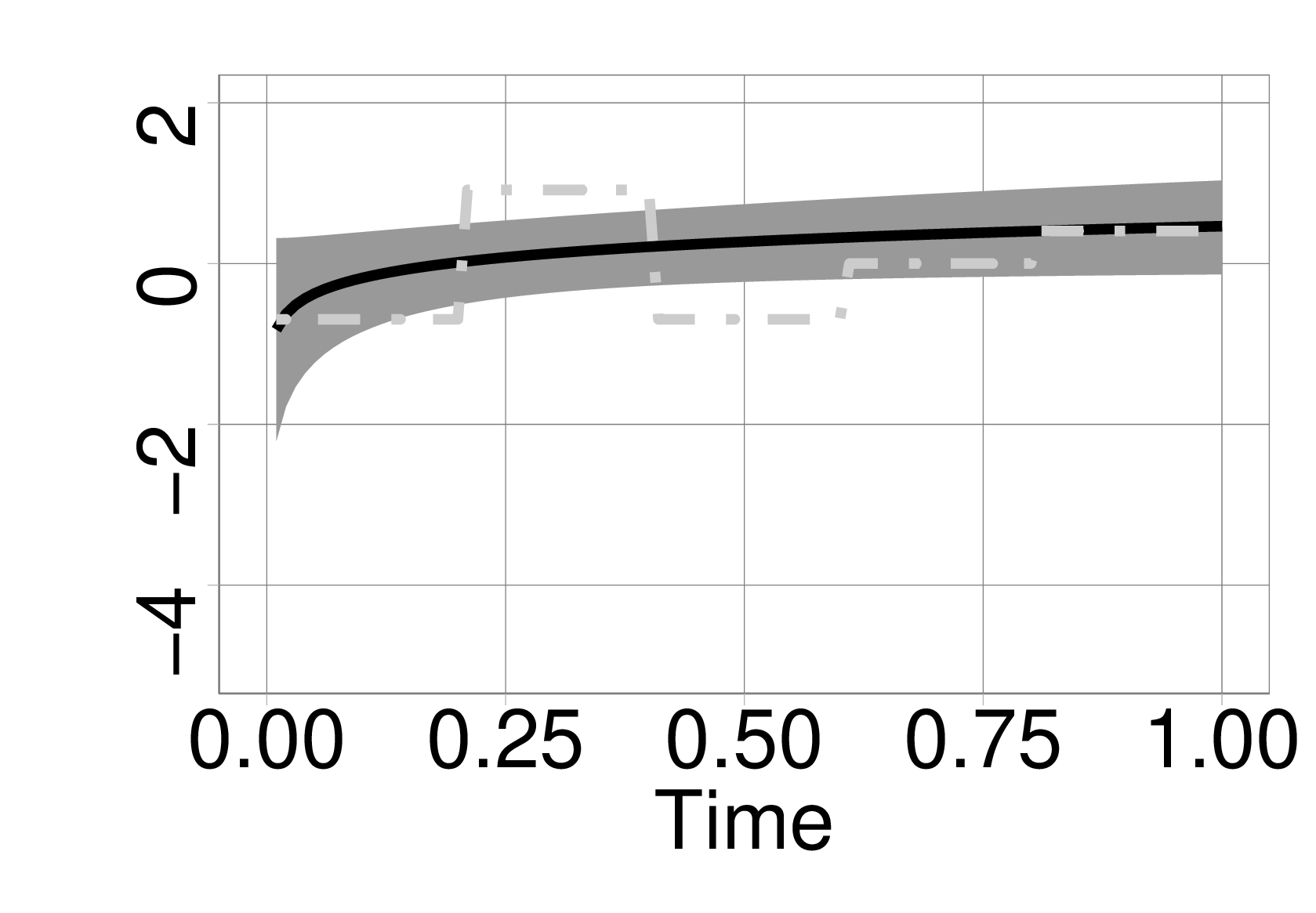}}\hspace*{-0.05cm}
 \subfigure[{\scriptsize \textit{PC4} ($K=5$). RMSD=0.095}] {\includegraphics[width=3.1cm, height=3.8cm, angle=270]{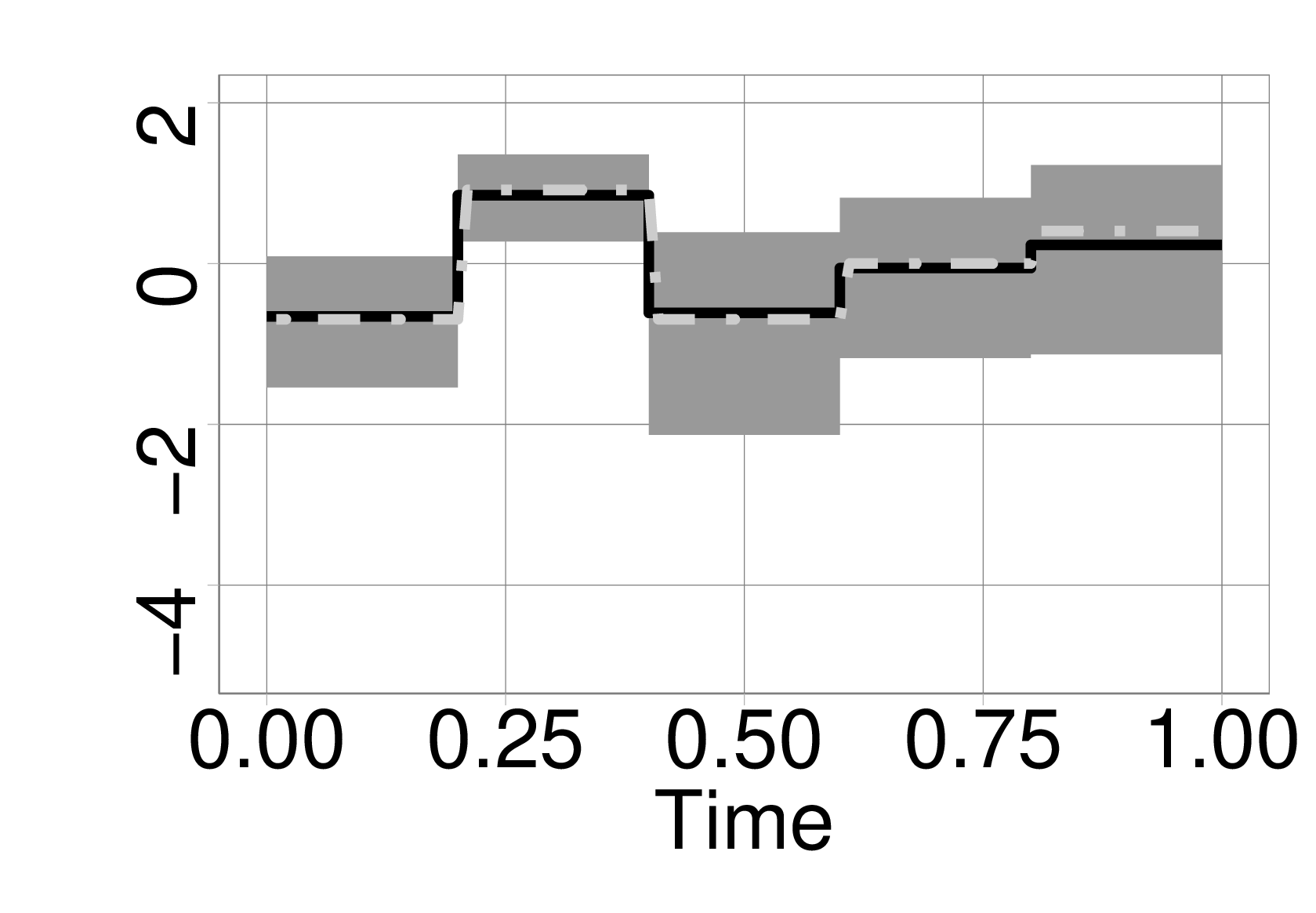}}\hspace*{-0.05cm}
 \subfigure[{\scriptsize \textit{PS2}($K=15$). RMSD=0.289}] {\includegraphics[width=3.1cm, height=3.8cm, angle=270]{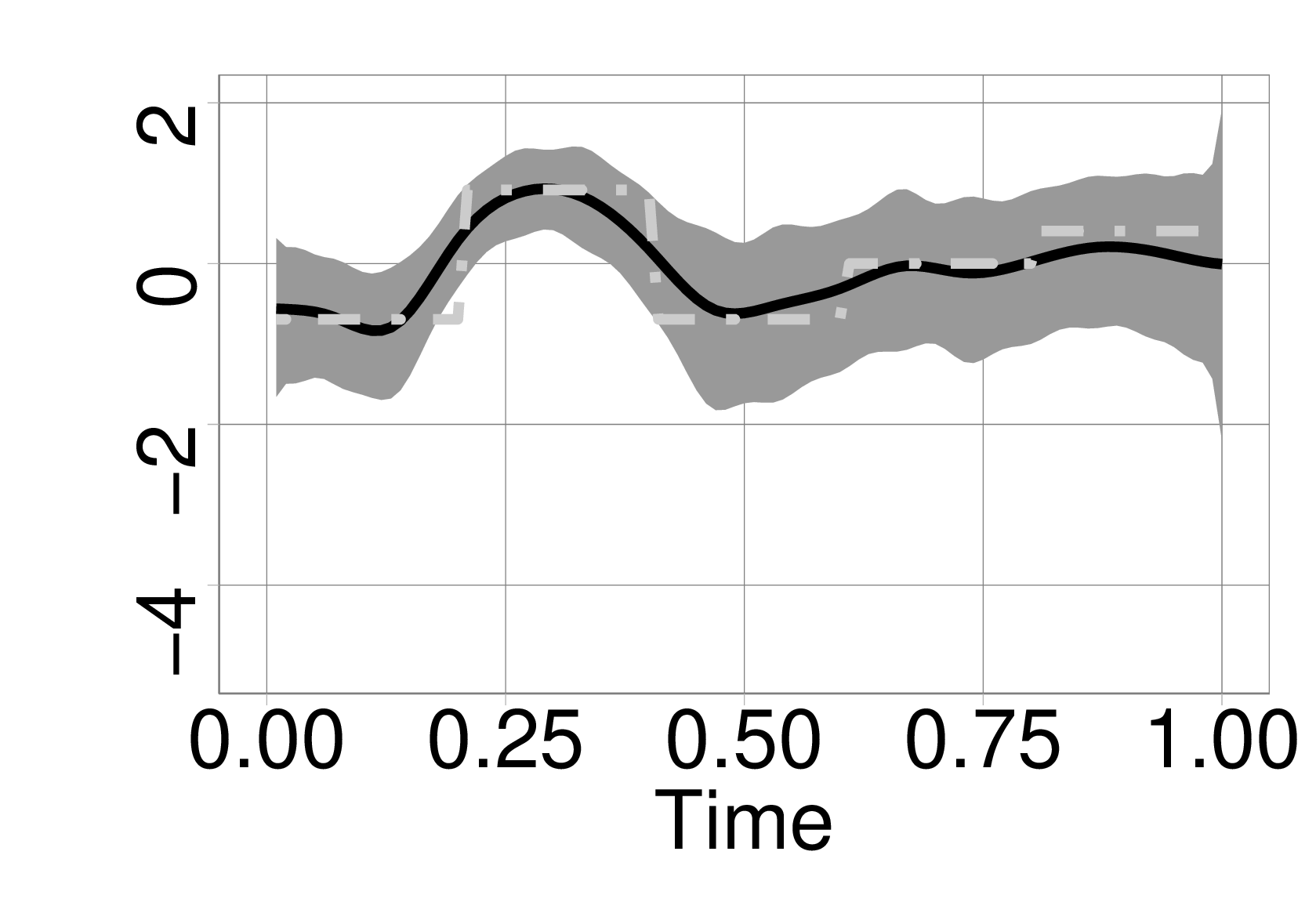}}\\\vspace{-0.30cm}
 \subfigure[{\scriptsize \textit{We}. RMSD=0.626}] {\includegraphics[width=3.1cm, height=3.8cm, angle=270]{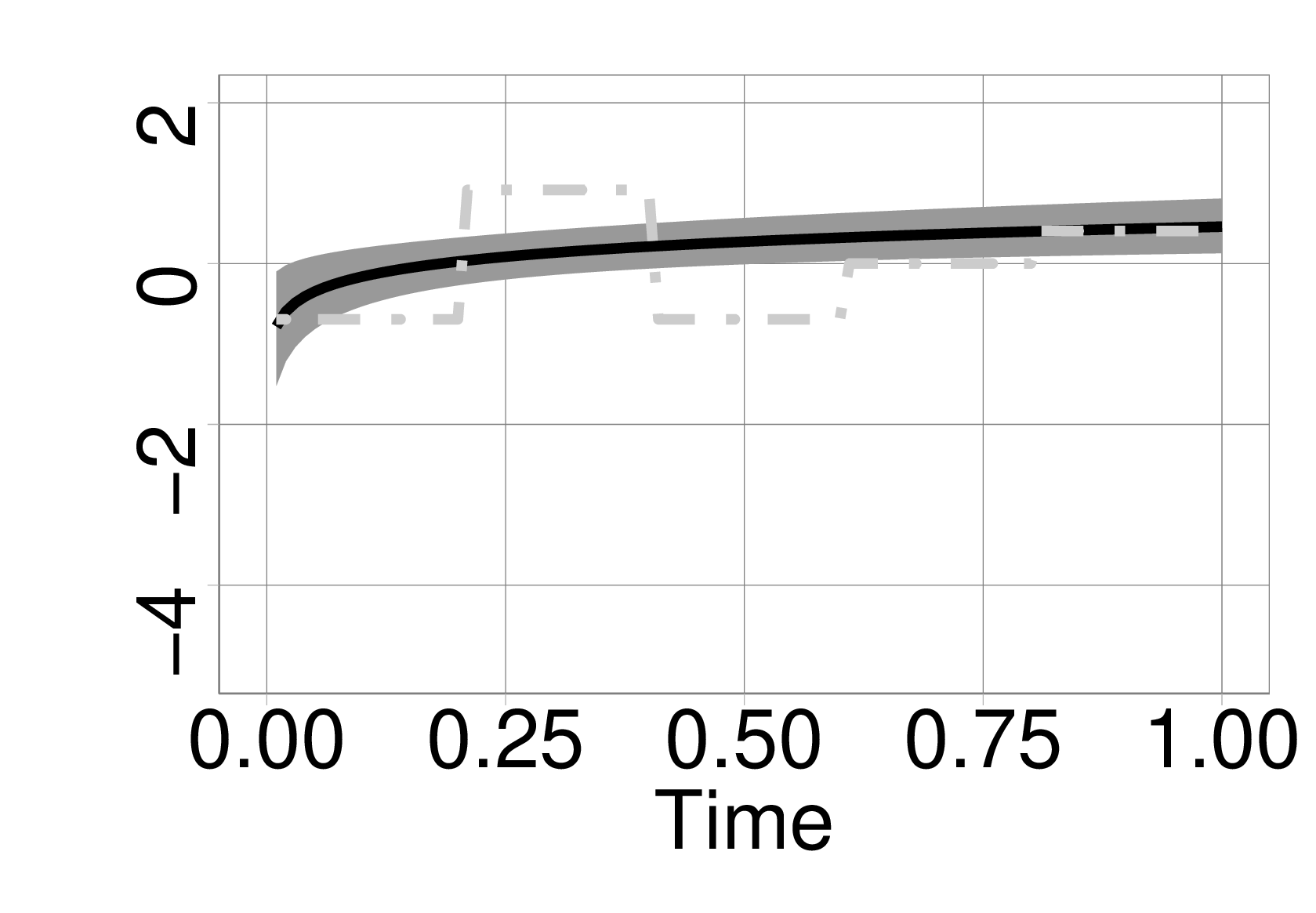}}\hspace*{-0.05cm}
 \subfigure[{\scriptsize \textit{PC3} ($K=5$). RMSD=0.042 }] {\includegraphics[width=3.1cm, height=3.8cm, angle=270]{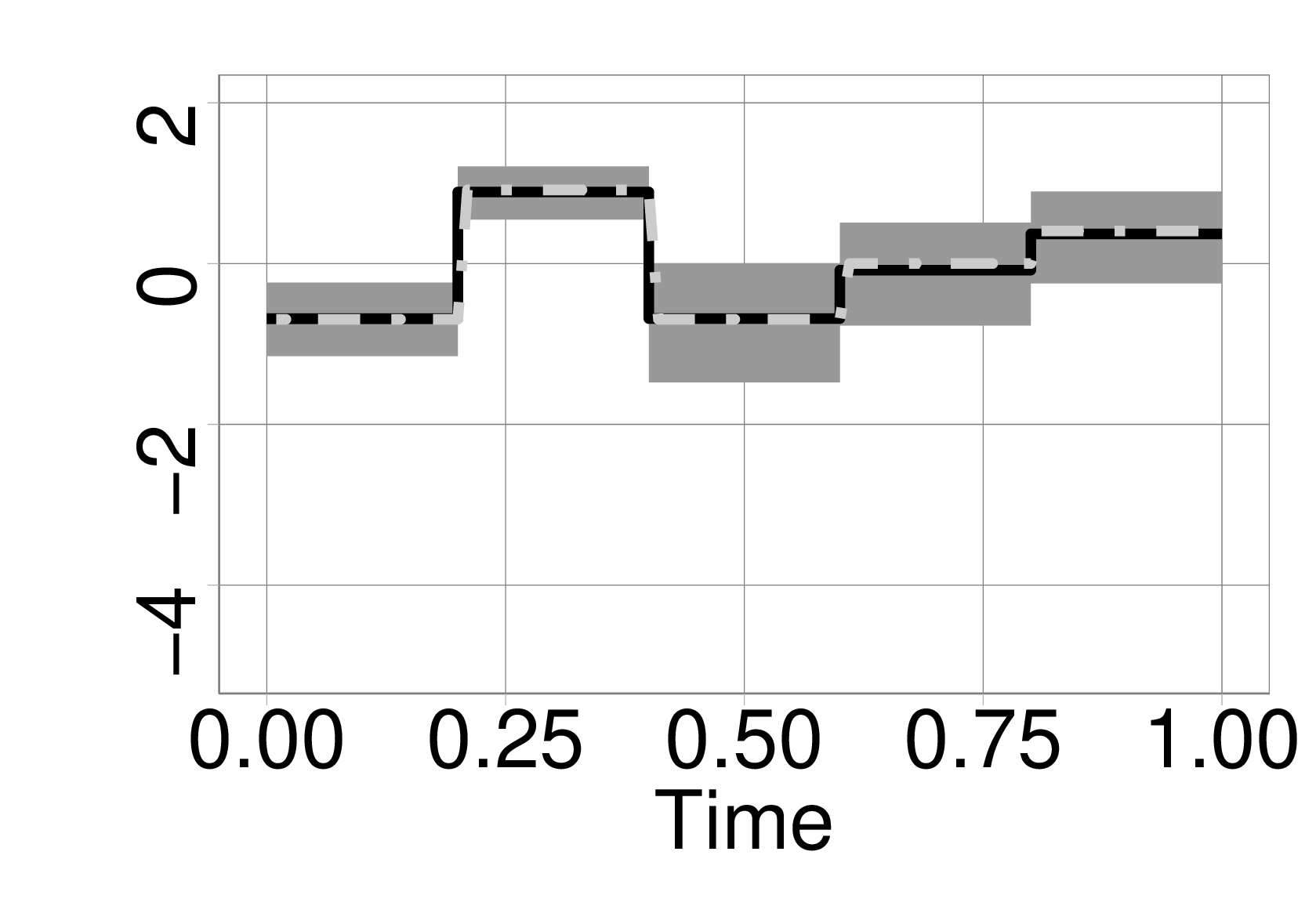}}\hspace*{-0.05cm}
 \subfigure[{\scriptsize\textit{PS2}($K=15$). RMSD=0.254}] {\includegraphics[width=3.1cm, height=3.8cm, angle=270]{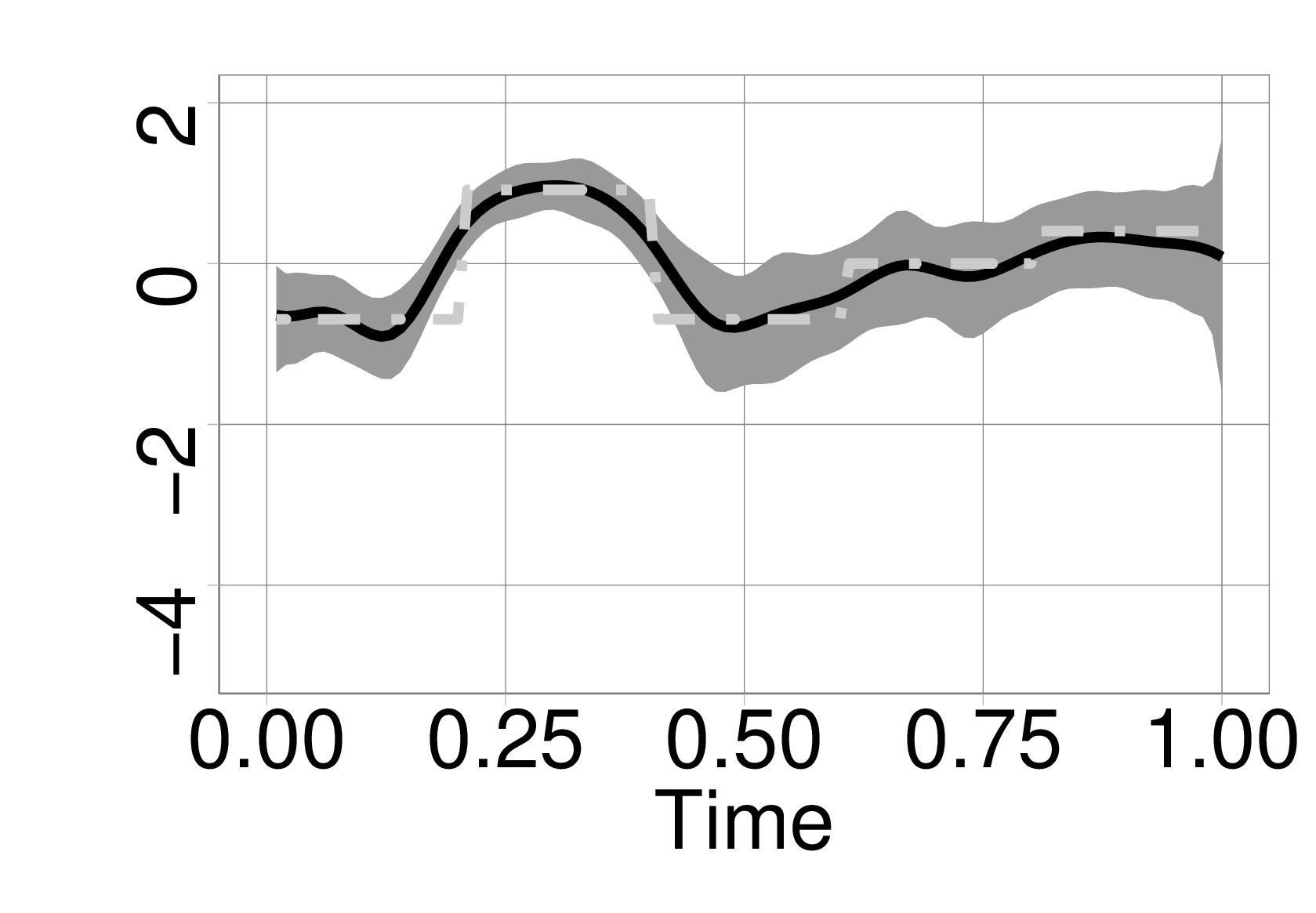}}\\\vspace{-0.30cm}

 \caption{Average replica pointwise of the posterior approximate means of the log-baseline hazard estimate (black solid line), average replica of the posterior 95\% credible intervals (grey area), and true log-baseline hazard function (grey dash-dotted line) in the simulated \textit{Scenario 2} under the $We$, $PC4$ ($K=5$), $PS2$ ($K=15$) for $N$ = 100 (row 1) and under the $We$, $PC3$ ($K=5$), $PS2$ ($K=15$) for $N$ = 300 (row 2).}

 \label{fig:6}
 \end{figure}
\begin{figure}[H]
\centering
 \subfigure[{\scriptsize\textit{We}. RMSD=0.131 }] {\includegraphics[width=3.1cm, height=3.8cm, angle=270]{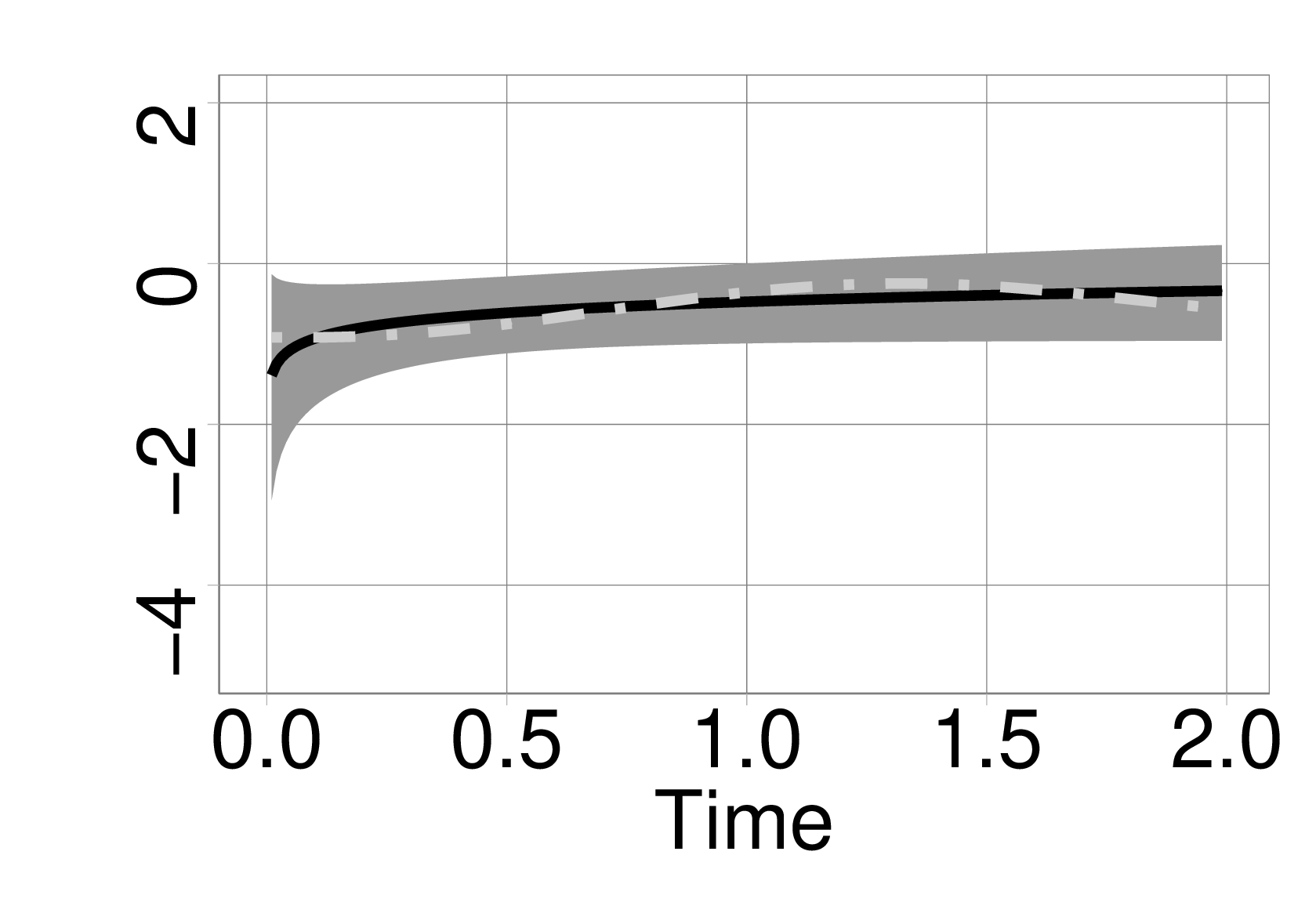}}\hspace*{-0.05cm}
 \subfigure[{\scriptsize \textit{PC4} ($K=5$). RMSD=0.116}] {\includegraphics[width=3.1cm, height=3.8cm, angle=270]{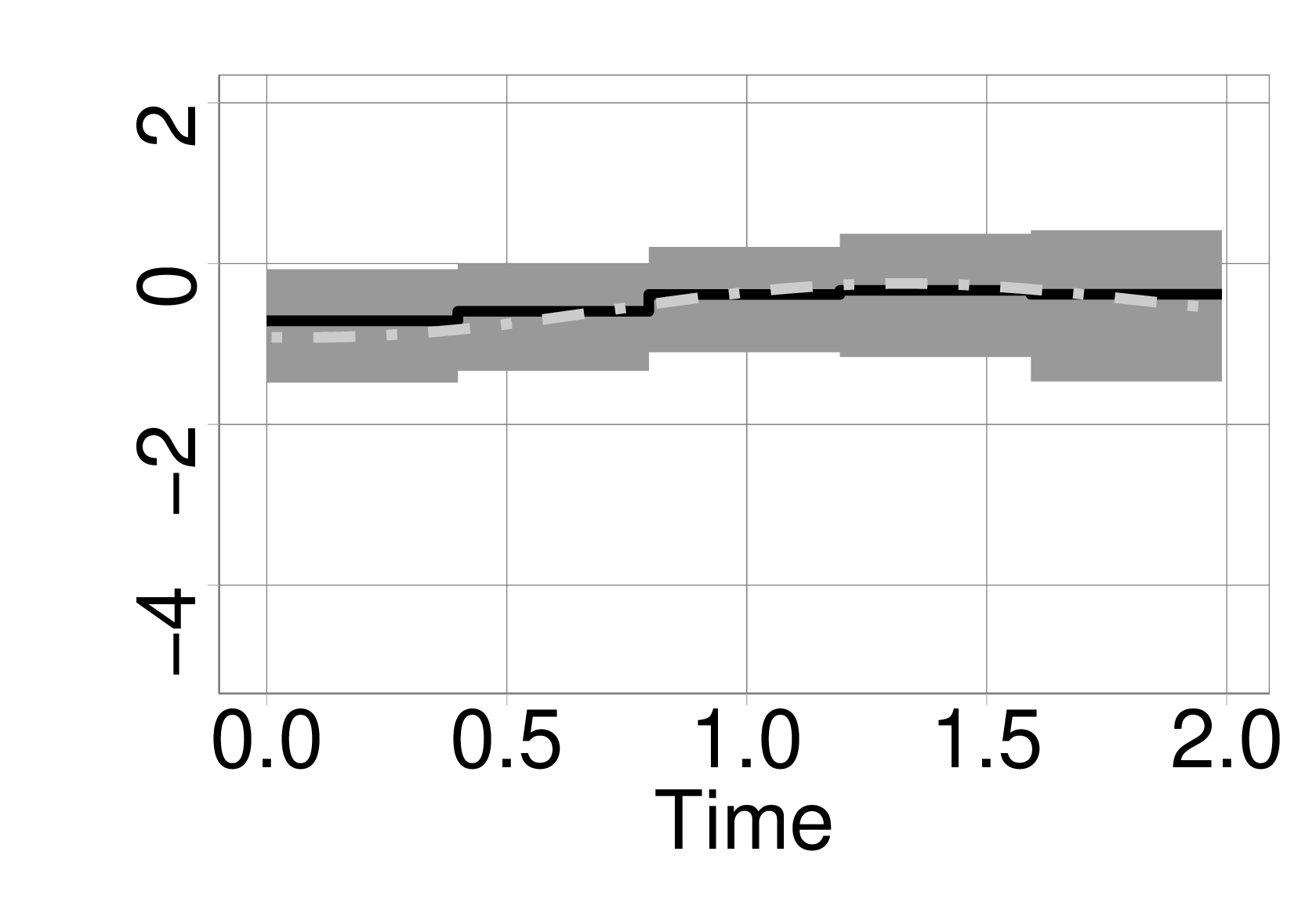}}\hspace*{-0.05cm}
 \subfigure[{\scriptsize \textit{PS3}($K=5$). RMSD=0.074}] {\includegraphics[width=3.1cm, height=3.8cm, angle=270]{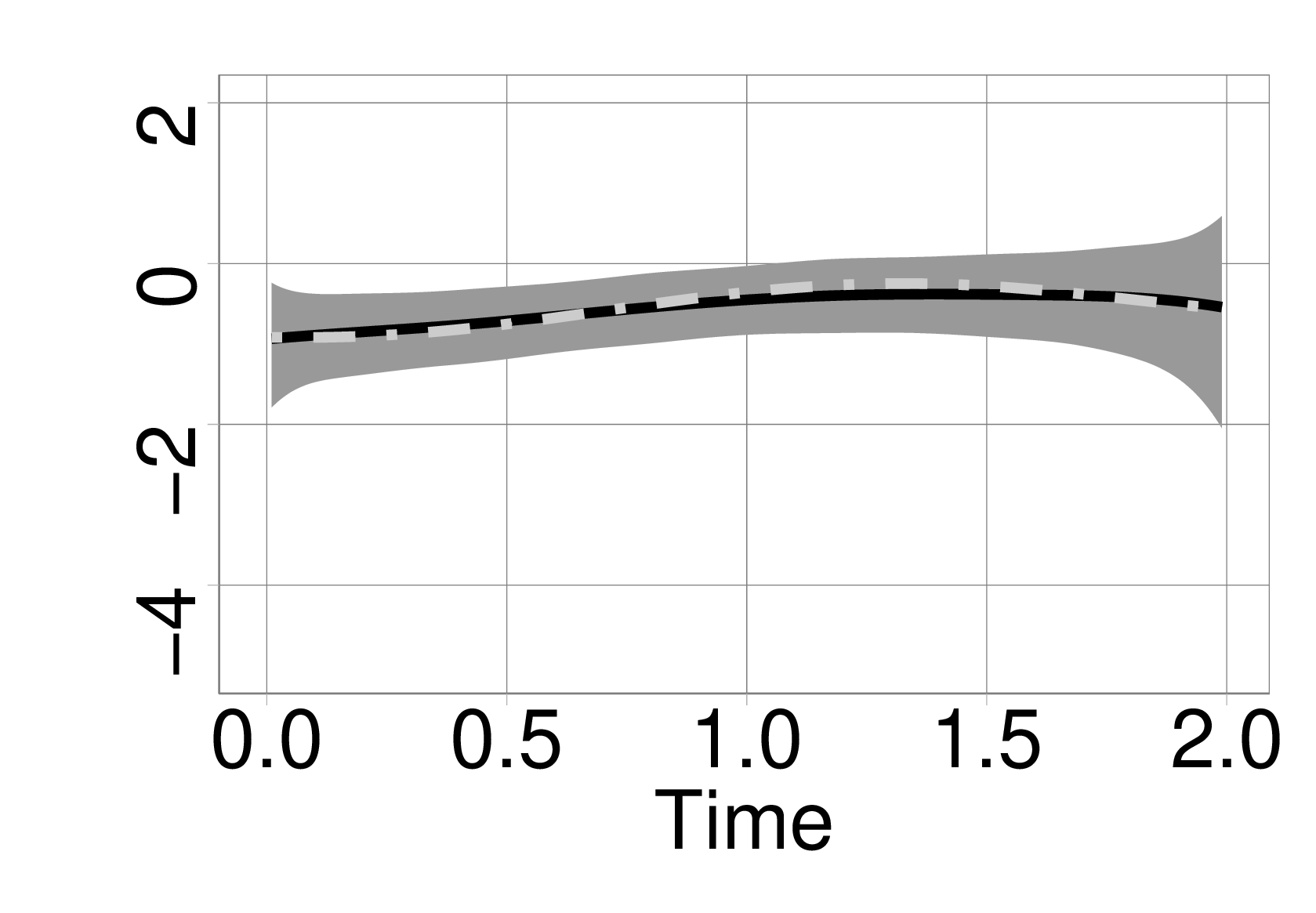}}\\\vspace{-0.30cm}
 \subfigure[{\scriptsize \textit{We}. RMSD=0.132}] {\includegraphics[width=3.1cm, height=3.8cm, angle=270]{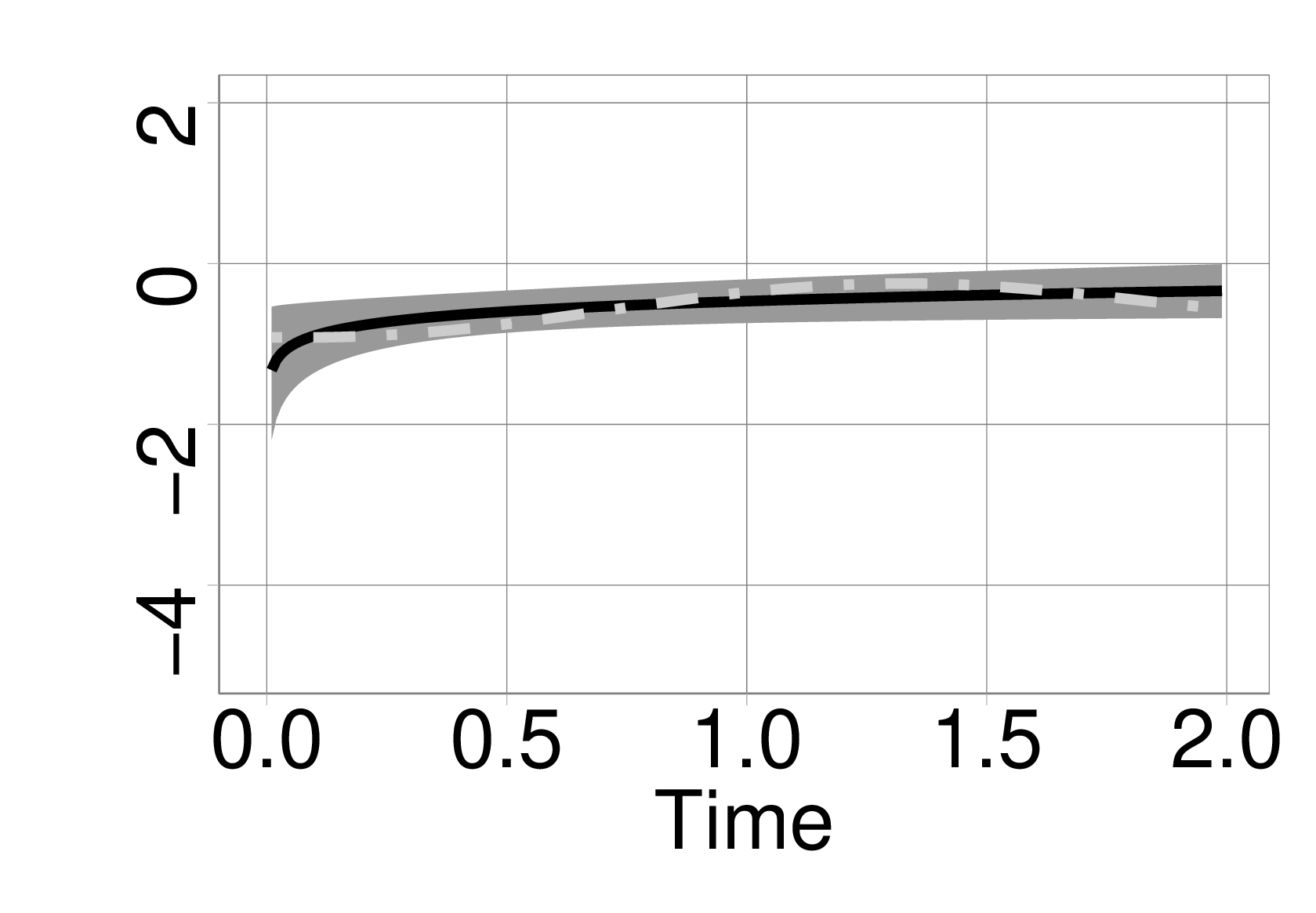}}\hspace*{-0.05cm}
 \subfigure[{\scriptsize \textit{PC1} ($K=5$). RMSD=0.066}] {\includegraphics[width=3.1cm, height=3.8cm, angle=270]{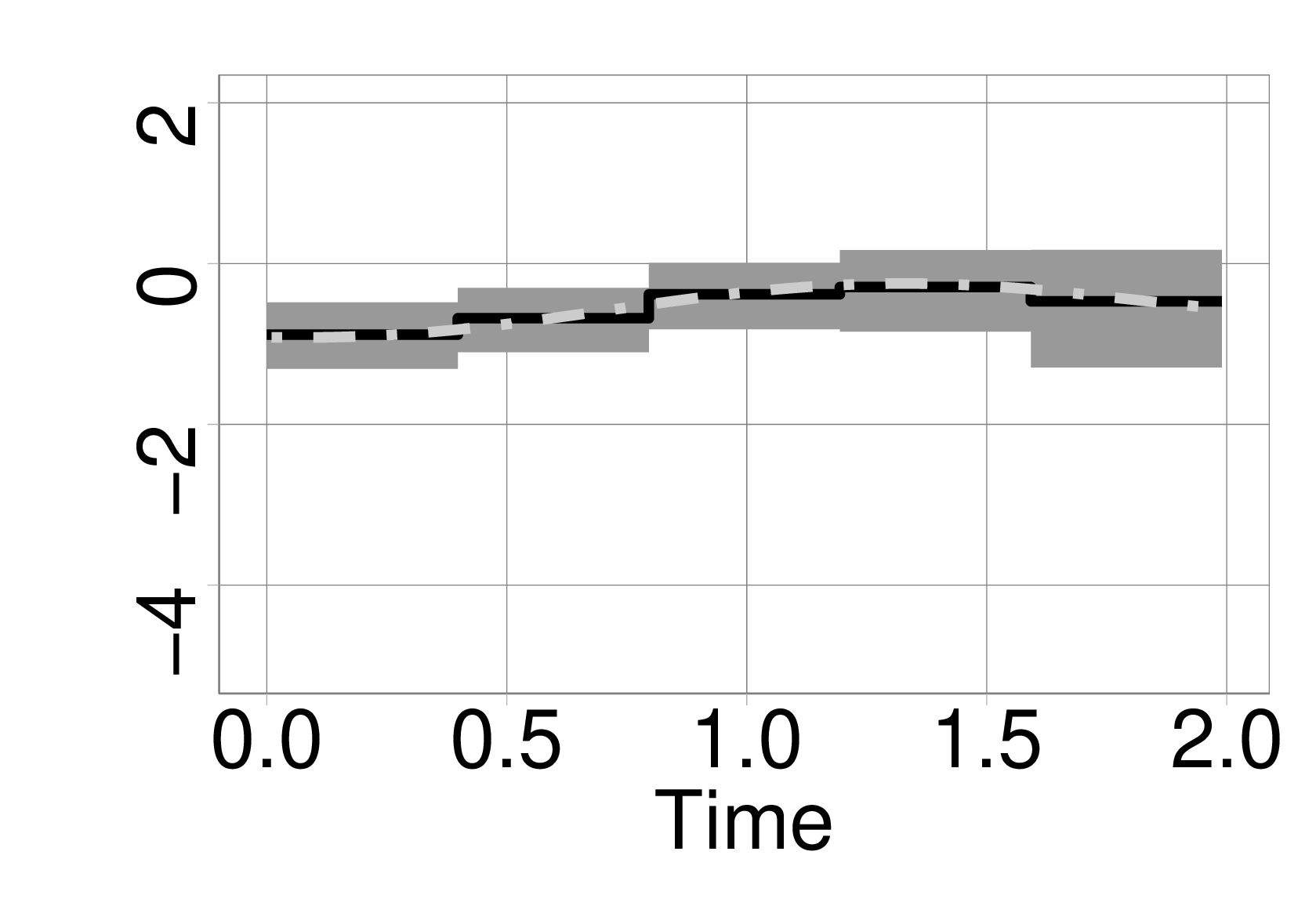}}\hspace*{-0.05cm}
 \subfigure[{\scriptsize \textit{PS3}($K=5$). RMSD=0.043}] {\includegraphics[width=3.1cm, height=3.8cm, angle=270]{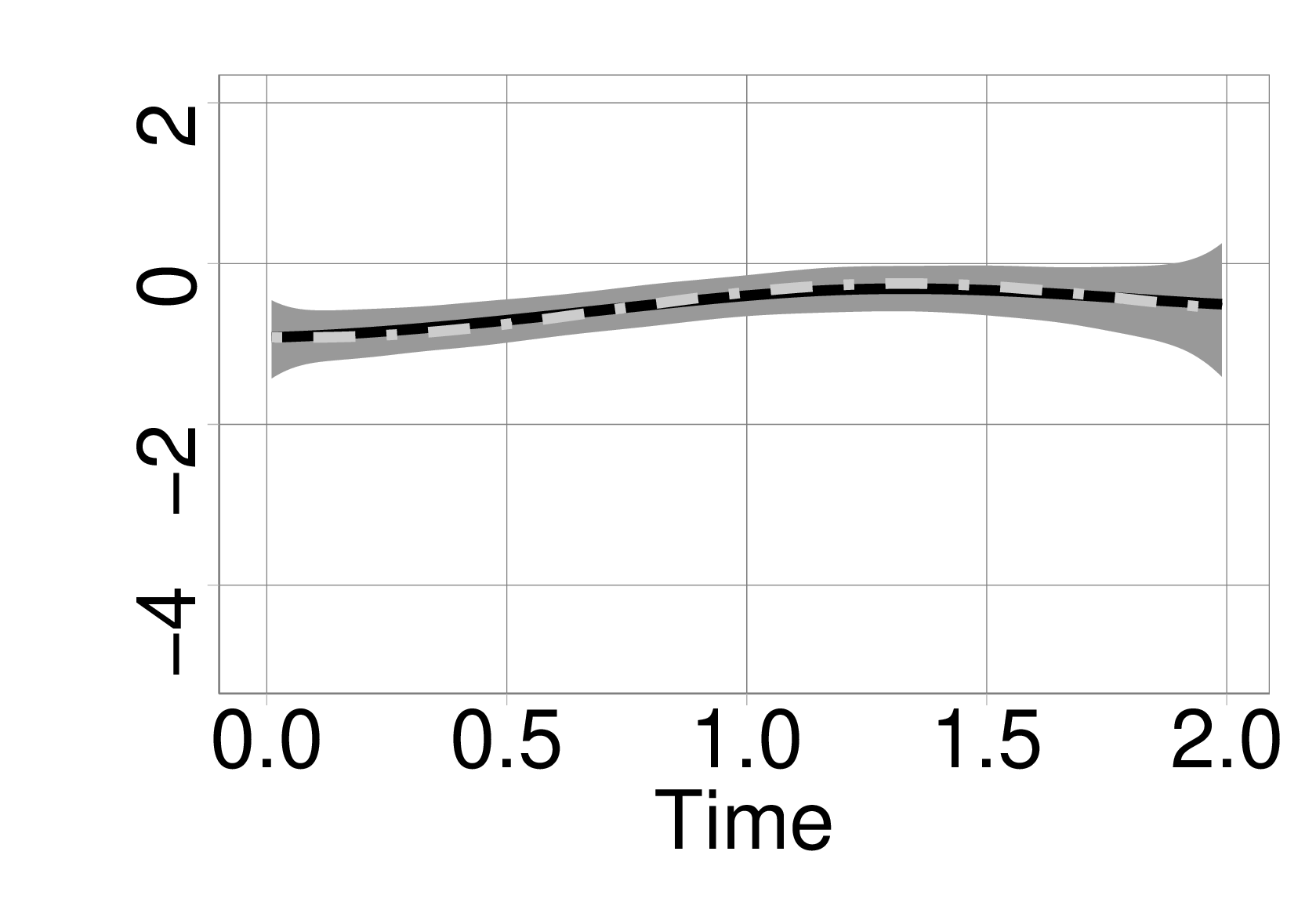}}\\\vspace{-0.30cm}

 \caption{Average replica pointwise of the posterior approximate means of the log-baseline hazard estimate (black solid line), average replica of the posterior 95\% credible intervals (grey area), and true log-baseline hazard function (grey dash-dotted line) in the simulated \textit{Scenario 3} under the $We$, $PC4$ ($K=5$), $PS3$ ($K=5$) for $N$ = 100 (row 1) and under the $We$, $PC1$ ($K=5$), $PS3$ ($K=5$) for $N$ = 300 (row 3).}

 \label{fig:7}
 \end{figure}

Figures~\ref{fig:5}, ~\ref{fig:6} and ~\ref{fig:7}  show the posterior mean of the baseline hazard function and a 95\% credible bound for the best models (based on RMSD criterion) between the three generic $h_{0}(t)$ specifications and for both $N$ values for \textit{Scenario 1}, \textit{Scenario 2} and \textit{Scenario 3}. In general, models under $N$ = 300 present lower RMSD values  than their $N$ = 100 counterparts as well as more accurate baseline hazard estimates (95\% of credible bounds are narrower).

\begin{center}
\textbf{FIGURES 5, 6, 7 AROUND HERE}\\
\end{center}

\section{Conclusions}

We have discussed different proposals for performing a fully time-to-event Bayesian analysis  in the context of the CPH model via parametric and semi-parametric definitions of the baseline hazard function. The Bayesian methodology allows the baseline hazard functions to be implemented in an easy conceptual way, even semi-parametric proposals that are necessary in contexts in which a certain complexity in the shape of the underlying function is expected. On this matter, we have examined some of the most popular proposals in the literature related to the subject: the Weibull distribution  as the most common parametric model, and piecewise constant
and  cubic B-spline baseline hazards as semi-parametric definitions. Flexibility and overfitting were discussed within both semi-parametric options with regard to different regularization schemes expressed in terms of prior distributions and  time axis partition configurations. These developments provide a unified framework to conduct a fully Bayesian analysis of complex survival data that will surely encourage more comprehensive analyses, which currently often rely on some versions of the CPH model without further examination. The flexibility of our approach allows for easy subsequent research on prior sensitivity, different criteria for determining the axis partition of non-parametric proposals and relationships between covariates and baseline hazard functions. Additionally, we have also incorporated a comparison with the frequentist approach to evaluate the performance of both methodologies under the CPH model.

The virulence database in Section 3 illustrates the main goals of this paper.  All inferential processes agree with the conclusions in~\cite{sanzpuig2017} that the cauliflower by-product infusion can be  an alternative preservation treatment. This fact evidences the robustness (regardless of the $h_{0}(t)$ specification) of the Cox model in estimating covariate effects. However, $PC$ models show a certain sensitivity to axis partition in estimating covariate effects. The outcomes also highlight the fact that piecewise constant and B-splines  specifications allow us to capture and introduce (dealing with different axis partition configurations) more flexibility in $h_{0}(t)$. However, piecewise constant options exhibit less flexibility, thus requiring a higher number of $K$ as well as a prior correlation specification to behave in a similar way to B-splines. Hence, in this illustrative example the $PC$ model underlines the efficacy of regularization Bayesian methods (based on defining correlation by means of prior definition) to overcome overfitting and instability in baseline hazard estimation under high $K$ values. In relation to the survival function estimation, this derived quantity shows greater robustness regardless of the baseline hazard specification. Both DIC and LPML reinforce the evidence observed in sensitivity analyses in which \textit{PS} models show  better behaviour than $PC$ models irrespective of the number of pre-fixed knots. Frequentist  methods showed similar performance to the Bayesian in the Cox inferential process within a framework of non-regularization in relation to Weibull and B-spline specification.\\
We have also exemplified our  proposals through different simulated data generated by Weibull, piecewise constant and  mixtures of  Weibull baseline hazard functions. In general, the outcomes indicate that moderate bias can be observed in estimates of the regression coefficient for a treatment effect when  the baseline hazard function specification does not match the origin specification. For baseline hazard estimates,  we appreciate small differences between the true baseline hazard and their point estimates, and lower RMSD values have a close relationship with the data-generating model. In terms of RMSD estimates the Weibull model provides the best results with Weibull  simulated data, although $PS$ models also exhibit good behaviour.  In the case of piecewise constant simulated data, the \textit{PC4} model is the best model, although \textit{PS} models present a very good behaviour in terms of RMSD values. \textit{PS3} models  provide the best estimates for the Weibull mixture data. In relation to the performance of the different number of knot configurations ($K$) explored, it is generally noticeable that PC models require a higher number of $K$ than $PS$ models within the same scenario. Thus, the need for regularization becomes more evident under $PC$ models. In all scenarios, the impact of the database size has generally been evident mainly in the estimation of the baseline hazard function, but has been less evident in the regression coefficient estimate.\\
Although in this article we have extolled the potential of Bayesian inference in dealing with semi-parametric specifications for the baseline hazard in the context of the CPH model, it must be stated that in many settings a simpler distribution may be suitable. However, using a more complex distribution can provide far more realistic  inferences in certain situations. Some interesting issues that  are beyond the scope of this paper deal with introducing uncertainty in the number of knots, including new regularization proposals such as penalized complexity priors, carrying out a sensitivity analysis within each scenario and also exploring in greater depth the performance of the frequentist approach under the ``semi-parametric'' specification of the baseline hazard function.

 \begin{table}[H]
  \centering
    \begin{tabular}{lrccc|lrccc}
    \hline
  \textbf{Model}&    $\boldsymbol{K}$ &\textbf{DIC}  &\textbf{pD}&\textbf{LPML} &   \textbf{Model}&$\boldsymbol{K}$ &\textbf{DIC} &\textbf{pD} &\textbf{LPML}\vspace*{0.05cm}  \vspace*{0.1cm} \\
\hline
\vspace*{-0.3cm}\\
$We$  & - &  4553.309 & 3.960& -2276.334             & & & & & \vspace*{0.1cm} \\
\hline
\vspace*{-0.3cm}\\
 $PC1$ &  5&    4484.455 &7.030 & -2241.921           & $PS1$  &  5&   4460.598 &9.930& -2230.660\\
       & 10&	4478.040 &12.067 & -2238.658           &        & 10&	 4462.866 &14.368& -2231.988    \\
       & 25&	4469.406 &27.313 &-2235.836           &        & 25&	 4462.494 &29.007& -2236.958    \\
       & 40&	4488.393 & 43.036&	-2249.157          &        & 40&	 4419.711 &42.537& -2230.357    \\
$PC2$   & 5&	4484.457 &7.030 & -2241.917           &   $PS2$     & 5&	 4460.024 &9.537& -2230.207    \\
       & 10&	4478.069 &12.081 &	-2238.661           &        & 10&    4462.249&13.831&	-2231.412    \\
       & 25&	4469.371 & 27.295& -2236.586           &        & 25&    4463.873&	26.345&-2233.509    \\
       & 40&   4488.417 &43.047&-2249.814         &        & 40&    4463.732&	38.084 &-2235.947   \\
$PC3$  & 5&	4484.439 &7.021 &-2241.905           &    $PS3$    &5&	4459.578 &8.572& -2229.787    \\
       & 10&		4477.979 & 12.036& -2238.632          &        & 10&    4458.998&10.467&	-2229.443    \\
       & 25&	4469.221 & 27.219& -2235.719           &        & 25&    4460.255&13.471&	-2230.112    \\
       & 40&	4487.049 & 42.356& -2245.979           &        & 40&    4458.403& 15.583&-2229.296    \\
$PC4$ & 5&	4484.445 & 7.014&	-2241.894          &        & &	 &&    \\
       & 10&	4477.070 & 11.508& -2238.193             & & & && \\
       & 25&	4463.265 & 22.566& -2231.649             & & & && \\
       & 40&	4471.340 & 29.782& -2235.798
\vspace*{0.1cm} \\
\hline
\vspace*{-0.3cm}
    \end{tabular}
    \caption{DIC, pD and LPML values for the  survival models defined by means of   Weibull, $PC$ and $PS$ specifications of the baseline hazard function with  number of knots $K$ = 5, 10, 25, and 40.}
\label{tab:1}
\end{table}

\begin{table}[H]
  \centering
    \begin{tabular}{|c|c|c|c|c|c|}
\cline{3-6}    \multicolumn{1}{r}{} &       & \multicolumn{2}{c|}{\textbf{Bayesian approach}} & \multicolumn{2}{c|}{\textbf{Frequentist approach}} \\
    \hline
    \textbf{Model} & $\boldsymbol K$  & \textbf{HR}$\boldsymbol{_{ST1}}$ & \textbf{HR}$\boldsymbol{_{ST3}}$ & \textbf{HR}$\boldsymbol{_{ST1}}$ & \textbf{HR}$\boldsymbol{_{ST3}}$\\
    \hline
    $We$ & --     & 0.640 (0.533, 0.760) & 0.654 (0.546, 0.774) & 0.637 (0.534, 0.760) & 0.652 (0.546, 0.777) \\
    \hline
             & 5     &  0.604 (0.503, 0.722) &  0.619 (0.515, 0.737) & 0.601 (0.503, 0.719) & 0.616 (0.515, 0.736) \\
      $PC1$ & 10    & 0.598 (0.498, 0.712) & 0.615 (0.513, 0.732) & 0.596 (0.498, 0.713) & 0.613 (0.512, 0.733) \\
       & 25    &  0.594 (0.495, 0.707) & 0.607 (0.505, 0.723) & 0.592 (0.495, 0.708) & 0.605 (0.506, 0.723) \\
             & 40    & 0.594 (0.494, 0.708) & 0.608 (0.507, 0.725) & 0.593 (0.496, 0.709) & 0.608 (0.508, 0.727) \\
    \hline
    \multirow{4}[2]{*}{$PS1$} & 5     & 0.596 (0.496, 0.709)  & 0.610 (0.508, 0.725) & 0.593 (0.495, 0.709) & 0.607 (0.508, 0.726) \\
          & 10    &  0.592 (0.493, 0.706) & 0.605 (0.505, 0.719) & 0.593 (0.495, 0.709) & 0.606 (0.506, 0.725) \\
          & 25    & 0.592 (0.493, 0.705) & 0.610 (0.509, 0.725) & 0.592 (0.495, 0.709) & 0.606 (0.507, 0.725) \\
          & 40    & 0.590 (0.491, 0.702) & 0.603 (0.501, 0.719) & 0.592 (0.495, 0.709) & 0.606 (0.507, 0.725) \\
    \hline
    \end{tabular}
\caption{HR$_{ST1}$ and HR$_{ST3}$: posterior mean and 95\% credible interval (Bayesian approach), and estimate and 95\% confidence intervals (Frequentist approach).}
  \label{tab:2}
\end{table}

\begin{table}[htbp]
  \centering
    \begin{tabular}{cccccccc|c}
    \toprule
    \multirow{2}[4]{*}{\textbf{Model}} & \multirow{2}[4]{*}{$\boldsymbol{N}$} & \multirow{2}[4]{*}{$\boldsymbol{K}$} & \multicolumn{5}{c|}{$\boldsymbol{\beta}$}   & \multicolumn{1}{c}{\textbf{log}$\boldsymbol{({h}_0(t))}$} \\
\cmidrule{4-9}          &       &       & \textbf{Average} & \textbf{Bias} & \textbf{SE} & \textbf{SD} & \textbf{CP} & \multicolumn{1}{l}{\textbf{RMSD}} \\
    \midrule
    \multirow{2}[4]{*}{$We$} & 100 & \textbf{--} & 1.035 & 0.035 & 0.230 & 0.211 & 0.97  &0.039 \\
\cmidrule{2-9}          & 300 & \textbf{--} & 1.008 & 0.008 & 0.132 & 0.136 & 0.95  & 0.007 \\
    \midrule
    \multirow{4}[8]{*}{$PC1$} & \multirow{2}[4]{*}{100} & 5 & 1.037 & 0.037 & 0.233 & 0.216 & 0.96  & 0.205 \\
\cmidrule{3-9}          &       & 15 & 1.049 & 0.049&0.234 & 0.216 & 0.97  &2.158 \\
\cmidrule{2-9}          & \multirow{2}[4]{*}{300} & 5 & 1.004 & 0.004 & 0.133 & 0.140 & 0.95  &0.198 \\
\cmidrule{3-9}          &       & 15 & 1.013 & 0.013 & 0.133 & 0.142 & 0.95  & 0.131 \\
    \midrule
    \multirow{4}[8]{*}{$PC2$} & \multirow{2}[4]{*}{100} & 5 & 1.038 &0.038& 0.233 & 0.215& 0.96  &0.205 \\
\cmidrule{3-9}          &       & 15 & 1.051 & 0.051 & 0.234 & 0.216 & 0.97 & 3.607 \\
\cmidrule{2-9}          & \multirow{2}[4]{*}{300} & 5 &1.004 &0.004 &0.133 & 0.140 & 0.96  &0.198 \\
\cmidrule{3-9}          &       & 15 &1.013 & 0.013 & 0.134 &0.141& 0.97  & 0.131 \\
    \midrule
    \multirow{4}[8]{*}{$PC3$} & \multirow{2}[4]{*}{100} & 5 & 1.037 & 0.037 & 0.234 & 0.216 & 0.95 & 0.205 \\
\cmidrule{3-9}          &       & 15 & 1.050 & 0.050 &0.234 & 0.216 & 0.96  & 1.083 \\
\cmidrule{2-9}          & \multirow{2}[4]{*}{300} & 5 & 1.004 &0.004 &0.133 & 0.140& 0.96  & 0.198 \\
\cmidrule{3-9}          &       & 15 & 1.014 &0.014 & 0.134 & 0.142 & 0.97 & 0.130 \\
    \midrule
    \multirow{4}[8]{*}{$PC4$} & \multirow{2}[4]{*}{100} & 5 & 0.946 & -0.054 & 0.234 & 0.210 & 0.97 & 0.212 \\
\cmidrule{3-9}          &       & 15 & 0.882 & -0.118 & 0.233 & 0.203 &0.96  & 0.206 \\
\cmidrule{2-9}          & \multirow{2}[4]{*}{300} & 5 & 0.970& -0.030 & 0.134 & 0.140 & 0.96 & 0.204 \\
\cmidrule{3-9}          &       & 15 & 0.944 &-0.056 & 0.133 & 0.139 & 0.93  & 0.145 \\
    \midrule
    \multirow{4}[8]{*}{$PS1$} & \multirow{2}[4]{*}{100} & 5 & 1.031 & 0.031 & 0.232 & 0.211 & 0.98  & 0.117 \\
\cmidrule{3-9}          &       & 15 & 0.996 & -0.004 & 0.228 & 0.203 & 0.97  & 0.205 \\
\cmidrule{2-9}          & \multirow{2}[4]{*}{300} & 5 & 1.010 & 0.010 & 0.133 & 0.140 & 0.95  & 0.063 \\
\cmidrule{3-9}          &       & 15 & 0.994 & -0.006 & 0.132 & 0.137 & 0.96  & 0.120 \\
    \midrule
    \multirow{4}[8]{*}{$PS2$} & \multirow{2}[4]{*}{100} & 5 & 0.925 & -0.075 & 0.231 & 0.205 & 0.96  & 0.095 \\
\cmidrule{3-9}          &       & 15 & 0.788 & -0.212 & 0.225 & 0.189 & 0.88  & 0.201 \\
\cmidrule{2-9}          & \multirow{2}[4]{*}{300} & 5 & 0.967 & -0.033 & 0.133 & 0.139 & 0.95  & 0.064 \\
\cmidrule{3-9}          &       & 15 & 0.902 & -0.098 & 0.131 & 0.134 & 0.86  & 0.116 \\
    \midrule
    \multirow{4}[8]{*}{$PS3$} & \multirow{2}[4]{*}{100} & 5 & 1.027 & 0.027 & 0.233 & 0.210 & 0.97  & 0.096 \\
\cmidrule{3-9}          &       & 15 & 1.023 & 0.023 & 0.234 & 0.209 & 0.97  & 0.121 \\
\cmidrule{2-9}          & \multirow{2}[4]{*}{300} & 5 & 1.007 & 0.007 & 0.134 & 0.140 & 0.97  & 0.071 \\
\cmidrule{3-9}          &       & 15 & 1.005 & 0.005 & 0.134 & 0.140 & 0.97  & 0.089 \\
    \bottomrule
    \end{tabular}%
      \caption{Average, bias, SE, SD and CP  of the regression coefficient $\beta$ and RMSD of the log$({h}_0(t))$ corresponding to all inferential and replicate processes for the \textit{Scenario 1} simulated data.}
  \label{tab:3}%
\end{table}%

\begin{table}[htbp]
  \centering
  \begin{tabular}{cccccccc|c}
    \toprule
    \multirow{2}[4]{*}{\textbf{Model}} & \multirow{2}[4]{*}{$\boldsymbol{N}$} & \multirow{2}[4]{*}{$\boldsymbol{K}$} & \multicolumn{5}{c|}{$\boldsymbol{\beta}$}   & \multicolumn{1}{c}{\textbf{log}$\boldsymbol{({h}_0(t))}$} \\
\cmidrule{4-9}          &       &       & \textbf{Average} & \textbf{Bias} & \textbf{SE} & \textbf{SD} & \textbf{CP} & \multicolumn{1}{l}{\textbf{RMSD}} \\
    \midrule
    \multirow{2}[4]{*}{$We$} & 100   & --    & 1.077 & 0.077 & 0.234 & 0.251 & 0.93  & 0.626 \\
\cmidrule{2-9}          & 300   & --    & 1.074 & 0.074 & 0.133 & 0.163 & 0.88  & 0.626 \\
    \midrule
    \multirow{4}[8]{*}{$PC1$} & \multirow{2}[4]{*}{100} & 5     &1.018 & 0.018 & 0.234 & 0.232 & 0.94  & 0.276 \\
\cmidrule{3-9}          &       & 15    & 1.018 &0.018 & 0.235 &0.229 & 0.95  & 7.933 \\
\cmidrule{2-9}          & \multirow{2}[4]{*}{300} & 5     & 1.012 &0.012& 0.133 & 0.149 &0.95  & 0.058 \\
\cmidrule{3-9}          &       & 15    & 1.013 & 0.013 & 0.134 & 0.150 & 0.92  & 0.889 \\
    \midrule
    \multirow{4}[8]{*}{$PC2$} & \multirow{2}[4]{*}{100} & 5     & 1.018 & 0.018& 0.234 & 0.232 & 0.94 & 0.760\\
\cmidrule{3-9}          &       & 15    & 1.017 & 0.017& 0.235 & 0.229 & 0.95 & 13.085\\
\cmidrule{2-9}          & \multirow{2}[4]{*}{300} & 5    & 1.011 & 0.011& 0.133 & 0.149 & 0.94 & 0.058\\
\cmidrule{3-9}          &       & 15    & 1.013 & 0.013& 0.134 & 0.151 & 0.92 & 1.291\\
    \midrule
    \multirow{4}[8]{*}{$PC3$} & \multirow{2}[4]{*}{100} & 5     & 1.017 & 0.017& 0.233 & 0.232 & 0.94 & 0.345\\
\cmidrule{3-9}          &       & 15   & 1.017 & 0.017& 0.235 &0.229 & 0.95 & 4.381\\
\cmidrule{2-9}          & \multirow{2}[4]{*}{300} & 5     & 1.012 & 0.012& 0.134 & 0.149 & 0.94 & 0.058\\
\cmidrule{3-9}          &       & 15    & 1.013 & 0.013& 0.134 &0.150 & 0.92 & 0.276\\
    \midrule
    \multirow{4}[8]{*}{$PC4$} & \multirow{2}[4]{*}{100} & 5    & 1.001 & 0.001& 0.230 & 0.226 & 0.94 & 0.095\\
\cmidrule{3-9}          &       & 15    & 0.973 & -0.027& 0.228 & 0.216 &0.95 & 0.202\\
\cmidrule{2-9}          & \multirow{2}[4]{*}{300} & 5   & 1.006 & 0.006& 0.133 & 0.148 & 0.94 & 0.042\\
\cmidrule{3-9}          &       & 15    & 0.996 & -0.004& 0.133 & 0.147 & 0.92 & 0.102\\
    \midrule
    \multirow{4}[8]{*}{$PS1$} & \multirow{2}[4]{*}{100} & 5     & 1.012 & 0.012 & 0.233 & 0.225 & 0.94  & 0.421 \\
\cmidrule{3-9}          &       & 15    & 0.992 & -0.008 & 0.231 & 0.223 & 0.95  & 0.402 \\
\cmidrule{2-9}          & \multirow{2}[4]{*}{300} & 5     & 1.013 & 0.013 & 0.134 & 0.150 & 0.92  & 0.387 \\
\cmidrule{3-9}          &       & 15    & 1.003 & 0.003 & 0.133 & 0.147 & 0.94  & 0.303 \\
    \midrule
    \multirow{4}[8]{*}{$PS2$} & \multirow{2}[4]{*}{100} & 5     & 1.001 & 0.001 & 0.226 & 0.211 & 0.96  & 0.405 \\
\cmidrule{3-9}          &       & 15    & 0.975 & -0.025 & 0.214 & 0.190 & 0.97  & 0.289 \\
\cmidrule{2-9}          & \multirow{2}[4]{*}{300} & 5     & 1.008 & 0.008 & 0.132 & 0.147 & 0.92  & 0.386 \\
\cmidrule{3-9}          &       & 15    & 0.993 & -0.007 & 0.128 & 0.137 & 0.94  & 0.254 \\
    \midrule
    \multirow{4}[8]{*}{$PS3$} & \multirow{2}[4]{*}{100} & 5     & 1.018 & 0.018 & 0.234 & 0.229 & 0.94  & 0.424 \\
\cmidrule{3-9}          &       & 15    & 1.015 & 0.015 & 0.235 & 0.229 & 0.94  & 0.305 \\
\cmidrule{2-9}          & \multirow{2}[4]{*}{300} & 5     & 1.014 & 0.014 & 0.134 & 0.151 & 0.92  & 0.388 \\
\cmidrule{3-9}          &       & 15    & 1.012 & 0.012 & 0.134 & 0.150 & 0.92  & 0.261 \\
    \bottomrule
    \end{tabular}%
     \caption{Average, bias, SE, SD and CP  of the regression coefficient $\beta$ and RMSD of the log$({h}_0(t))$ corresponding to all inferential and replicate processes for the \textit{Scenario 2} simulated data.}
  \label{tab:4}%
\end{table}%

\begin{table}[htbp]
  \centering
  \begin{tabular}{cccccccc|c}
    \toprule
    \multirow{2}[4]{*}{\textbf{Model}} & \multirow{2}[4]{*}{$\boldsymbol{N}$} & \multirow{2}[4]{*}{$\boldsymbol{K}$} & \multicolumn{5}{c|}{$\boldsymbol{\beta}$}   & \multicolumn{1}{c}{\textbf{log}$\boldsymbol{({h}_0(t))}$} \\
\cmidrule{4-9}          &       &       & \textbf{Average} & \textbf{Bias} & \textbf{SE} & \textbf{SD} & \textbf{CP} & \multicolumn{1}{l}{\textbf{RMSD}} \\
    \midrule
    \multirow{2}[4]{*}{$We$} & 100   & --    & 0.955 & -0.045 & 0.230 & 0.234 & 0.93  & 0.131 \\
\cmidrule{2-9}          & 300   & --    & 0.960 & -0.040 & 0.131 & 0.119 & 0.94  & 0.132 \\
    \midrule
    \multirow{4}[8]{*}{$PC1$} & \multirow{2}[4]{*}{100} & 5     & 0.983 & -0.017 & 0.234 &0.254 & 0.93  &0.309 \\
\cmidrule{3-9}          &       & 15     & 0.989 & -0.011 & 0.235 & 0.254 & 0.93 & 4.524\\
\cmidrule{2-9}          & \multirow{2}[4]{*}{300} & 5     & 0.979 & -0.021 & 0.133 & 0.120 & 0.95  & 0.066\\
\cmidrule{3-9}          &       & 15     & 0.984 &-0.016 & 0.133 & 0.121 & 0.95  & 0.245\\
    \midrule
    \multirow{4}[8]{*}{$PC2$} & \multirow{2}[4]{*}{100} & 5      & 0.985 &-0.015 & 0.234 & 0.255 & 0.93  & 0.831\\
\cmidrule{3-9}          &       & 15    & 0.992 & -0.008 & 0.235 & 0.255 &0.93  & 7.012\\
\cmidrule{2-9}          & \multirow{2}[4]{*}{300} & 5     & 0.980 & -0.020 & 0.133 & 0.121 & 0.95  & 0.066\\
\cmidrule{3-9}          &       & 15     & 0.984 & -0.016 & 0.133 & 0.122 & 0.96  & 0.313\\
    \midrule
    \multirow{4}[8]{*}{$PC3$} & \multirow{2}[4]{*}{100} & 5     & 0.984 & -0.016 &0.234 & 0.254 & 0.93  & 0.466\\
\cmidrule{3-9}          &       & 15     & 0.991 & -0.009 & 0.235 & 0.255 & 0.94  &3.962\\
\cmidrule{2-9}          & \multirow{2}[4]{*}{300} & 5     & 0.979& -0.021 &0.133 & 0.120 & 0.95  & 0.066\\
\cmidrule{3-9}          &       & 15     &0.984 & -0.016 & 0.133 & 0.122 & 0.96  & 0.102\\
    \midrule
    \multirow{4}[8]{*}{$PC4$} & \multirow{2}[4]{*}{100} & 5      & 0.865 & -0.135 & 0.236 & 0.251 & 0.88  & 0.116\\
\cmidrule{3-9}          &       & 15     & 0.802 & -0.198 & 0.232 & 0.240 & 0.83  & 0.141\\
\cmidrule{2-9}          & \multirow{2}[4]{*}{300} & 5     & 0.938 & -0.062 & 0.133 & 0.121 & 0.94  & 0.077\\
\cmidrule{3-9}          &       & 15    & 0.902 & -0.098 & 0.133 & 0.118 & 0.91  & 0.075\\
    \midrule
    \multirow{4}[8]{*}{$PS1$} & \multirow{2}[4]{*}{100} & 5     & 0.978 & -0.022 & 0.232 & 0.251 & 0.93  & 0.136 \\
\cmidrule{3-9}          &       & 15    & 0.941 & -0.059 & 0.228 & 0.243 & 0.93  & 0.224 \\
\cmidrule{2-9}          & \multirow{2}[4]{*}{300} & 5     & 0.980 & -0.020 & 0.133 & 0.121 & 0.96  & 0.053 \\
\cmidrule{3-9}          &       & 15    & 0.967 & -0.033 & 0.132 & 0.121 & 0.94  & 0.129 \\
    \midrule
    \multirow{4}[8]{*}{$PS2$} & \multirow{2}[4]{*}{100} & 5     & 0.822 & -0.178 & 0.233 & 0.252 & 0.84  & 0.127 \\
\cmidrule{3-9}          &       & 15    & 0.675 & -0.325 & 0.223 & 0.236 & 0.71  & 0.235 \\
\cmidrule{2-9}          & \multirow{2}[4]{*}{300} & 5     & 0.917 & -0.083 & 0.133 & 0.123 & 0.91  & 0.058 \\
\cmidrule{3-9}          &       & 15    & 0.844 & -0.156 & 0.132 & 0.122 & 0.80  & 0.114 \\
    \midrule
    \multirow{4}[8]{*}{$PS3$} & \multirow{2}[4]{*}{100} & 5     & 0.966 & -0.034 & 0.233 & 0.244 & 0.92  & 0.074 \\
\cmidrule{3-9}          &       & 15    & 0.964 & -0.036 & 0.233 & 0.242 & 0.92  & 0.084 \\
\cmidrule{2-9}          & \multirow{2}[4]{*}{300} & 5     & 0.974 & -0.026 & 0.133 & 0.120 & 0.95  & 0.043 \\
\cmidrule{3-9}          &       & 15    & 0.973 & -0.027 & 0.133 & 0.119 & 0.95  & 0.048 \\
    \bottomrule
    \end{tabular}%
    \caption{Average, bias, SE, SD and CP  of the regression coefficient $\beta$ and RMSD of the log$({h}_0(t))$ corresponding to all inferential and replicate processes for the \textit{Scenario 3} simulated data.}
  \label{tab:5}%
\end{table}%

\begin{acknowledgement}
\noindent L\'azaro's work was supported by a predoctoral FPU fellowship (FPU2013/02042) from the Spanish Ministry of Education, Culture and Sports. This research work was funded by grant MTM2016-77501-P from the Spanish Ministry of Economy and Competitiveness co-financed with FEDER funds.

\end{acknowledgement}
\vspace*{1pc}

\noindent {\bf{Conflict of Interest}}

\noindent {\it{The authors have declared no conflict of interest. }}



\bibliographystyle{spr-chicago}
\bibliography{ch4}

\end{document}